%% file: Chapter_MasterFile.tex
\documentclass[a4paper,twoside,nofootinbib]{revtex4}
\usepackage{lmodern}

\usepackage[T1]{fontenc}
\usepackage[utf8]{inputenc}
\usepackage{verbatim}
\usepackage{bm}
\usepackage{amsmath}
\usepackage{amssymb}
\usepackage{cancel}
\usepackage{graphicx}
\PassOptionsToPackage{normalem}{ulem}
\usepackage{ulem}

\makeatletter

\pdfpageheight\paperheight
\pdfpagewidth\paperwidth

\@ifundefined{textcolor}{}
{%
 \definecolor{BLACK}{gray}{0}
 \definecolor{WHITE}{gray}{1}
 \definecolor{RED}{rgb}{1,0,0}
 \definecolor{GREEN}{rgb}{0,1,0}
 \definecolor{BLUE}{rgb}{0,0,1}
 \definecolor{CYAN}{cmyk}{1,0,0,0}
 \definecolor{MAGENTA}{cmyk}{0,1,0,0}
 \definecolor{YELLOW}{cmyk}{0,0,1,0}
}

\usepackage{siunitx}
\usepackage{placeins}

\makeatother

\begin{document}
\global\long\def\ket#1{\left|#1\right\rangle }%

\global\long\def\bra#1{\left\langle #1\right|}%

\global\long\def\braket#1{\left\langle #1\right\rangle }%

\global\long\def\Tr{\text{Tr}}%

\global\long\def\sgn{\text{sgn}}%

\global\long\def\Re{\text{Re}}%

\global\long\def\Im{\text{Im}}%

\title{Twisted bilayer graphene: low-energy physics, electronic and optical
properties}
\author{Gonçalo Catarina$^{1}$, Bruno Amorim$^{2}$, Eduardo V. Castro$^{2,3,4}$,
João M. V. P. Lopes$^{4,5}$, Nuno M. R. Peres$^{6}$ }
\affiliation{$^{1}$QuantaLab, International Iberian Nanotechnology Laboratory
(INL), 4715-330 Braga, Portugal\\
$^{2}$CeFEMA, Instituto Superior Técnico, Universidade de Lisboa,
1049-001 Lisboa, Portugal\\
$^{3}$Beijing Computational Science Research Center, 100084 Beijing,
China\\
$^{4}$Centro de Física das Universidades do Minho e Porto and Departamento
de Física e Astronomia, Faculdade de Ciências, Universidade do Porto,
4169-007 Porto, Portugal\\
$^{5}$Centro de Física das Universidades do Minho e Porto and Departamento
de Engenharia Física, Faculdade de Engenharia, Universidade do Porto,
4200-465 Porto, Portugal\\
$^{6}$Centro de Física das Universidades do Minho e Porto and Departamento
de Física and QuantaLab, Universidade do Minho, Campus de Gualtar,
4710-057 Braga, Portugal}
\begin{abstract}
Van der Waals (vdW) heterostructures —formed by stacking or growing
two-dimensional (2D) crystals on top of each other— have emerged as
a new promising route to tailor and engineer the properties of 2D
materials. Twisted bilayer graphene (tBLG), a simple vdW structure
where the interference between two misaligned graphene lattices leads
to the formation of a moiré pattern, is a test bed to study the effects
of the interaction and misalignment between layers, key players for
determining the electronic properties of these stackings. In this
chapter, we present in a pedagogical way the general theory used to
describe lattice mismatched and misaligned vdW structures. We apply
it to the study of tBLG in the limit of small rotations and see how
the coupling between the two layers leads both to an angle dependent
renormalization of graphene's Fermi velocity and appearance of low-energy
van Hove singularities. The optical response of this system is then
addressed by computing the optical conductivity and the dispersion
relation of tBLG surface plasmon-polaritons.%
\end{abstract}
\maketitle
\input{sections/Section_Intro.tex}

\input{sections/Section_Basics.tex}

\input{sections/Section_tBLG.tex}

\input{sections/Section_Optical.tex}

\input{sections/Section_Conclusions.tex}

\bibliographystyle{plunsrt}

\end{document}

%% file: sections/Section_Intro.tex
\section{Introduction}

\label{section:motivation}

Two-dimensional (2D) crystals are a new family of promising materials,
with graphene being the first and most well known example of this
large class. Having as a common feature their low dimensionality,
2D materials display a plethora of physical properties, ranging from
the insulating to the superconducting, having a high potential for
many technological applications \citep{Geim2009,Butler2013,Das2015}.
Van der Waals (vdW) heterostructures —formed by stacking or growing
2D crystals on top of each other— have emerged as a new promising
route to tailor and engineer the properties of 2D materials \citep{NovoselovNeto2012,Novoselov2016}.
The variety of possible structures generated seems to be practically
unlimited but, at the same time, their behavior is expected to be
hard to predict due to the complexity of the layered structure. In
order to create structures with tailored properties, one must first
be able to model and predict the properties of a given vdW structure.
These are determined not only by the properties of the individual
2D layers, but also by the mutual interaction between them when brought
into close proximity.

The focus of this chapter is on one of the simplest vdW structures,
the twisted bilayer graphene (tBLG): a graphene sheet on top of other
graphene sheet, with a twist angle. By understanding and modeling
the properties of this simple stacking, we are taking a step into
the ultimate goal of understanding and predicting the behavior of
arbitrary vdW heterostructures, which will, in principle, allow us
to create revolutionary new materials with tailored properties. We
investigate, within a theoretical framework, the electronic spectrum
reconstruction and the optical response.

\begin{figure}
\centering{}\includegraphics[width=0.6\textwidth]{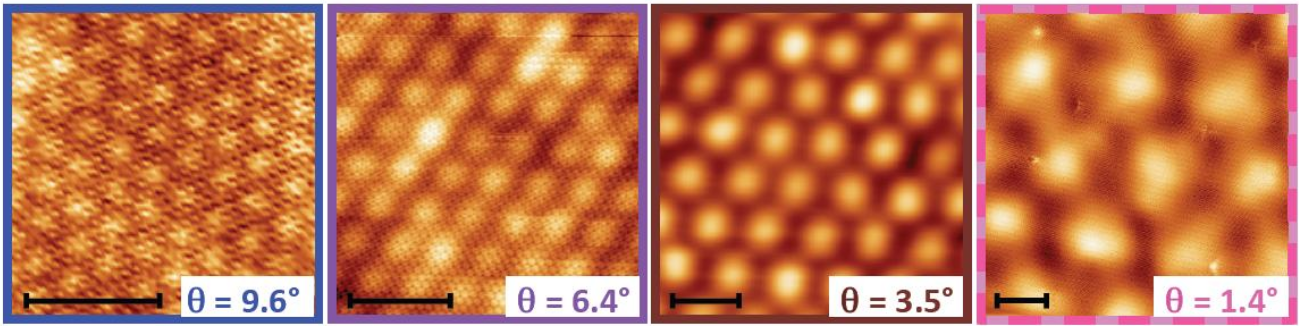}
\caption{Scanning tunneling microscope images of tBLG moiré patterns. All scale
bars are $5\,\text{nm}$. Source: Ref. \citep{Brihuega2012}.}
\label{fig:moire_exp}
\end{figure}

The complex geometry of the tBLG affects significantly its electronic
properties, making even the single-particle models quite involved.
Before moving onto a review of these models, we thus devote some attention
to the crystal structure. The twist angle, $\theta$, between one
graphene layer with respect to the other gives origin to a competition
between different periodicities of the individual layers, which manifests
itself in the appearance of a moiré pattern that can be visualized
experimentally (Fig. \ref{fig:moire_exp}). This pattern displays
a periodicity (or quasiperiodicity), forming a lattice, which is referred
to as moiré superlattice, with a large multiatomic supercell. While
the moiré pattern exists for any $\theta$, a strictly periodic commensurate
superstructure only occurs for the so-called commensurate angles.
Commensurate angles are given by the expression \citep{Santos2012}
\begin{equation}
\cos(\theta)=\frac{3m^{2}+3mr+r^{2}/2}{3m^{2}+3mr+r^{2}},\quad0^{\circ}<\theta<30^{\circ},
\end{equation}
where $m$ and $r$ are coprime positive integers.

For commensurate structures, \textit{ab initio} numerical studies
based on density functional theory have been performed \citep{Latil2007,Morell2010,LaissardiereMayouMagaud2012}.
However, since the unit cell of the tBLG superlattice contains a large
number of sites, especially at small $\theta$, these \textit{ab initio}
calculations incur a significant computational cost and are therefore
rather unpractical. To avoid this difficulty, semi-analytical theories
have been developed in order to describe the low-energy electronic
properties of the tBLG. These theories focus mainly on the low-energy
electronic states near the individual layer Dirac cones in a way that
the model Hamiltonian describes Dirac electrons moving in each layer
and hybridized by interlayer hopping. The first low-energy theory
of this kind, which focused on the limit of small misalignment, was
proposed by Lopes dos Santos et al. \citep{Santos2007}, and further
developed in Ref. \citep{Santos2012}. A similar treatment based on
a continuum approximation was done by Bistritzer and MacDonald \citep{Bistritzer2011},
generalizing the method to incommensurate structures. In Ref. \citep{Gail2011},
the authors made further simplifications to these low-energy Hamiltonians
and derived an effective $2\times2$ Hamiltonian, from which analytical
expressions for the electronic spectrum can be obtained. A general
description of incommensurate double layers, formed by any 2D materials
and valid for arbitrary misalignment, was developed in Ref. \citep{Koshino2015}.
This theory reduces to previous ones in the case of tBLG at small
twist angle. More recently, in Ref. \citep{Weckbecker2016}, the authors
proposed a model which is identical to that derived by Bistritzer
and MacDonald, but with a rescaling in the coupling momentum scale,
in better agreement with tight-binding \textit{ab initio} calculations.

The chapter is organized as follows: in section \ref{chapter:model},
we introduce basic concepts related to the theoretical description
of graphene systems. Section \ref{chapter:tBLG} contains the derivation
of a low-energy effective model for the tBLG, which is the starting
point for the remaining work. In section \ref{chapter:opticalresponse},
we compute the optical conductivity within the linear response theory
and apply this result to the study of the spectrum of tBLG surface
plasmon-polaritons. Finally, in section \ref{chapter:conclusions},
we present our main conclusions. %

%% file: sections/Section_Basics.tex
\section{Basics of monolayer and bilayer graphene}

\label{chapter:model}

In this section, we start with a review of the tight-binding model
for single layer graphene (SLG). This allows us to introduce general
concepts and fix notation. We also analyze the description of SLG
within a folded zone scheme, which will provide us a better understanding
of the tBLG system. Finally, we briefly describe the properties of
a particular stacking of bilayer graphene (BLG), the Bernal stacking.
The description of an arbitrary arrangement of BLG, the tBLG, is left
for the next section.

\subsection{Single layer graphene}

\label{section:SLGbasics}

\subsubsection{Lattice geometry}

\label{subsection:SLGgeometry}

\begin{figure}
\centering{}\includegraphics[width=0.4\textwidth]{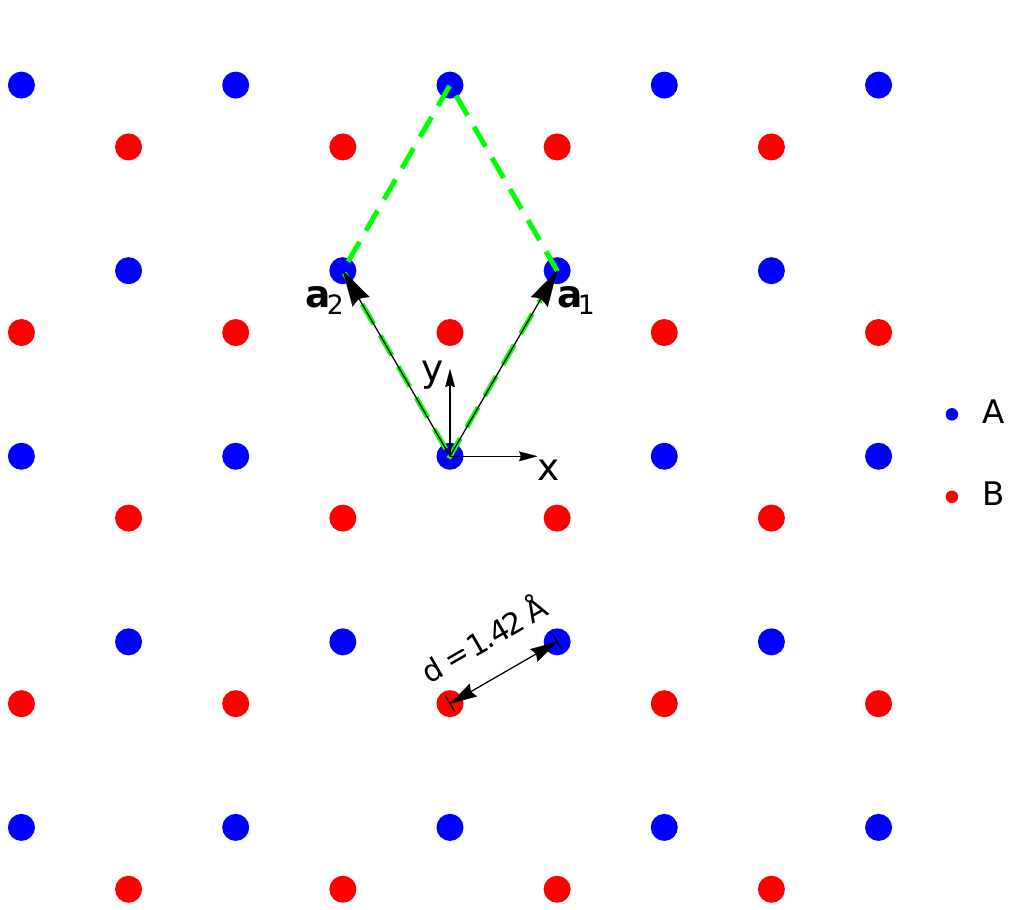}\caption{SLG geometry. The honeycomb structure can be seen as two interpenetrating
hexagonal lattices, $A$ (blue) and $B$ (red). The dashed green line
marks a unit cell of this system, which contains 2 atoms. The coordinate
system is chosen to be centered at a carbon of sublattice $A$.}
\label{fig:SLGgeometry}
\end{figure}

A SLG is a 2D layer made out of carbon atoms arranged into a honeycomb
structure. We choose the coordinate system depicted in Fig. \ref{fig:SLGgeometry},
such that the zig-zag direction is aligned with the $x$-axis and
the armchair direction with the $y$-axis. Each unit cell contains
two carbon atoms that belong to different sublattices, $A$ and $B$.
The unit cells form a hexagonal Bravais lattice $\left\{ \boldsymbol{R}\right\} $,
with positions 
\begin{equation}
\boldsymbol{R}=n_{1}\bm{a}_{1}+n_{2}\bm{a}_{2},\quad n_{1},n_{2}\in\mathbb{Z},\label{eq:SLGlattice}
\end{equation}
where the basis vectors $\bm{a}_{1}$ and $\bm{a}_{2}$ are given
by 
\begin{equation}
\bm{a}_{1}=a\left(1/2,\sqrt{3}/2\right),\quad\bm{a}_{2}=a\left(-1/2,\sqrt{3}/2\right),
\end{equation}
and $a\simeq2.46\si{\angstrom}$ is the lattice parameter \citep{GeimMacDonald2007},
which is related to the carbon-carbon distance, $d$, by $a=\sqrt{3}d$.
The area of the unit cell is 
\begin{equation}
A_{u.c.}=\left|\bm{a}_{1}\times\bm{a}_{2}\right|=\frac{\sqrt{3}}{2}a^{2}.
\end{equation}

We will focus on systems with periodic boundary conditions, $N=\mathcal{N}_{1}\mathcal{N}_{2}$
unit cells (such that $n_{i}=0,1,...,\mathcal{N}_{i}-1$), in the
limit of $\mathcal{N}_{i}\rightarrow\infty$.

\subsubsection{Tight-binding model}

\label{subsection:SLGTB}

We intend to describe the physical properties of a SLG. An isolated
carbon atom has electronic configuration 1s$^{2}$2s$^{2}$2p$^{2}$.
In graphene, from the four outer electrons, three of them are arranged
in a $sp_{2}$ hybridization and form in-plane covalent $\sigma$
bonds between nearest neighbor carbon atoms. The remaining $p_{z}$
electron is delocalized. Most of the electronic properties of graphene
are governed by the delocalized $p_{z}$ electrons. The relevant dynamics
of these electrons can be accurately modeled within a simple single-orbital,
nearest-neighbor tight-binding model \citep{NetoGuineaPeresEtAl2009},
which is the approach we shall also adopt here.

In the tight-binding approximation, we represent the electronic Hamiltonian
in terms of an orthonormal atomic-like basis, the so-called Wannier
states. In the second quantization formalism, a general tight-binding
Hamiltonian can be written as 
\begin{equation}
H=\sum_{\bm{R},\bm{\delta},\alpha,\beta}c_{\alpha}^{\dagger}\left(\bm{R}\right)h_{\bm{\delta}}^{\alpha\beta}c_{\beta}\left(\bm{R}+\bm{\delta}\right).\label{eq:general_TB_Hamiltonian}
\end{equation}
In this expression, $c_{\alpha}^{\dagger}(\bm{R})$ $\left(c_{\alpha}(\bm{R})\right)$
are creation (annihilation) operators for an electron in an atomic-like
state of kind $\alpha$, which is centered at $\bm{R}+\bm{\tau}_{\alpha}$,
where $\bm{R}$ is the position of the unit cell and $\bm{\tau}_{\alpha}$
is the relative position of the orbital center inside the unit cell.
We will focus on spin independent models and therefore we have omitted
the spin degree of freedom. Alternatively, this can be included into
the index $\alpha$. We represent a state created by $c_{\alpha}^{\dagger}(\bm{R})$
as $\left|\bm{R},\alpha\right\rangle $ and we write the orbital in
real space as $w_{\alpha}\left(\bm{r}-\bm{R}-\bm{\tau}_{\alpha}\right)$
(with $\bm{r}$ the position). $h_{\bm{\delta}}^{\alpha\beta}$ are
hopping integrals, given by 
\begin{equation}
h_{\bm{\delta}}^{\alpha\beta}=\left\langle \bm{R},\alpha\right|H\left|\bm{R}+\bm{\delta},\beta\right\rangle ,
\end{equation}
where $\bm{\delta}$ runs over neighboring unit cells. Translational
invariance of the system has been assumed, which is manifest in the
assumption that $h_{\bm{\delta}}^{\alpha\beta}$ is independent of
$\bm{R}$. Due to the localization of the atomic-like orbitals, $h_{\bm{\delta}}^{\alpha\beta}$
decays very fast for large values of $\left|\bm{\delta}\right|$ and,
therefore, we usually need to consider just a few hoppings to describe
the electronic properties of the system.

In the single-orbital tight-binding model for graphene, we have two
kinds of orbitals, the $p_{z}$ orbitals located at the $A$ and $B$
sites ($\alpha=A,B$), which, in the coordinate system of Fig.~\ref{fig:SLGgeometry},
are centered at positions $\bm{\tau}_{A}=(0,0)$ and $\bm{\tau}_{B}=(0,d)$.
In the nearest-neighbor approximation, we only keep the on-site and
nearest-neighbor hoppings, 
\begin{align}
h_{\bm{0}}^{AA} & =h_{\bm{0}}^{BB}\equiv\epsilon_{p_{z}},\\
h_{\bm{\delta}_{NN}}^{AB} & =h_{-\bm{\delta}_{NN}}^{BA}\equiv-t,
\end{align}
where $\bm{\delta}_{NN}$ are the vectors that, for any $A$ site,
link its unit cell to the one of the corresponding nearest-neighbor
$B$ sites, $\bm{\delta}_{NN}=\boldsymbol{0},\,-\bm{a}_{1},\,-\bm{a}_{2}$,
and neglect all other hoppings. According to \textit{ab initio} calculations,
$t=2.97\si{\electronvolt}$ \citep{Reich2002}. Without loss of generality,
we can redefine the zero of energy to coincide with the on-site energy
and therefore set $\epsilon_{p_{z}}=0$. The tight-binding Hamiltonian
for SLG is thus written as 
\begin{equation}
H=-t\sum_{\bm{R}}c_{A}^{\dagger}\left(\bm{R}\right)\left(c_{A}\left(\bm{R}\right)+c_{B}\left(\bm{R}-\bm{a}_{1}\right)+c_{B}\left(\bm{R}-\bm{a}_{2}\right)\right)+\text{h.c.},\label{eq:SLG_tb_Hamiltonian}
\end{equation}
where h.c. stands for hermitian conjugate.

In order to diagonalize the Hamiltonian, we make use of Bloch's theorem.
Bloch's theorem states that, in a periodic system, the electron wavefunction
has the form of a Bloch wave, 
\begin{equation}
\psi_{\bm{k},n}(\bm{r})=e^{i\bm{k}\cdot\bm{r}}u_{\bm{k},n}(\bm{r}),
\end{equation}
where $\bm{k}$ is the crystal or Bloch momentum, $n$ is a band index
and $u_{\bm{k},n}(\bm{r})$ a periodic function with the same periodicity
of the crystal, i.e., $u_{\bm{k},n}(\bm{r})=u_{\bm{k},n}(\bm{r}+\bm{R})$
for all crystal lattice vectors $\bm{R}$. An equivalent statement
of Bloch's theorem is that electronic states in a periodic system
satisfy 
\begin{equation}
\psi_{\bm{k},n}(\bm{r}+\bm{R})=e^{i\bm{k}\cdot\bm{R}}\psi_{\bm{k},n}(\bm{r}),\label{eq:Bloch_theorem_2}
\end{equation}
being eigenstates of the lattice translation operator with corresponding
eigenvalue $e^{i\bm{k}\cdot\bm{R}}$. Graphene wavefunctions that
satisfy Bloch's theorem can be written in the localized basis as 
\begin{equation}
\psi_{\bm{k},\alpha}(\bm{r})=\frac{1}{\sqrt{N}}\sum_{\bm{R}}e^{i\bm{k}\cdot\left(\bm{R}+\bm{\tau}_{\alpha}\right)}w_{\alpha}\left(\bm{r}-\bm{R}-\bm{\tau}_{\alpha}\right),
\end{equation}
or, in bra-ket notation,
\begin{equation}
\left|\psi_{\bm{k},\alpha}\right\rangle =\frac{1}{\sqrt{N}}\sum_{\bm{R}}e^{i\bm{k}\cdot\left(\bm{R}+\bm{\tau}_{\alpha}\right)}\left|\bm{R},\alpha\right\rangle .\label{eq:SLGpsiket}
\end{equation}

In general, eigenstates will be a superposition of states involving
all atomic-like orbitals. Therefore, we look for eigenstates of the
SLG Hamiltonian, Eq.~(\ref{eq:SLG_tb_Hamiltonian}), in the general
form 
\begin{equation}
\left|\psi_{\bm{k}}\right\rangle =\sum_{\alpha}u_{\alpha}(\bm{k})\left|\psi_{\bm{k},\alpha}\right\rangle ,\label{eq:SLGpsiket2}
\end{equation}
where $u_{\alpha}(\bm{k})$ are complex amplitudes. Note that there
is some arbitrariness in these expressions, since we can drop the
phase $e^{i\bm{k}\cdot\bm{\tau}_{\alpha}}$ in Eq.~(\ref{eq:SLGpsiket})
and include it in the complex amplitudes $u_{\alpha}(\bm{k})$ in
Eq.~(\ref{eq:SLGpsiket2}) \citep{BenaMontambaux2009}. Obviously,
no physical quantity can depend on this choice, but the representation
of operators can. The convention used in Eq.~(\ref{eq:SLGpsiket})
simplifies the representation of the current operator within the tight-binding
model \citep{Indranil2003} and we will therefore stick to it.

From the time-independent single-particle Schrödinger equation, 
\begin{equation}
H\left|\psi\right\rangle =E\left|\psi\right\rangle ,\label{eq:Schrodinger}
\end{equation}
where $E$ is the energy, choosing $\ket{\psi}$ of the form of Eq.~(\ref{eq:SLGpsiket})
and applying the bras $\left\langle \bm{R},A\right|$ and $\left\langle \bm{R},B\right|$
(for any $\bm{R}$), we end up with a closed system of equations that
we conveniently write in a matrix form, 
\begin{equation}
H(\bm{k})\cdot\begin{bmatrix}u_{A}(\bm{k})\\
u_{B}(\bm{k})
\end{bmatrix}=E\begin{bmatrix}u_{A}(\bm{k})\\
u_{B}(\bm{k})
\end{bmatrix},\label{eq:SLGeqsmatrixform}
\end{equation}
where $H(\bm{k})$ is the Hamiltonian in the $\left|\psi_{\bm{k},A}\right\rangle ,\left|\psi_{\bm{k},B}\right\rangle $
basis, 
\begin{equation}
H(\bm{k})=\begin{bmatrix}0 & -tf(\bm{k})\\
-tf^{*}(\bm{k}) & 0
\end{bmatrix},\label{eq:SLGmatrix}
\end{equation}
with 
\begin{equation}
f(\bm{k})=\sum_{i=1}^{3}e^{i\bm{k}\cdot\bm{d}_{i}},\label{eq:f(k)}
\end{equation}
in which $\bm{d}_{1}=\left(\bm{a}_{1}+\bm{a}_{2}\right)/3$, $\bm{d}_{2}=\left(-2\bm{a}_{1}+\bm{a}_{2}\right)/3$,
$\bm{d}_{3}=\left(\bm{a}_{1}-2\bm{a}_{2}\right)/3$ are the positions
of the three nearest neighboring $B$ sites to an $A$ site and $^{*}$
stands for complex conjugate. The eigenvalues of $H(\bm{k})$ are
given by
\begin{equation}
E_{\pm}(\bm{k})=\pm t\sqrt{4\cos\left(\frac{\sqrt{3}}{2}dk_{x}\right)\cos\left(\frac{3}{2}dk_{y}\right)+2\cos\left(\sqrt{3}dk_{x}\right)+3}.\label{eq:SLGeigenvalues}
\end{equation}
This spectrum is represented in Fig.~\ref{fig:SLGkspaceandspectrum}(b).

\subsubsection{Low-energy Dirac Hamiltonian}

\label{subsection:SLGDirac}

If we are only interested in the low-energy properties of graphene,
which are the most relevant experimentally, a simplified Hamiltonian
can be obtained. As we can see in Fig.~\ref{fig:SLGkspaceandspectrum}(b),
the spectrum of SLG is gapless with the two bands touching at the
two inequivalent corners of the Brillouin zone (BZ): the K and K$^{\prime}=-$K
points, with 
\begin{equation}
\text{K}=\frac{4\pi}{3\sqrt{3}d}\left(1,0\right).
\end{equation}

In neutral graphene, we have one $p_{z}$ electron, per carbon atom,
contributing to the electronic structure. Also, we know that we have
as many bands as atoms in the unit cell and that every state gets
filled with two electrons, due to spin degeneracy. Therefore, the
neutral configuration corresponds to the situation where half of the
bands are filled, by increasing order of energy. This implies that,
in neutral graphene, the band $E_{-}(\bm{k})$ is completely full
and the band $E_{+}(\bm{k})$ is empty, with the Fermi level lying
at $E=0$ and intersecting the bands at K and K$^{\prime}$. The physics
of graphene is thus dominated by electronic states close to these
points. Writing the electronic Bloch-momentum as $\bm{k}=\pm\text{K}+\bm{q}$
and Taylor expanding to lowest order in $\bm{q}$, we obtain the low-energy
Hamiltonian 
\begin{equation}
H^{\pm\text{K}}(\bm{q})=\hbar v_{F}\begin{bmatrix}0 & \pm q_{x}-iq_{y}\\
\pm q_{x}+iq_{y} & 0
\end{bmatrix}=\hbar v_{F}\bm{q}\cdot\left(\pm\sigma_{x},\sigma_{y}\right),\label{eq:SLGmatrixDirac}
\end{equation}
where the Fermi velocity, $v_{F}$, is identified as $v_{F}=\frac{3td}{2\hbar}$,
$\hbar$ is the reduced Planck constant, $\sigma_{x}$ and $\sigma_{y}$
are Pauli matrices and the $\pm$ sign indicates the point around
which the expansion is made. This low-energy Hamiltonian is recognized
as a (massless) Dirac Hamiltonian and for this reason the K and K$^{\prime}$
points are called Dirac points.

\subsubsection{Reciprocal space and folded band description}

\label{subsection:SLGkspace}

\begin{figure}
\centering{}\includegraphics[width=0.8\textwidth]{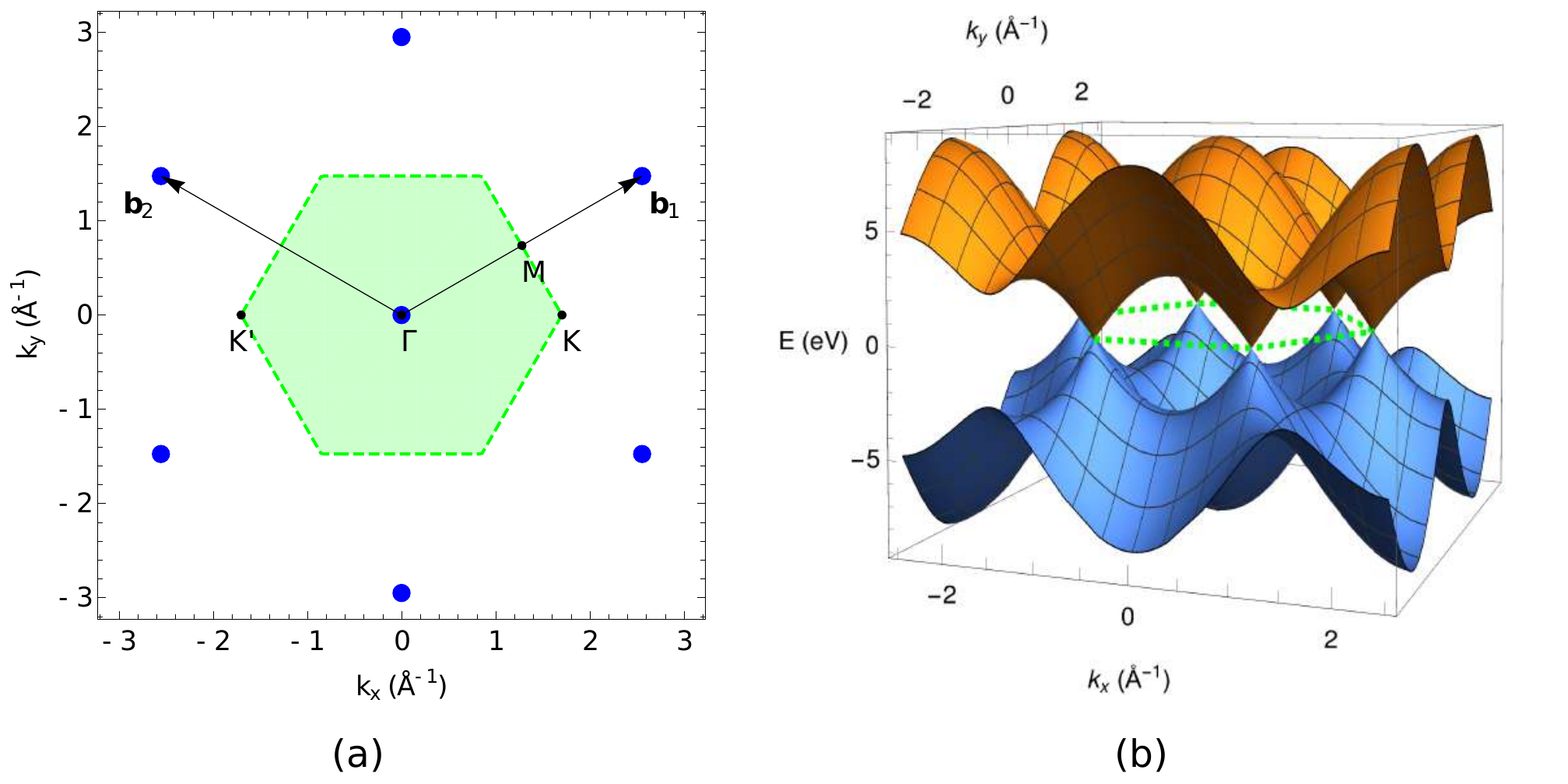}\caption{SLG reciprocal space (a) and electronic spectrum (b). In (a), the
blue circles represent points in the reciprocal lattice; just like
the direct lattice, the reciprocal one is also hexagonal, though rotated
and with a different lattice parameter. The green primitive unit cell
marks the first BZ; some relevant points are represented in it: $\text{\ensuremath{\Gamma}}=(0,0)$,
$\text{M}=\left(1,1/\sqrt{3}\right)\frac{\pi}{a}$, $\text{K}=\left(\frac{4\pi}{3a},0\right)$,
$\text{K}^{\prime}=-\text{K}$. The green dashed line marks the first
BZ boundaries.}
\label{fig:SLGkspaceandspectrum}
\end{figure}

Given the real space direct lattice, Eq.~(\ref{eq:SLGlattice}),
we can define a set of points $\left\{ \bm{G}\right\} $ such that
$e^{i\bm{G}\cdot\bm{R}}=1$. These points also form a lattice, which
is referred to as reciprocal lattice. The points of the reciprocal
lattice $\left\{ \bm{G}\right\} $ can be written in terms of a basis
as 
\begin{equation}
\bm{G}=m_{1}\bm{b}_{1}+m_{2}\bm{b}_{2},\quad m_{1},m_{2}\in\mathbb{Z},
\end{equation}
where the reciprocal lattice basis vectors $\bm{b}_{1}$ and $\bm{b}_{2}$
obey, by definition, the relation 
\begin{equation}
\bm{a}_{i}\cdot\bm{b}_{j}=2\pi\delta_{i,j}.\label{eq:kspacedef}
\end{equation}
For graphene, this leads to 
\begin{equation}
\bm{b}_{1}=\frac{4\pi}{3d}\left(\sqrt{3}/2,1/2\right),\quad\bm{b}_{2}=\frac{4\pi}{3d}\left(-\sqrt{3}/2,1/2\right).
\end{equation}
The reciprocal lattice for SLG is shown in Fig.~\ref{fig:SLGkspaceandspectrum}(a).

By definition, Bloch states are unchanged under shifts of the crystal-momentum
by a reciprocal lattice vector, $\bm{k}\rightarrow\bm{k}+\bm{G}$.
This means that the electronic properties of a periodic system are
completely characterized if we focus on crystal-momenta that are restricted
to a unit cell in reciprocal space, the first BZ.

\begin{figure}
\centering{}\includegraphics[width=0.8\textwidth]{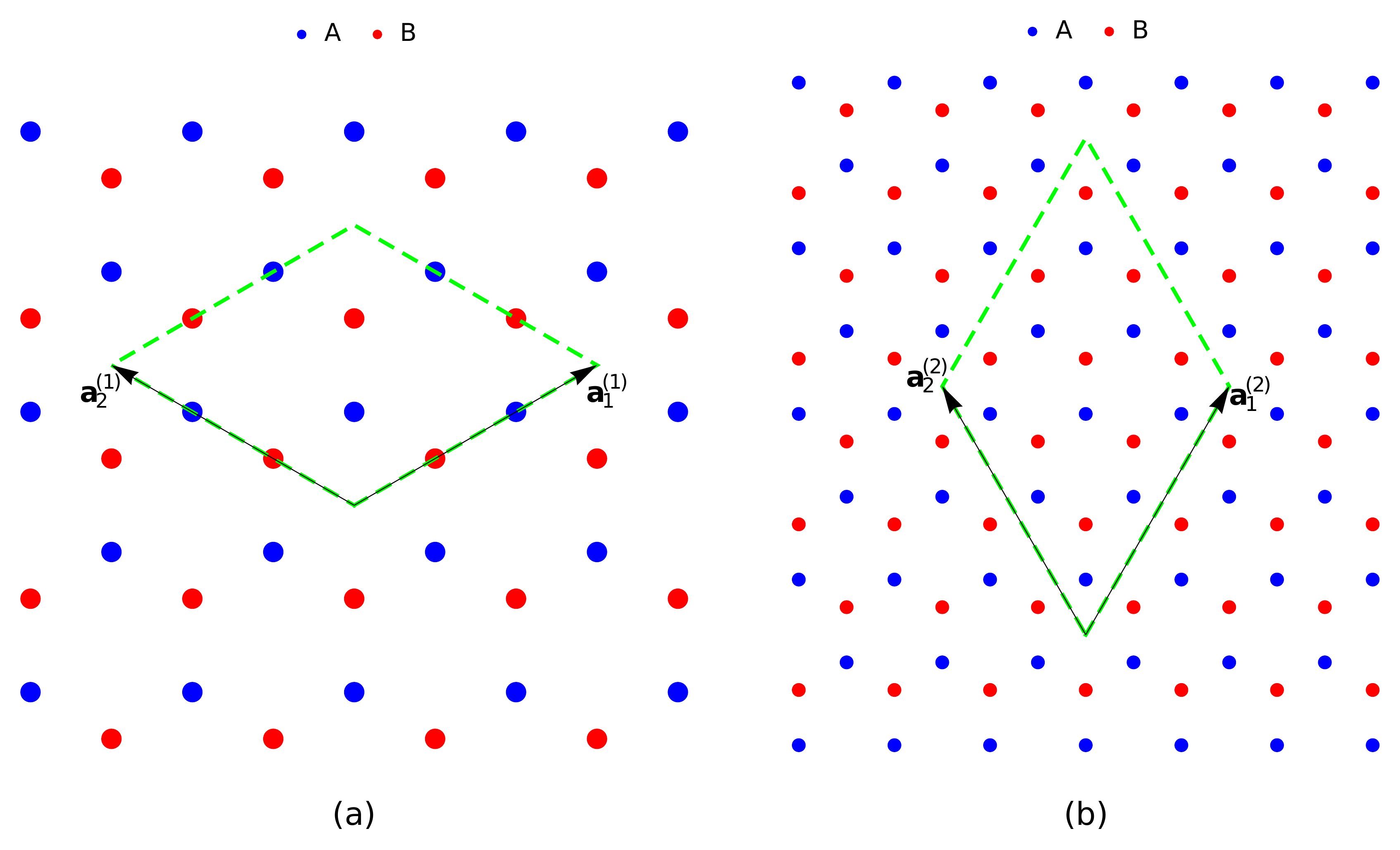}\caption{Basis vectors and unit cells for a folded band description of SLG
with (a) $p=1$ (6-atom unit cell), (b) $p=2$ (18-atom unit cell).}
\label{fig:SLGgeometry_folded}
\end{figure}

We now notice that, for the geometry described in Fig.~\ref{fig:SLGgeometry},
we are free to pick a larger unit cell, with a corresponding smaller
BZ, provided that this cell captures the periodicity of the system.
As an example, we consider unit cells with the shape of a rhombus
containing $2\times3^{p}$ ($p\in\mathbb{N}$) carbon atoms. The basis
vectors for the corresponding lattice are given by 
\begin{align}
\bm{a}_{1}^{(p)} & =\sqrt{3}^{p+1}d\begin{cases}
\left(1/2,\sqrt{3}/2\right) & \text{if }p\text{ is even}\\
\left(\sqrt{3}/2,1/2\right) & \text{if }p\text{ is odd}
\end{cases},\\
\bm{a}_{2}^{(p)} & =\sqrt{3}^{p+1}d\begin{cases}
\left(-1/2,\sqrt{3}/2\right) & \text{if }p\text{ is even}\\
\left(-\sqrt{3}/2,1/2\right) & \text{if }p\text{ is odd}
\end{cases}.
\end{align}
For $p=0$ we recover the minimal unit cell. The enlarged unit cells
for $p=1$ and $p=2$ are shown in Fig.~\ref{fig:SLGgeometry_folded}.
The corresponding reciprocal lattice vectors are 
\begin{align}
\bm{b}_{1}^{(p)} & =\frac{4\pi}{\sqrt{3}^{p}3d}\begin{cases}
\left(\sqrt{3}/2,1/2\right) & \text{if }p\text{ is even}\\
\left(1/2,\sqrt{3}/2\right) & \text{if }p\text{ is odd}
\end{cases},\\
\bm{b}_{2}^{(p)} & =\frac{4\pi}{\sqrt{3}^{p}3d}\begin{cases}
\left(-\sqrt{3}/2,1/2\right) & \text{if }p\text{ is even}\\
\left(-1/2,\sqrt{3}/2\right) & \text{if }p\text{ is odd}
\end{cases}.
\end{align}
It is apparent that, as the unit cell size and $\left|\bm{a}_{i}^{(p)}\right|$
increase, $\left|\bm{b}_{i}^{(p)}\right|$ and the corresponding BZ
become smaller. At the same time, the number of sublattice sites in
the unit cell increases from $2$ to $2\times3^{p}$, which leads
to $2\times3^{p}$ bands. Since the system being described is always
the same, these additional bands are obtained by folding the original
bands into the smaller BZ.

\begin{figure}
\centering{}\includegraphics[width=0.8\columnwidth]{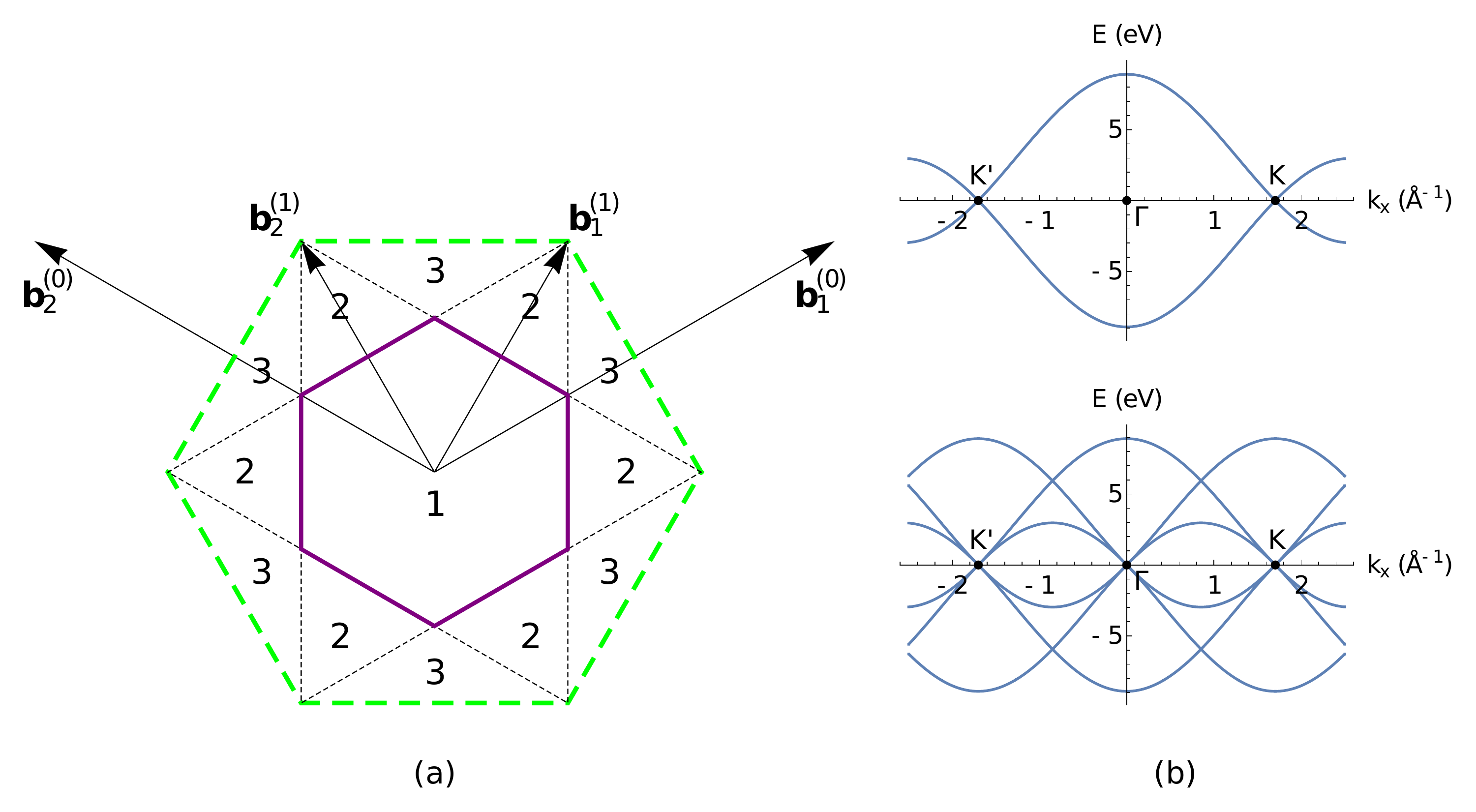}\caption{(a) Reciprocal space folding scheme. The green dashed line marks the
original BZ, while the purple line marks the BZ for a $p=1$ folding.
Regions labeled by $1$, $2$ and $3$ correspond to the first, second
and third BZs for the folded case. (b) Electronic spectrum of SLG
for $p=0$ (top) and $p=1$ (bottom). Plots with $k_{y}=0$.}
\label{fig:SLGfolded_kspacepictureandspectrum}
\end{figure}

We now wish to write the Hamiltonian in reciprocal space for the case
of an enlarged unit cell. We could always rewrite the Hamiltonian
for the larger unit cell in direct space and then follow the same
procedure as in section~\ref{subsection:SLGTB}. However, we will
follow an alternative approach. We expect that it should be possible
to write the folded Hamiltonian directly in reciprocal space in terms
of the Bloch waves defined for the unfolded one. We first discuss
the $p=1$ case. By inspecting Fig.~\ref{fig:SLGfolded_kspacepictureandspectrum}(a),
and according to the previous discussion, we see that, when using
the enlarged unit cell, we are reducing the size of the BZ by $1/3$.
Although the description is different, the overall system is the same.
Hence, the information from regions $2$ and $3$ of the original
BZ must be encoded into the reduced BZ (region $1$). Let us now imagine
that we already have the Hamiltonian for the folded case. Since we
have six atoms per unit cell, we must have six bands. If we then represent
the spectrum using an extended zone scheme —the first two bands in
the first BZ, the second ones in the second BZ and the third ones
in the third BZ— we obtain a spectrum that coincides exactly with
the unfolded one. This provides a way of putting the information from
regions $2$ and $3$ into $1$. We observe that, for each $\bm{k}$
in region $1$, we can get to regions $2$ and $3$ (or equivalent
regions) by translations of $\bm{b}_{1}^{(1)}$ and $\bm{b}_{2}^{(1)}$.

Recalling the unfolded original Hamiltonian, 
\begin{equation}
H_{\bm{k}}^{(0)}=\begin{bmatrix}0 & -tf(\bm{k})\\
-tf^{*}(\bm{k}) & 0
\end{bmatrix},
\end{equation}
we may now write the folded Hamiltonian in the enlarged basis, $\ket{\bm{k}}$,
$\ket{\bm{k}+\bm{b}_{1}^{(1)}}$, $\ket{\bm{k}+\bm{b}_{2}^{(1)}}$,
as 
\begin{equation}
H_{\bm{k}}^{(1)}=\begin{bmatrix}H_{\bm{k}}^{(0)} & 0 & 0\\
0 & H_{\bm{k}+\bm{b}_{1}^{(1)}}^{(0)} & 0\\
0 & 0 & H_{\bm{k}+\bm{b}_{2}^{(1)}}^{(0)}
\end{bmatrix}.
\end{equation}
For a given $p>0$, it is straightforward to generalize and write
\begin{equation}
H_{\bm{k}}^{(p)}=\begin{bmatrix}H_{\bm{k}}^{(p-1)} & 0 & 0\\
0 & H_{\bm{k}+\bm{b}_{1}^{(p)}}^{(p-1)} & 0\\
0 & 0 & H_{\bm{k}+\bm{b}_{2}^{(p)}}^{(p-1)}
\end{bmatrix}.
\end{equation}
Note that inside $H^{(p)}$, we have information of all Hamiltonians
back to the original one, $H^{(0)}$.

In Fig.~\ref{fig:SLGfolded_kspacepictureandspectrum}(b), we plot
the eigenvalues for both the original and $1/3$ folded Hamiltonians.
This construction will be useful to understand the tBLG, as we will
see in section \ref{subsection:tBLGmatrix}.

\subsubsection{Density of states and carrier density profile}

\label{subsection:SLGDOScarrierdensity}

We finish the discussion of the SLG addressing two quantities —the
density of states (DOS) and the carrier density profile— that help
to characterize the electronic structure of the system when doped
with electrons or holes. By definition, the DOS describes the number
of states, per interval of energy, at each energy level, available
to be occupied. As for the carrier density profile, it defines the
relation between the density of carriers $n$ (positive for electrons,
negative for holes) that is needed to reach a given Fermi level $\mu$;
this is a useful quantity since the carrier density is the parameter
well defined in experimental results. Given the electronic spectrum,
both the DOS and the carrier density profile can be calculated in
a straightforward manner.

\begin{figure}
\centering{}\includegraphics[width=0.8\textwidth]{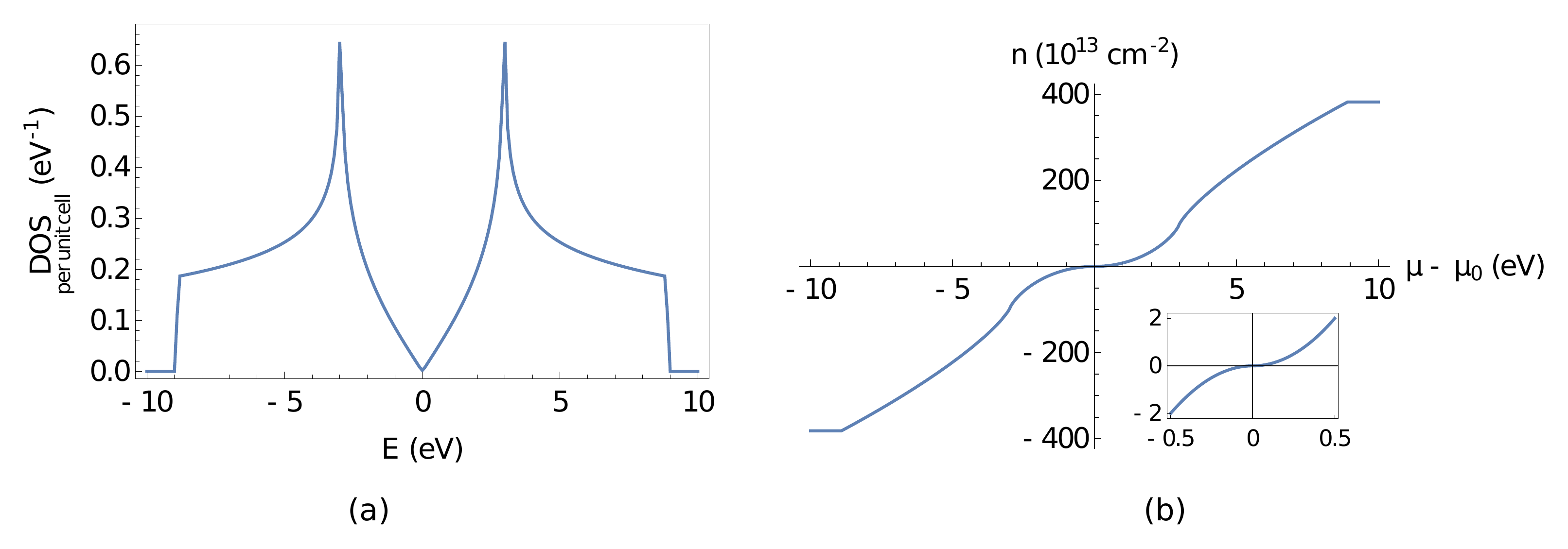}\caption{DOS (a) and carrier density profile (b) for SLG. (a) shows the DOS
per unit cell. In (b), $\mu_{0}$ is the Fermi level for neutral graphene.}
\label{fig:SLGDOScarrierdensity} 
\end{figure}

Results for the DOS and carrier density profile in SLG are presented
in Fig. \ref{fig:SLGDOScarrierdensity}. We first address the carrier
density. Experimentally, record values up to $\left|n\right|\sim4\times10^{14}\si{\per\centi\meter\squared}$
have been reported \citep{Efetov2010}. Nevertheless, under ambient
conditions, typical values for doping are one order of magnitude below
\citep{Das2008,Mak2009}. We will stick within this range, which corresponds
to the zoomed region in Fig. \ref{fig:SLGDOScarrierdensity}(b). As
can be seen from this inset, the corresponding Fermi level is far
away from what is needed to reach the peaks in the DOS —the so-called
van Hove singularities—, making them inaccessible. This is a big downside
since electronic instabilities that can lead to new phases of matter
are expected when we cross a van Hove singularity \citep{Fleck1997,Gonzalez2008,Nandkishore2012}.
One of the reasons that motivates the study of tBLG systems is precisely
the fact that we can bring van Hove singularities to arbitrarily low
energies by varying the twist angle \citep{Li2010}.

\subsection{Introduction to bilayers: Bernal-stacked bilayer graphene}

\label{section:BLG_AB}

\subsubsection{Structure}

\label{subsection:ABgeometry}

\begin{figure}
\centering{}\includegraphics[width=0.4\textwidth]{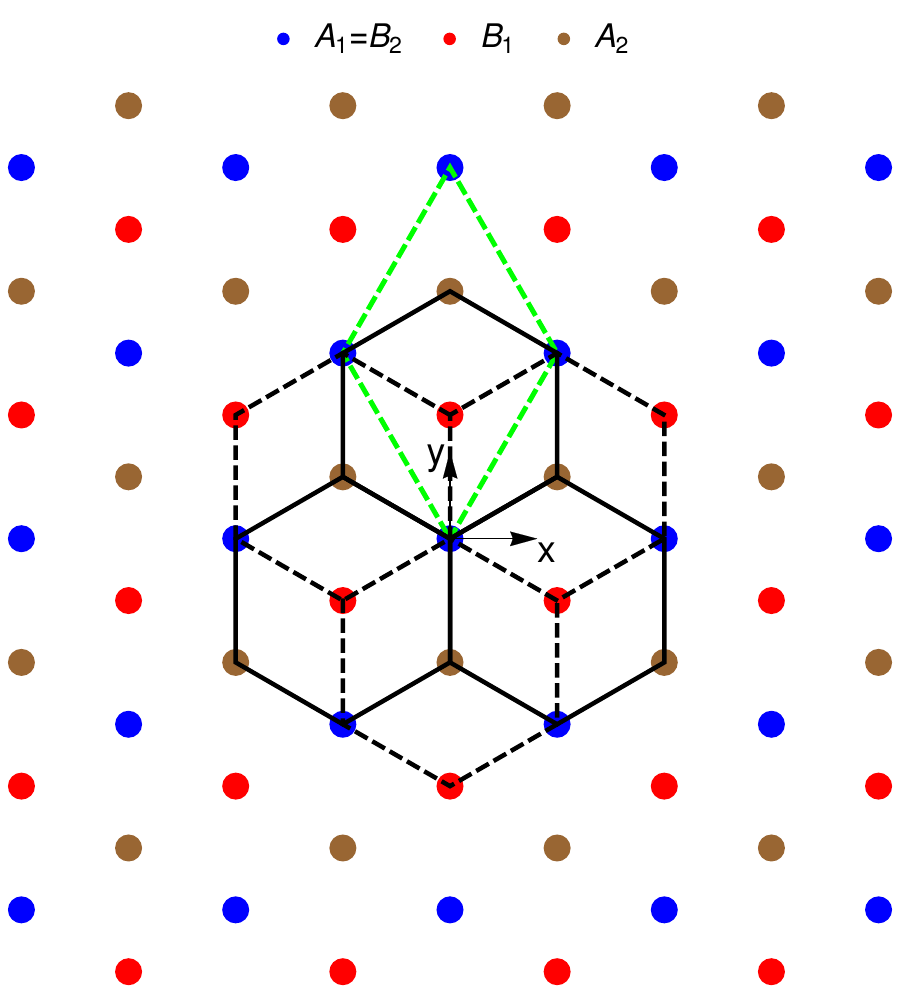}\caption{Bernal-stacked BLG geometry (top view). We label the bottom layer
(dashed black lines) as 1 and the top layer (solid black lines) as
2. The unit cell used for the SLG (green dashed line) is maintained,
keeping both direct and reciprocal space descriptions identical as
before, except that each unit cell now contains four atoms.}
\label{fig:ABgeometry} 
\end{figure}

A BLG is a stacking of two SLGs, where the typical experimental interlayer
distance is $d_{\perp}=3.35\si{\angstrom}$ \citep{RozhkovSboychakovRakhmanovEtAl2015}.
Among the possible stacking arrangements, two are worth pointing out:
1) AA stacking, where each carbon atom from the top layer is placed
exactly above its correspondent in the bottom layer; 2) AB stacking,
or Bernal stacking, which is obtained by sliding one of the layers
with respect to the other along the armchair direction, such that
the atoms of sublattice $A$ from one layer are aligned with the atoms
of sublattice $B$ from the other layer, implying the remaining to
be located in the center of the hexagons (Fig.~\ref{fig:ABgeometry}).
Both AA and AB stacking share the same Bravais lattice structure with
SLG, having the same unit cell. Experimentally, the AA stacking is
considered metastable, while both the Bernal stacking and the tBLG
are found to be stable \citep{RozhkovSboychakovRakhmanovEtAl2015}.
In this section, we analyze the electronic properties of Bernal-stacked
BLG.

\subsubsection{Tight-binding model}

\label{ABmodel}

To model this system, we retain the approximations used before for
each individual layer; in addition, we take into account interlayer
hopping, in a transversal tight-binding approximation between nearest-neighbors.
We start by writing the Hamiltonian for the bilayer as a sum of three
terms, 
\begin{equation}
H=H_{1}+H_{2}+H_{\perp},
\end{equation}
where $H_{\ell}$ is the Hamiltonian for each individual layer $\ell=1,2$,
while $H_{\perp}$ takes into account interlayer coupling. In the
second quantized formalism, using the same approximations as for the
SLG case (Eq.~(\ref{eq:SLG_tb_Hamiltonian})), we obtain
\begin{align}
H_{1} & =-t\sum_{\bm{R}}c_{1,A}^{\dagger}\left(\bm{R}\right)\left[c_{1,B}\left(\bm{R}\right)+c_{1,B}\left(\bm{R}-\bm{a}_{1}\right)+c_{1,B}\left(\bm{R}-\bm{a}_{2}\right)\right]+\text{h.c.},\\
H_{2} & =-t\sum_{\bm{R}}c_{2,A}^{\dagger}\left(\bm{R}\right)\left[c_{2,B}\left(\bm{R}\right)+c_{2,B}\left(\bm{R}-\bm{a}_{1}\right)+c_{2,B}\left(\bm{R}-\bm{a}_{2}\right)\right]+\text{h.c.},
\end{align}
where $c_{\ell,\alpha}^{\dagger}\left(\bm{R}\right)/c_{\ell,\alpha}\left(\bm{R}\right)$
is the creation/annihilation fermionic operator for an electron in
a atomic-like state $\left|\ell,\bm{R},\alpha\right\rangle $ located
at cell $\bm{R}$, sublattice $\alpha$ and layer $\ell$. For $H_{\perp}$,
we consider a homogeneous interlayer hopping, $t_{\perp}$, between
nearest neighbors only, 
\begin{equation}
\left\langle 1,\bm{R},A\right|H_{\perp}\left|2,\bm{R},B\right\rangle =t_{\perp},
\end{equation}
and set $t_{\perp}=0.33\si{\electronvolt}$, which is compatible with
the range of estimated values \citep{RozhkovSboychakovRakhmanovEtAl2015}.
In the second quantized formalism, we can thus write 
\begin{equation}
H_{\perp}=t_{\perp}\sum_{\bm{R}}c_{1,A}^{\dagger}\left(\bm{R}\right)c_{2,B}\left(\bm{R}\right)+\text{h.c.}.
\end{equation}

We now move to the reciprocal space and write the Hamiltonian in terms
of fermionic operators of electronic states of the Bloch form, 
\begin{equation}
\left|\psi_{\ell,\bm{k},\alpha}\right\rangle =\frac{1}{\sqrt{N}}\sum_{\bm{R}}e^{i\bm{k}\cdot\left(\bm{R}+\bm{\tau}_{\ell,\alpha}\right)}\left|\ell,\bm{R},\alpha\right\rangle ,\label{eq:SLGpsiket_bilayer}
\end{equation}
where $\bm{\tau}_{\ell,\alpha}$ are the in-plane positions of the
four carbon atoms in the unit cell, which read as $\bm{\tau}_{1,A}=\bm{\tau}_{2,B}=\left(0,0\right)$
and $\bm{\tau}_{1,B}=-\bm{\tau}_{2,A}=\left(0,d\right)$. The corresponding
creation operators can be written as 
\begin{equation}
c_{\ell,\alpha}^{\dagger}(\bm{k})=\frac{1}{\sqrt{N}}\sum_{\bm{R}}e^{i\bm{k}\cdot\left(\bm{R}+\bm{\tau}_{\ell,\alpha}\right)}c_{\ell,\alpha}^{\dagger}(\bm{R}),\label{eq:operator_dFT}
\end{equation}
which can be understood as a discrete Fourier Transform of the operators
$c_{\ell,\alpha}^{\dagger}(\bm{R})$. Using the property 
\begin{equation}
\sum_{\bm{R}}e^{i\bm{R}\cdot\left(\bm{k}-\bm{k}^{\prime}\right)}=N\sum_{\bm{G}}\delta_{\bm{k}-\bm{k}^{\prime},\bm{G}},
\end{equation}
which for $\bm{k},\bm{k}^{\prime}\in\text{BZ}$ yields $\sum_{\bm{R}}e^{i\bm{R}\cdot\left(\bm{k}-\bm{k}^{\prime}\right)}=N\delta_{\bm{k},\bm{k}^{\prime}}$,
we can invert Eq.~(\ref{eq:operator_dFT}), obtaining 
\begin{equation}
c_{\ell,\alpha}^{\dagger}(\bm{R})=\frac{1}{\sqrt{N}}\sum_{\bm{k}}e^{-i\bm{k}\cdot\left(\bm{R}+\bm{\tau}_{\ell,\alpha}\right)}c_{\ell,\alpha}^{\dagger}(\bm{k}),\label{eq:localized_to_Bloch_operator}
\end{equation}
where the sum (which becomes an integral in the limit of an infinite
crystal) is restricted to the first BZ. Therefore, we can write the
Hamiltonian in a second quantized form as 
\begin{equation}
H=\sum_{\bm{k}}\Psi^{\dagger}(\bm{k})\cdot H(\bm{k})\cdot\Psi(\bm{k}),
\end{equation}
where we have introduced $\Psi^{\dagger}(\bm{k})=\left[\begin{array}{cccc}
c_{1,A}^{\dagger}(\bm{k}) & c_{1,B}^{\dagger}(\bm{k}) & c_{2,A}^{\dagger}(\bm{k}) & c_{2,B}^{\dagger}(\bm{k})\end{array}\right]$ and 
\begin{equation}
H(\bm{k})=\left[\begin{array}{cccc}
0 & -tf(\bm{k}) & 0 & t_{\perp}\\
-tf^{*}(\bm{k}) & 0 & 0 & 0\\
0 & 0 & 0 & -tf(\bm{k})\\
t_{\perp} & 0 & -tf^{*}(\bm{k}) & 0
\end{array}\right].\label{eq:AB_Hmatrix}
\end{equation}

\begin{figure}
\centering{}\includegraphics[width=0.5\textwidth]{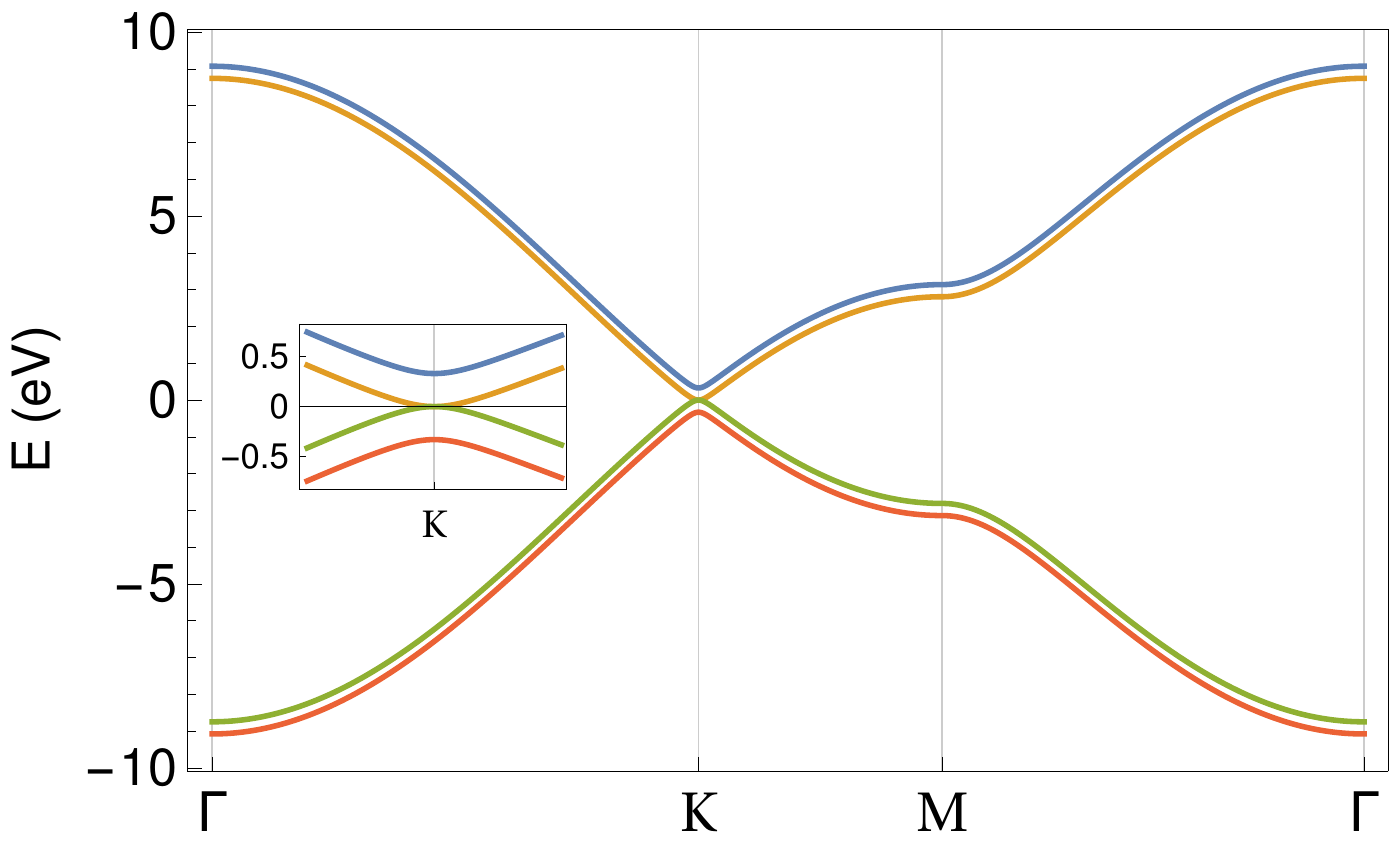}\caption{Electronic spectrum for Bernal-stacked BLG, along the $\bm{k}$-space
trajectory $\text{\ensuremath{\Gamma}}\rightarrow\text{K}\rightarrow\text{M}\rightarrow\text{\ensuremath{\Gamma}}$.}
\label{fig:ABspectrum}
\end{figure}

Diagonalizing $H(\bm{k})$, we obtain the electronic spectrum for
Bernal-stacked BLG as four bands, 
\begin{equation}
E_{\pm,\pm}(\bm{k})=\pm t\sqrt{\left(\frac{t_{\perp}}{2t}\right)^{2}+4\cos\left(\frac{\sqrt{3}}{2}dk_{x}\right)\cos\left(\frac{3}{2}dk_{y}\right)+2\cos\left(\sqrt{3}dk_{x}\right)+3}\pm\frac{t_{\perp}}{2}.
\end{equation}
The pair of bands $E_{+,-}(\bm{k})$ and $E_{-,+}(\bm{k})$ are gapless
and touch at the K and K$^{\prime}$ points of the BZ. We show the
obtained band structure along a representative path in first BZ in
Fig. \ref{fig:ABspectrum}.

%% file: sections/Section_tBLG.tex
\section{Twisted bilayer graphene}

\label{chapter:tBLG}

In this section, we aim at deriving a model for the tBLG system. We
follow the work done by Bistritzer and MacDonald \citep{Bistritzer2011}
and construct a continuum low-energy effective Hamiltonian, which
is valid for twist angles $\theta\lesssim10^{\circ}$ and independent
of the structure being commensurate or incommensurate. The electronic
properties of this system are then addressed.

\subsection{Geometry and moiré pattern}

\label{subsection:tBLGgeometry}

We begin by establishing a general geometry for a tBLG. A completely
arbitrary arrangement can be achieved in the following manner: we
start with a perfectly aligned BLG (for concreteness we take this
to be Bernal-stacked) and, with one of the layers fixed, which we
will refer to as layer 1, we translate the second, layer 2, by a vector
$\bm{\tau}_{0}$ and then rotate it by an angle $\theta$ (anti-clockwise
and about the origin). This way, each layer $\ell=1,2$ is described
by the following lattice points:
\begin{equation}
\bm{R}_{\ell}=n_{1}\bm{a}_{\ell,1}+n_{2}\bm{a}_{\ell,2},\quad n_{1},n_{2}\in\mathbb{Z},\label{eq:tBLG_lattice_sites}
\end{equation}
where $\bm{a}_{\ell,1}$ and $\bm{a}_{\ell,2}$ are the basis vectors
of each layer, which are related by $\bm{a}_{2,i}=\mathcal{R}_{\theta}\cdot\bm{a}_{1,i}$,
where $\mathcal{R}_{\theta}$ is the rotation matrix that describes
an anti-clockwise rotation by $\theta$ about the origin of a 2D coordinate
system, 
\begin{equation}
\mathcal{R}_{\theta}=\begin{bmatrix}\cos(\theta) & -\sin(\theta)\\
\sin(\theta) & \cos(\theta)
\end{bmatrix}.
\end{equation}
The positions of the $A$ and $B$ sites for each layer are given
by

\begin{align}
 & \bm{\tau}_{1,A}=\left(0,0\right), &  & \bm{\tau}_{2,A}=\mathcal{R}_{\theta}\cdot\left[\left(0,-d\right)+\bm{\tau}_{0}\right],\\
 & \bm{\tau}_{1,B}=\left(0,d\right), &  & \bm{\tau}_{2,B}=\mathcal{R}_{\theta}\cdot\left[\left(0,0\right)+\bm{\tau}_{0}\right].
\end{align}

Associated to the lattices $\left\{ \bm{R}_{\ell}\right\} $, we have
the corresponding reciprocal lattices $\left\{ \bm{G}_{\ell}\right\} $
which are spanned by the vectors $\bm{b}_{\ell,1}$, and $\bm{b}_{\ell,2}$.
The reciprocal lattice basis vectors are also related via rotation
as $\bm{b}_{2,i}=\mathcal{R}_{\theta}\cdot\bm{b}_{1,i}$.

The distinct periodicity of the two layers gives origin to an interference
effect that leads to the formation of a moiré pattern. The moiré pattern
is nothing more than a beat effect \citep{SanJose2014} and can be
understood as follows. Let us consider two functions $h_{1}(\bm{r})$
and $h_{2}(\bm{r})$ with the same periodicity as the layers $1$
and $2$, respectively. We choose these functions as 
\begin{equation}
h_{\ell}(\bm{r})=\sum_{k=1}^{3}\cos\left(\bm{G}_{\ell,k}\cdot\bm{r}\right),
\end{equation}
where we have written $\bm{G}_{\ell,1}=\bm{b}_{\ell,1}$, $\bm{G}_{\ell,2}=\bm{b}_{\ell,2}$,
$\bm{G}_{\ell,3}=\bm{b}_{\ell,1}-\bm{b}_{\ell,2}$. We can study the
interference effects between the two layers by studying the function
$h_{m}(\bm{r})=h_{1}(\bm{r})+h_{2}(\bm{r})$. Standard manipulation
allows us to write
\begin{equation}
h_{m}(\bm{r})=\sum_{k=1}^{3}2\cos\left(\frac{\bm{G}_{1,k}+\bm{G}_{2,k}}{2}\cdot\bm{r}\right)\cos\left(\frac{\bm{G}_{1,k}-\bm{G}_{2,k}}{2}\cdot\bm{r}\right).
\end{equation}
Therefore, we see that the function will have fast oscillations controlled
by $\left(\bm{G}_{1,k}+\bm{G}_{2,k}\right)/2$, which are modulated
by a slowly oscillating envelop function that oscillates with $\left(\bm{G}_{1,k}-\bm{G}_{2,k}\right)/2$.
It is this envelop function that is responsible for the moiré pattern.
Given that only the amplitude (and not the sign) of the envelop affects
the visibility of the moiré pattern, this appears to oscillate with
$\bm{G}_{1,k}-\bm{G}_{2,k}$. For this same reason, the moiré pattern
is not affected by the translations of one layer with respect to the
other. Therefore, the function $h_{m}(\bm{r})$ will display a quasi-periodic
pattern, with an associated reciprocal lattice $\left\{ \bm{G}^{m}\right\} $
that is spanned by the moiré basis vectors 
\begin{equation}
\bm{b}_{1}^{m}=\bm{b}_{1,1}-\bm{b}_{2,1},\quad\bm{b}_{2}^{m}=\bm{b}_{1,2}-\bm{b}_{2,2}.
\end{equation}
In the coordinate system where layer 2 is rotated by $\theta/2$ and
layer 1 is rotated by $-\theta/2$, these are given by

\begin{equation}
\bm{b}_{1}^{m}=\sqrt{3}\left|\Delta\text{K}\right|\left(\frac{1}{2},-\frac{\sqrt{3}}{2}\right),\quad\bm{b}_{2}^{m}=\sqrt{3}\left|\Delta\text{K}\right|\left(\frac{1}{2},\frac{\sqrt{3}}{2}\right),
\end{equation}
where $\left|\Delta\text{K}\right|=2\text{\ensuremath{\left|\text{K}\right|}}\sin\left(\theta/2\right)$
is the separation between the Dirac points of the two layers, with
$\text{\ensuremath{\left|\text{K}\right|}}=4\pi/\left(3\sqrt{3}d\right)$.

Associated to the reciprocal lattice $\left\{ \bm{G}^{m}\right\} $,
we can define a moiré real lattice $\left\{ \bm{R}^{m}\right\} $,
spanned by basis vectors $\bm{a}_{1}^{m}$ and $\bm{a}_{2}^{m}$,
such that $\bm{a}_{i}^{m}\cdot\bm{b}_{j}^{m}=2\pi\delta_{i,j}$, which
are explicitly given by 
\begin{equation}
\bm{a}_{1}^{m}=\frac{4\pi}{3\left|\Delta\text{K}\right|}\left(\frac{\sqrt{3}}{2},-\frac{1}{2}\right),\quad\bm{a}_{2}^{m}=\frac{4\pi}{3\left|\Delta\text{K}\right|}\left(\frac{\sqrt{3}}{2},\frac{1}{2}\right).
\end{equation}
The unit cell of the moiré lattice has area 
\begin{equation}
A_{m.u.c.}=\left|\bm{a}_{1}^{m}\times\bm{a}_{2}^{m}\right|=\frac{\sqrt{3}}{2}\left(\frac{4\pi}{3\left|\Delta\text{K}\right|}\right)^{2}=\frac{3\sqrt{3}d^{2}}{8\sin^{2}(\theta/2)}.
\end{equation}

\begin{figure}
\begin{centering}
\includegraphics[width=0.8\textwidth]{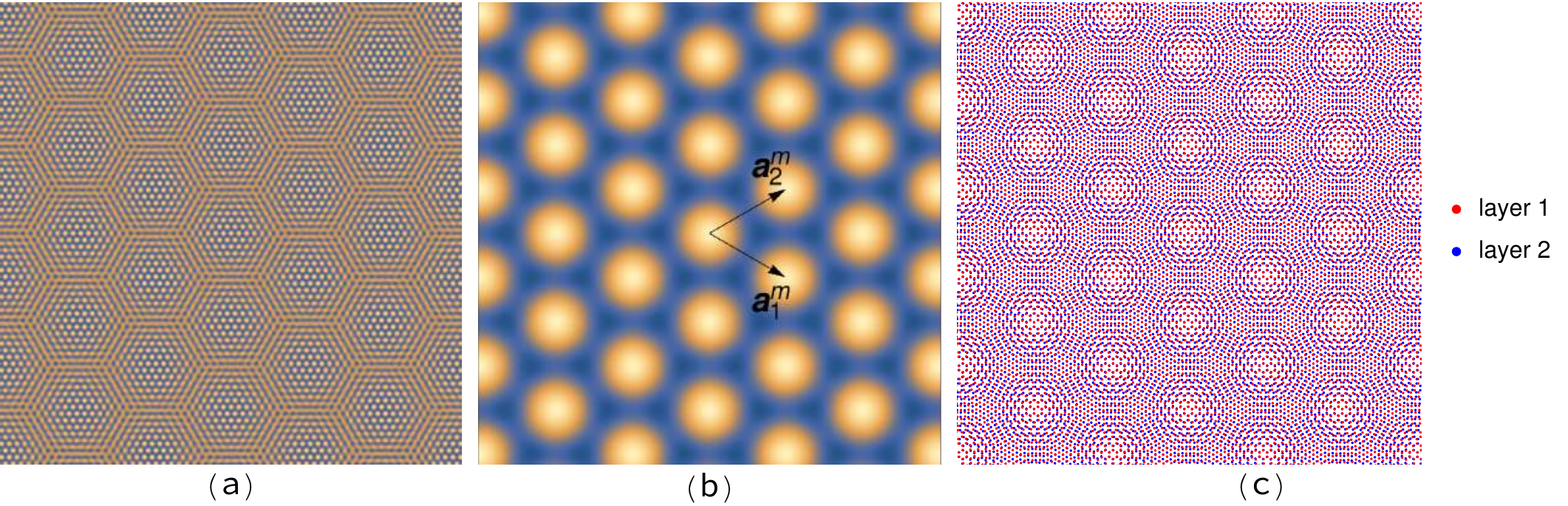} 
\par\end{centering}
\caption{Formation of moiré pattern due to the interference of two periodic
structures. (a) Density plot of the function $h_{m}(\bm{r})$. (b)
Density plot of the envelop function $\tilde{h}_{m}(\bm{r})=\sum_{k=1}^{3}2\cos\left[\left(\bm{G}_{1,k}-\bm{G}_{2,k}\right)\cdot\bm{r}\right]$.
The basis vectors of the moiré lattice are also shown. (c) Representation
of the structure of tBLG. The emergence of the moiré pattern is clear.
In all plots it was considered a twist angle $\theta=5^{\circ}$ (layer
2 rotated by $\theta/2$, layer 1 rotated by $-\theta/2$) and $\bm{\tau}_{0}=\left(0,d\right)$,
such that in the unrotated limit, AB stacked BLG is recovered. The
consideration of different values of $\bm{\tau}_{0}$ just leads to
a shift in real space of the moiré pattern.}
\label{fig:tBLGgeometry} 
\end{figure}

In Fig. \ref{fig:tBLGgeometry}, we show an example of the emergence
of a moiré pattern in the function $h_{m}(\bm{r})$ and compare it
to the moiré pattern that appears when the two lattices that form
a tBLG structure are superimposed.

\subsection{Model Hamiltonian}

We will now see how a tBLG structure can be modeled. The approach
described here closely follows the work of Ref.~\citep{Bistritzer2011}
for tBLG in the small angle limit, which was generalized for other
materials and arbitrary angles in Ref.~\citep{Koshino2015}. The
starting point of the method is a tight-binding representation of
the Hamiltonian, which is partitioned as 
\begin{equation}
H=H_{1}+H_{2}+H_{\perp},
\end{equation}
where $H_{\ell}$ is the Hamiltonian of the isolated layer $\ell=1,2$
and $H_{\perp}=V_{12}+V_{21}$ is the interlayer Hamiltonian, which
describes hybridization between the two layers. $V_{12}$ describes
electron hopping from layer $2$ to layer $1$ and $V_{21}=V_{12}^{\dagger}$
describes the inverse process. The general approach is based on a
two-center approximation for the interlayer Hamiltonian and in an
expansion of the full Hamiltonian in terms of Bloch waves of the individual
layers.

\subsubsection{Hamiltonian for rotated graphene monolayers}

We want to express the full Hamiltonian in terms of Bloch waves for
the individual layers of the form of Eqs.~(\ref{eq:SLGpsiket}) and
(\ref{eq:SLGpsiket_bilayer}),
\begin{equation}
\left|\psi_{\ell,\bm{k},\alpha}\right\rangle =\frac{1}{\sqrt{N_{\ell}}}\sum_{\bm{R}_{\ell}}e^{i\bm{k}\cdot\left(\bm{R}_{\ell}+\bm{\tau}_{\ell,\alpha}\right)}\left|\ell,\bm{R}_{\ell},\alpha\right\rangle ,
\end{equation}
where $\ell=1,2$ labels the layer, $\alpha=A,B$ is the sublattice,
$N_{\ell}$ is the number of unit cells of each layer, $\bm{R}_{\ell}$
are the lattice sites, $\bm{\tau}_{\ell,\alpha}$ are the positions
of the orbital centers in the unit cell and $\left|\ell,\bm{R}_{\ell},\alpha\right\rangle $
are localized atomic-like Wannier states. In this basis, and in the
single-orbital nearest-neighbor approximation, the Hamiltonian for
each layer reads as
\begin{equation}
H_{\ell}(\bm{k})=\begin{bmatrix}0 & -tf_{\ell}(\bm{k})\\
-tf_{\ell}^{*}(\bm{k}) & 0
\end{bmatrix},\label{eq:SLG_hamiltonian_rotated}
\end{equation}
where $f_{\ell}(\bm{k})=\sum_{i=1}^{3}e^{i\bm{k}\cdot\bm{d}_{\ell,i}}$,
with $\bm{d}_{\ell,1}=\left(\bm{a}_{\ell,1}+\bm{a}_{\ell,2}\right)/3$,
$\bm{d}_{\ell,2}=\left(-2\bm{a}_{\ell,1}+\bm{a}_{\ell,2}\right)/3$,
$\bm{d}_{\ell,3}=\left(\bm{a}_{\ell,1}-2\bm{a}_{\ell,2}\right)/3$.

If we are interested in low-energy states, we can describe each layer
with a Dirac Hamiltonian by writing $\bm{k}=\pm\text{K}_{\ell}+\bm{q}$,
where $\pm\text{K}_{\ell}$ points are the Dirac points of each layer
with $\text{K}_{1}=\left(\frac{4\pi}{3a},0\right)$ and $\text{K}_{2}=\mathcal{R}_{\theta}\cdot\text{K}_{1}$,
and expanding to lowest order in $\bm{q}$. The obtained Hamiltonians
can be written in a unified way as 
\begin{equation}
H_{\ell}^{\pm\text{K}}(\bm{q})=\pm\hbar v_{F}\left|\bm{q}\right|\left[\begin{array}{cc}
0 & e^{\mp i\left(\theta_{\bm{q}}-\theta_{\ell}\right)}\\
e^{\pm i\left(\theta_{\bm{q}}-\theta_{\ell}\right)} & 0
\end{array}\right],\label{eq:RSLGmatrix}
\end{equation}
where $\theta_{1}=0$, $\theta_{2}=\theta$ and $\theta_{\bm{q}}$
is the angle that the momentum $\bm{q}$ makes with the $x$ axis,
such that $\bm{q}=\left|\bm{q}\right|\left(\cos\theta_{\bm{q}},\sin\theta_{\bm{q}}\right)$.
The above equation can also be written in a compact form as 
\begin{equation}
H_{\ell}^{\pm\text{K}}(\bm{q})=v_{F}\hbar\bm{q}\cdot\left(\pm\sigma_{x}^{\theta_{\ell}},\sigma_{y}^{\theta_{\ell}}\right),\label{eq:rotated_Dirac_Hamiltonian}
\end{equation}
where $\sigma_{x}^{\theta}=\sigma_{x}\cos\theta-\sigma_{y}\sin\theta$
and $\sigma_{y}^{\theta}=\sigma_{x}\sin\theta+\sigma_{y}\cos\theta$
are rotated Pauli matrices.

\subsubsection{General interlayer Hamiltonian in terms of Bloch waves\label{subsec:General-interlayer-Hamiltonian}}

We write the interlayer Hamiltonian in second quantization in the
basis of atomic-like localized states of each layer as
\begin{equation}
V_{12}=\sum_{\bm{R}_{1},\alpha,\bm{R}_{2},\beta}c_{1,\alpha}^{\dagger}\left(\bm{R}_{1}\right)t_{12}^{\alpha\beta}\left(\bm{R}_{1},\bm{R}_{2}\right)c_{2,\beta}\left(\bm{R}_{2}\right),
\end{equation}
where 
\begin{equation}
t_{12}^{\alpha\beta}\left(\bm{R}_{1},\bm{R}_{2}\right)=\left\langle 1,\bm{R}_{1},\alpha\right|H_{\perp}\left|2,\bm{R}_{2},\beta\right\rangle 
\end{equation}
is the interlayer hopping in the tight-binding basis. Writing the
operators in terms of Bloch waves,
\begin{equation}
c_{\ell,\alpha}^{\dagger}\left(\bm{R}_{\ell}\right)=\frac{1}{\sqrt{N_{\ell}}}\sum_{\bm{k}_{\ell}}e^{-i\bm{k}_{\ell}\cdot\left(\bm{R}_{\ell}+\bm{\tau}_{\ell,\alpha}\right)}c_{\ell,\alpha}^{\dagger}\left(\bm{k}_{\ell}\right),
\end{equation}
with the sum over $\bm{k}_{\ell}$ restricted to the BZ of layer $\ell$,
we obtain 
\begin{equation}
V_{12}=\sum_{\bm{k}_{1},\alpha,\bm{k}_{2},\beta}c_{1,\alpha}^{\dagger}\left(\bm{k}_{1}\right)T_{12}^{\alpha\beta}\left(\bm{k}_{1},\bm{k}_{2}\right)c_{2,\beta}^{\dagger}\left(\bm{k}_{2}\right),
\end{equation}
where 
\begin{equation}
T_{12}^{\alpha\beta}\left(\bm{k}_{1},\bm{k}_{2}\right)=\frac{1}{\sqrt{N_{1}N_{2}}}\sum_{\bm{R}_{1},\bm{R}_{2}}e^{-i\bm{k}_{1}\cdot\left(\bm{R}_{1}+\bm{\tau}_{1,\alpha}\right)}t_{12}^{\alpha\beta}\left(\bm{R}_{1},\bm{R}_{2}\right)e^{i\bm{k}_{2}\cdot\left(\bm{R}_{2}+\bm{\tau}_{2,\beta}\right)}.\label{eq:interlayer_Bloch}
\end{equation}

The previous change of basis does not lead to a great simplification.
Progress can be made if, in the spirit of a two-center approximation,
we assume that the interlayer hopping $t_{12}^{\alpha\beta}\left(\bm{R}_{1},\bm{R}_{2}\right)$
is only a function of the separation between the center of the two
orbitals, i.e.
\begin{equation}
t_{12}^{\alpha\beta}\left(\bm{R}_{1},\bm{R}_{2}\right)=t_{12}^{\alpha\beta}\left(\bm{R}_{1}+\bm{\tau}_{1,\alpha}-\bm{R}_{2}-\bm{\tau}_{2,\beta}\right).
\end{equation}
We now write the interlayer hopping in terms of a 2D Fourier transform,
\begin{equation}
t_{12}^{\alpha\beta}\left(\bm{R}_{1}+\bm{\tau}_{1,\alpha}-\bm{R}_{2}-\bm{\tau}_{2,\beta}\right)=\int_{\mathbb{R}^{2}}\frac{d^{2}\bm{p}}{\left(2\pi\right)^{2}}e^{i\bm{p}\cdot\left(\bm{R}_{1}+\bm{\tau}_{1,\alpha}-\bm{R}_{2}-\bm{\tau}_{2,\beta}\right)}t_{12}^{\alpha\beta}\left(\bm{p}\right).\label{eq:hopping_FT}
\end{equation}
Provided $t_{12}^{\alpha\beta}\left(\bm{r}\right)$ is known, where
$\bm{r}$ is the in-plane separation between the two orbitals, we
can evaluate $t_{12}^{\alpha\beta}\left(\bm{p}\right)$ by inverting
the Fourier transform, 
\begin{equation}
t_{12}^{\alpha\beta}\left(\bm{p}\right)=\int_{\mathbb{R}^{2}}d^{2}\bm{r}e^{-i\bm{p}\cdot\bm{r}}t_{12}^{\alpha\beta}\left(\bm{r}\right).\label{eq:invFT_tp}
\end{equation}

Inserting Eq.~(\ref{eq:hopping_FT}) into Eq.~(\ref{eq:interlayer_Bloch}),
we obtain 
\begin{equation}
T_{12}^{\alpha\beta}\left(\bm{k}_{1},\bm{k}_{2}\right)=\frac{1}{\sqrt{N_{1}N_{2}}}\int_{\mathbb{R}^{2}}\frac{d^{2}\bm{p}}{\left(2\pi\right)^{2}}\sum_{\bm{R}_{1}}e^{-i\left(\bm{k}_{1}-\bm{p}\right)\cdot\left(\bm{R}_{1}+\bm{\tau}_{1,\alpha}\right)}t_{12}^{\alpha\beta}\left(\bm{p}\right)\sum_{\bm{R}_{2}}e^{i\left(\bm{k}_{2}-\bm{p}\right)\cdot\left(\bm{R}_{2}+\bm{\tau}_{2,\beta}\right)}.
\end{equation}
Using the sum rule $\sum_{\bm{R}_{\ell}}e^{i\bm{k}\cdot\bm{R}_{\ell}}=N_{\ell}\sum_{\bm{G}_{\ell}}\delta_{\bm{k},\bm{G}_{\ell}}$,
this can be written as 
\begin{equation}
T_{12}^{\alpha\beta}\left(\bm{k}_{1},\bm{k}_{2}\right)=\sqrt{N_{1}N_{2}}\int_{\mathbb{R}^{2}}\frac{d^{2}\bm{p}}{\left(2\pi\right)^{2}}\sum_{\bm{G}_{1},\bm{G}_{2}}e^{-i\bm{G}_{1}\cdot\bm{\tau}_{1,\alpha}}t_{12}^{\alpha\beta}\left(\bm{p}\right)e^{i\bm{G}_{2}\cdot\bm{\tau}_{2,\beta}}\delta_{\bm{k}_{1}-\bm{p},\bm{G}_{1}}\delta_{\bm{k}_{2}-\bm{p},\bm{G}_{2}}.
\end{equation}
Now, we use the relation between a $\delta$-Kronecker and a $\delta$-Dirac
function, $\delta_{\bm{k},\bm{k}^{\prime}}=\delta\left(\bm{k}-\bm{k}^{\prime}\right)\left(2\pi\right)^{2}/A$,
where $A$ is the total area of the system, to perform the integration
over $\bm{p}$, obtaining 
\begin{equation}
T_{12}^{\alpha\beta}\left(\bm{k}_{1},\bm{k}_{2}\right)=\sqrt{\frac{N_{1}N_{2}}{A^{2}}}\sum_{\bm{G}_{1},\bm{G}_{2}}e^{-i\bm{G}_{1}\cdot\bm{\tau}_{1,\alpha}}t_{12}^{\alpha\beta}\left(\bm{k}_{1}+\bm{G}_{1}\right)e^{-i\bm{G}_{2}\cdot\bm{\tau}_{2,\beta}}\delta_{\bm{k}_{1}+\bm{G}_{1},\bm{k}_{2}+\bm{G}_{2}},\label{eq:interlayer_hopping_Bloch_0}
\end{equation}
where we also made the redefinition $\bm{G}_{2}\rightarrow-\bm{G}_{2}$
. Noticing that the total area can be written as $A=A_{u.c.1}N_{1}=A_{u.c.2}N_{2}$,
where $A_{u.c.\ell}$ is the area of the unit cell of layer $\ell$
($A_{u.c.1}=A_{u.c.2}=A_{u.c.}=\sqrt{3}a^{2}/2$), the above equation
can be written as 
\begin{equation}
T_{12}^{\alpha\beta}\left(\bm{k}_{1},\bm{k}_{2}\right)=\frac{1}{\sqrt{A_{u.c.1}A_{u.c.2}}}\sum_{\bm{G}_{1},\bm{G}_{2}}e^{-i\bm{G}_{1}\cdot\bm{\tau}_{1,\alpha}}t_{12}^{\alpha\beta}\left(\bm{k}_{1}+\bm{G}_{1}\right)e^{-i\bm{G}_{2}\cdot\bm{\tau}_{2,\beta}}\delta_{\bm{k}_{1}+\bm{G}_{1},\bm{k}_{2}+\bm{G}_{2}}.\label{eq:interlayer_hopping_Bloch}
\end{equation}
This equation shows that two states of layer 1 and 2 with respective
crystal-momentum $\bm{k}_{1}$ and $\bm{k}_{2}$ are only coupled
if reciprocal lattice vectors $\bm{G}_{1}$ and $\bm{G}_{2}$ of each
layer exist such that 
\begin{equation}
\bm{k}_{1}+\bm{G}_{1}=\bm{k}_{2}+\bm{G}_{2}.
\end{equation}
This is the so-called generalized umklapp condition \citep{Koshino2015}.

\subsubsection{Interlayer hopping for $p_{z}$ orbitals}

To make further progress, we must specify the functional form of $t_{12}^{\alpha\beta}\left(\bm{r}\right)$.
First, since in graphene both $A$ and $B$ sites correspond to the
same $p_{z}$ orbital of carbon, we assume $t_{12}^{AA}\left(\bm{r}\right)=t_{12}^{BB}\left(\bm{r}\right)=t_{12}^{AB}\left(\bm{r}\right)=t_{12}^{BA}\left(\bm{r}\right)=t_{\perp}\left(\bm{r}\right)$.
In the two-center approximation, we express $t_{\perp}\left(\bm{r}\right)$
in terms of Slater-Koster parameters \citep{SlaterKoster1954}, $V_{pp\sigma}$
and $V_{pp\pi}$, as follows: 
\begin{equation}
t_{\perp}\left(\bm{r}\right)=\cos^{2}(\gamma)\ V_{pp\sigma}\left(\sqrt{d_{\perp}^{2}+\left|\bm{r}\right|^{2}}\right)+\sin^{2}(\gamma)\ V_{pp\pi}\left(\sqrt{d_{\perp}^{2}+\left|\bm{r}\right|^{2}}\right),
\end{equation}
where $d_{\perp}=3.35\si{\angstrom}$ (assuming the same interlayer
distance as in Bernal-stacked BLG) and $\gamma$ is the angle between
the $z$ axis and the line connecting the two orbital centers, which
leads to 
\begin{equation}
\cos^{2}(\gamma)=\frac{d_{\perp}^{2}}{d_{\perp}^{2}+\left|\bm{r}\right|^{2}},\quad\sin^{2}(\gamma)=\frac{\left|\bm{r}\right|^{2}}{d_{\perp}^{2}+\left|\bm{r}\right|^{2}}.
\end{equation}

In order to evaluate $t_{\perp}\left(\bm{p}\right)$, we still need
to model the dependency of the Slater-Koster parameters on the separation.
In Ref. \citep{LaissardiereMayouMagaud2012}, the authors explored
an exponentially decreasing model for $V_{pp\sigma}$ and $V_{pp\pi}$,
which we shall adopt: 
\begin{equation}
V_{pp\sigma}(r)=t_{\perp}\ \text{exp}\left[q_{\sigma}(1-r/d_{\perp})\right],\quad V_{pp\pi}(r)=-t\ \text{exp}\left[q_{\pi}(1-r/d)\right].
\end{equation}
We stress that $V_{pp\pi}(d)=-t$ and $V_{pp\sigma}(d_{\perp})=t_{\perp}$,
which recovers the values for both SLG and Bernal-stacked BLG. To
fix $q_{\pi}$, the authors took the characteristic second nearest-neighbor
hopping amplitude in SLG, $t^{\prime}\approx0.1t$ \citep{KretininYuJalilEtAl2013},
and obtained 
\begin{equation}
\frac{V_{pp\pi}(d)}{V_{pp\pi}(\sqrt{3}d)}=\frac{t}{t'}\Leftrightarrow q_{\pi}\simeq3.15.
\end{equation}
The remaining parameter, $q_{\sigma}$, was fixed assuming equal spatial
exponentially decreasing coefficients, i.e., 
\begin{equation}
\frac{q_{\pi}}{d}=\frac{q_{\sigma}}{d_{\perp}}\Leftrightarrow q_{\sigma}\simeq7.42.
\end{equation}

\begin{figure}
\centering{}\includegraphics[width=0.4\textwidth]{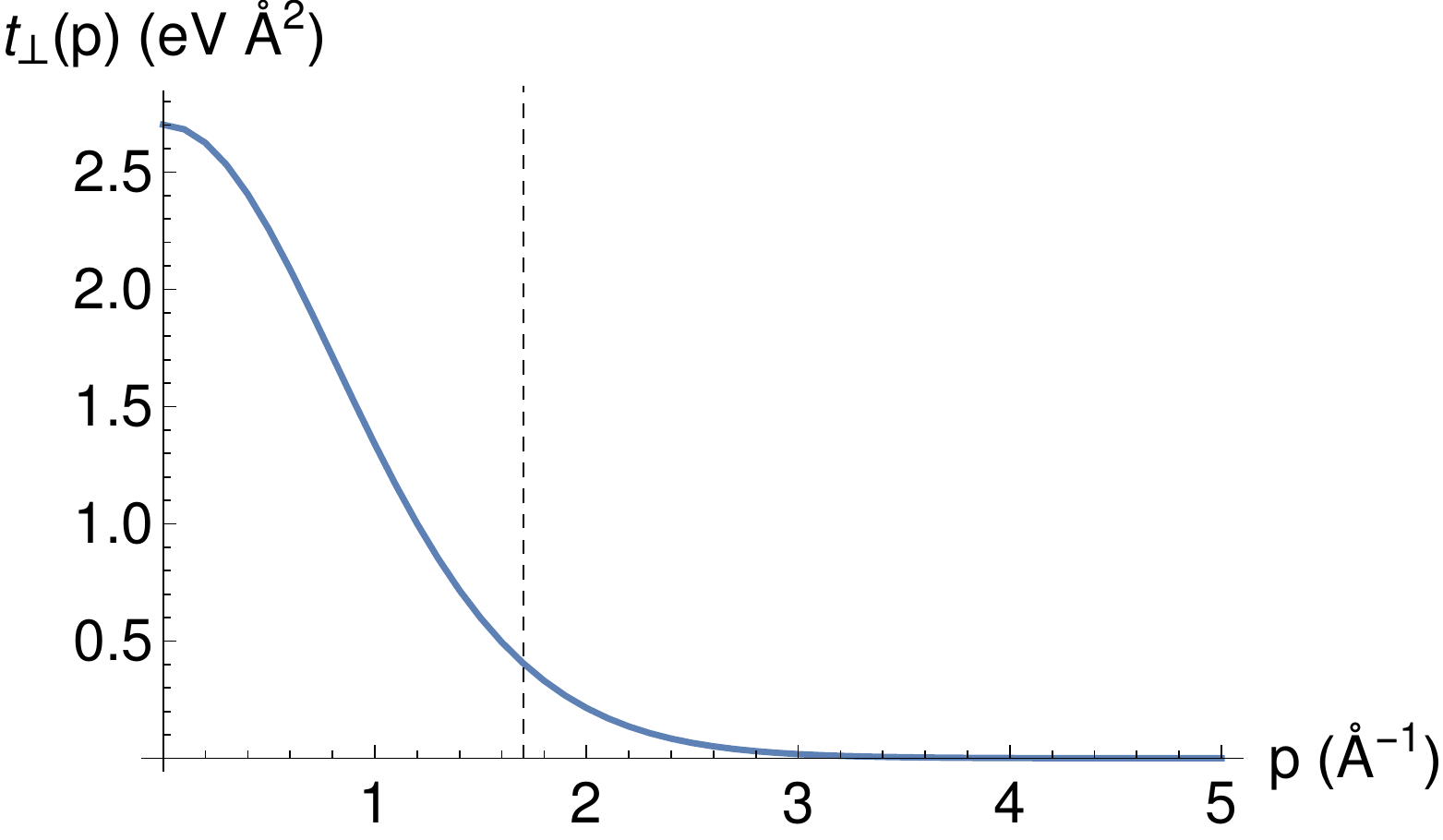}
\caption{Fourier transform for the interlayer hopping in tBLG. The vertical
dashed line marks the position of the Dirac point: $p=\left|\text{K}\right|$.}
\label{fig:tperp(k)}
\end{figure}

Using this model, we can evaluate $t_{\perp}\left(\bm{p}\right)$
by evaluating numerically the integral 
\begin{equation}
t_{\perp}\left(\bm{p}\right)=2\pi\int_{0}^{\infty}drrJ_{0}\left(\left|\bm{p}\right|r\right)t_{\perp}\left(r\right),
\end{equation}
where $J_{0}(x)$ is a Bessel function of the first kind, which results
from the angular integration in Eq. \eqref{eq:invFT_tp}. From the
above equation, it is clear that $t_{\perp}\left(\bm{p}\right)$ is
actually just a function of $\left|\bm{p}\right|$. In addition, we
can anticipate that $t_{\perp}\left(\bm{p}\right)$ should decay very
rapidly with $\left|\bm{p}\right|$ on the reciprocal lattice scale.
Intuitively, since $d_{\perp}>d$ by more than a factor of 2, the
two-center interlayer hopping term, $t_{\perp}\left(\bm{r}\right)$,
which depends on the three-dimensional separation, $\sqrt{\bm{r}^{2}+d_{\perp}^{2}}$,
will be weakly dependent on $\bm{r}$ for values $\left|\bm{r}\right|\lesssim d_{\perp}$,
which determine the dominant interlayer hopping. Therefore, $t_{\perp}(\bm{r})$
has a broadened distribution and its Fourier transform, $t_{\perp}(\bm{p})$,
must be sharp and decline very rapidly for $\left|\bm{p}\right|d_{\perp}>1$.
This expectation is proven correct in Fig. \ref{fig:tperp(k)}, where
we plot the numerical result obtained for $t_{\perp}\left(\bm{p}\right)$.

The fact that $t_{\perp}\left(\bm{p}\right)$ decays rapidly for large
values of $\left|\bm{p}\right|$ has important consequences, as it
means that only a few umklapp processes contribute significantly to
the interlayer coupling.

\subsubsection{Interlayer Hamiltonian for small twist angles}

We now wish to specialize to the case of tBLG in the limit of small
twist angles. For small $\theta$, the Dirac points $\text{K}_{1}$
and $\text{K}_{2}$ are close to each other and we can neglect coupling
between $\text{K}_{\ell}$ and $-\text{K}_{\ell}$ points. If we are
only interested in low-energy physics we can expand all quantities
around these points. Therefore, close to the $\text{K}_{\ell}$ points,
we can write 
\begin{equation}
\bm{k}_{\ell}=\text{K}_{\ell}+\bm{q}_{\ell}.
\end{equation}
As a result, the interlayer coupling, Eq.~(\ref{eq:interlayer_hopping_Bloch}),
becomes

\begin{equation}
T_{12}^{\alpha\beta}\left(\bm{q}_{1},\bm{q}_{2}\right)=\frac{1}{A_{u.c.}}\sum_{\bm{G}_{1},\bm{G}_{2}}e^{-i\bm{G}_{1}\cdot\bm{\tau}_{1,\alpha}}t_{12}^{\alpha\beta}\left(\text{K}_{1}+\bm{q}_{1}+\bm{G}_{1}\right)e^{-i\bm{G}_{2}\cdot\bm{\tau}_{2,\beta}}\delta_{\text{K}_{1}+\bm{q}_{1}+\bm{G}_{1},\text{K}_{2}+\bm{q}_{2}+\bm{G}_{2}}.
\end{equation}

\begin{figure}
\centering{}\includegraphics[width=0.6\columnwidth]{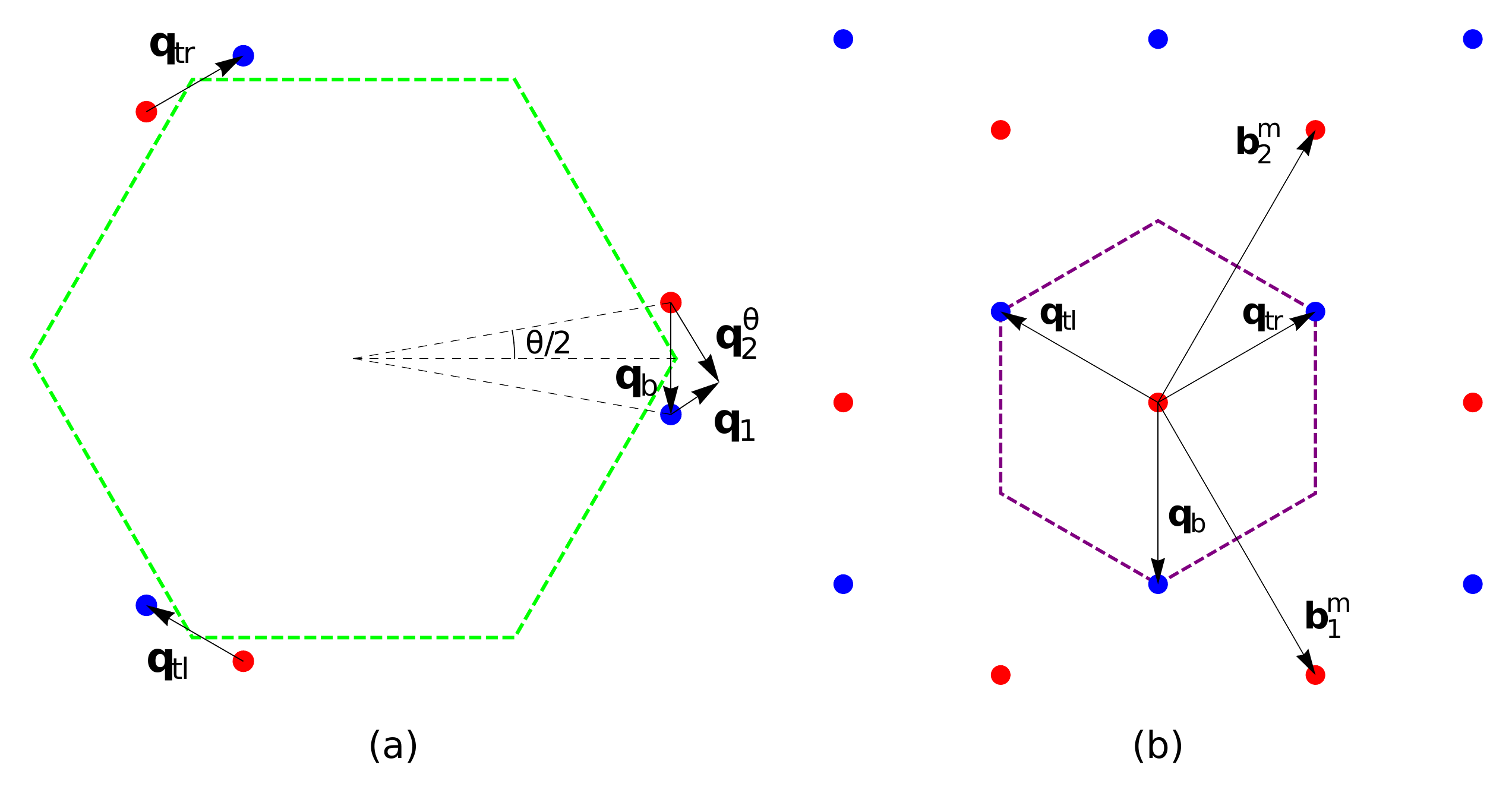}\caption{Momentum-space geometrical picture for the interlayer hopping on a
tBLG. (a) The green dashed line marks the first BZ for an unrotated
SLG; the red (blue) circles mark the three equivalent Dirac points
for layer 1 (2). States from both layers couple when $\bm{q}_{2}-\bm{q}_{1}=\bm{q}_{b},\bm{q}_{tr},\bm{q}_{tl}$.
(b) The three equivalent Dirac points in the first BZ result in three
distinct hopping processes in reciprocal space; when we capture processes
to all orders in the interlayer hopping, but considering only the
transference of momentum by $\bm{q}_{b},\bm{q}_{tr},\bm{q}_{tl}$,
we obtain this $\bm{k}$-space honeycomb structure, which captures
the periodicity of the moiré pattern. The purple dashed line marks
a moiré unit cell in reciprocal space.}
\label{fig:tBLG_hopping_picture}
\end{figure}

For states close to the Dirac points, we have $\left|\bm{q}_{1}\right|,\left|\bm{q}_{2}\right|\ll\left|\text{K}\right|$,
and we can approximate $t_{\perp}\left(\text{K}_{1}+\bm{q}_{1}+\bm{G}_{1}\right)\simeq t_{\perp}\left(\text{K}_{1}+\bm{G}_{1}\right)$.
As previously discussed, $t_{\perp}\left(\bm{p}\right)$ decays rapidly
as a function of $\left|\bm{p}\right|$ and we can thus keep only
the three most relevant processes, which correspond to interlayer
hopping terms with momentum close to the three equivalent Dirac points.
Therefore, we restrict the sum to $\bm{G}_{\ell}=\bm{g}_{\ell,1},\,\bm{g}_{\ell,2},\,\bm{g}_{\ell,3}$,
with $\bm{g}_{\ell,1}=\bm{0}$, $\bm{g}_{\ell,2}=\bm{b}_{\ell,2}$
and $\bm{g}_{\ell,3}=-\bm{b}_{\ell,1}$. This leads to
\begin{equation}
T_{12}^{\alpha\beta}\left(\bm{q}_{1},\bm{q}_{2}\right)=T_{\bm{q}_{b}}^{\alpha\beta}\delta_{\bm{q}_{1}-\bm{q}_{2},-\bm{q}_{b}}+T_{\bm{q}_{tr}}^{\alpha\beta}\delta_{\bm{q}_{1}-\bm{q}_{2},-\bm{q}_{tr}}+T_{\bm{q}_{tl}}^{\alpha\beta}\delta_{\bm{q}_{1}-\bm{q}_{2},-\bm{q}_{tl}},\label{eq:Dirac_interlayer}
\end{equation}
where we used that fact that $\left|\text{K}_{1}+\bm{g}_{1,n}\right|=\left|\text{K}\right|=4\pi/\left(3a\right)$
for $n=1(b),2(tr),3(tl)$. In the above equation, we have defined
$T_{\boldsymbol{q}_{n}}^{\alpha\beta}=\frac{t_{\perp}\left(\left|\text{K}\right|\right)}{A_{u.c.}}e^{i\bm{g}_{1,n}\cdot\bm{\tau}_{1,\alpha}}e^{-i\bm{g}_{2,n}\cdot\bm{\tau}_{2,\beta}}$,
which, in the $A,B$ basis, can be written in the following matrix
form:
\begin{align}
T_{\bm{q}_{b}} & =\frac{t_{\perp}\left(\left|\text{K}\right|\right)}{A_{u.c.}}\left[\begin{array}{cc}
1 & 1\\
1 & 1
\end{array}\right],\\
T_{\bm{q}_{tr}} & =\frac{t_{\perp}\left(\left|\text{K}\right|\right)}{A_{u.c.}}e^{-i\bm{g}_{1,2}\cdot\bm{\tau}_{0}}\left[\begin{array}{cc}
e^{i\phi} & 1\\
e^{-i\phi} & e^{i\phi}
\end{array}\right],\\
T_{\bm{q}_{tl}} & =\frac{t_{\perp}\left(\left|\text{K}\right|\right)}{A_{u.c.}}e^{-i\bm{g}_{1,3}\cdot\bm{\tau}_{0}}\left[\begin{array}{cc}
e^{-i\phi} & 1\\
e^{i\phi} & e^{-i\phi}
\end{array}\right],
\end{align}
with $\phi=2\pi/3$. In addition, we have also introduced the vectors
\begin{align}
\bm{q}_{b} & =\text{K}_{1}-\text{K}_{2},\label{eq:q_b}\\
\bm{q}_{tr} & =\text{K}_{1}+\bm{g}_{1,2}-\text{K}_{2}-\bm{g}_{2,2},\label{eq:q_tr}\\
\bm{q}_{tl} & =\text{K}_{1}+\bm{g}_{1,3}-\text{K}_{2}-\bm{g}_{2,3}.\label{eq:q_tl}
\end{align}
In the coordinate system where layer $2$ is rotated by $\theta/2$
and layer $1$ by $-\theta/2$, these three vectors are given explicitly
given by 
\begin{align}
\bm{q}_{b} & =\left|\Delta\text{K}\right|\left(0,-1\right),\\
\bm{q}_{tr} & =\left|\Delta\text{K}\right|\left(\frac{\sqrt{3}}{2},\frac{1}{2}\right),\\
\bm{q}_{tl} & =\left|\Delta\text{K}\right|\left(-\frac{\sqrt{3}}{2},\frac{1}{2}\right).
\end{align}
In Fig.~\ref{fig:tBLG_hopping_picture}, we represent the different
transfered momenta in reciprocal space.

The interlayer hopping from layer $2$ to layer $1$ can be obtained
from the hermitian conjugate of Eq.~(\ref{eq:Dirac_interlayer}),
\begin{equation}
T_{21}^{\alpha\beta}\left(\bm{q}_{2},\bm{q}_{1}\right)=\left(T_{\bm{q}_{b}}^{\beta\alpha}\right)^{*}\delta_{\bm{q}_{2}-\bm{q}_{1},\bm{q}_{b}}+\left(T_{\bm{q}_{tr}}^{\beta\alpha}\right)^{*}\delta_{\bm{q}_{2}-\bm{q}_{1},\bm{q}_{tr}}+\left(T_{\bm{q}_{tl}}^{\beta\alpha}\right)^{*}\delta_{\bm{q}_{2}-\bm{q}_{1},\bm{q}_{tl}}.\label{eq:Dirac_interlayer_21}
\end{equation}

If we set $\theta=0$ and $\bm{\tau}_{0}=\bm{0}$, we recover Bernal-stacked
BLG, with the interlayer coupling given by
\begin{equation}
T_{12}^{\alpha\beta}\left(\bm{q}_{2},\bm{q}_{1}\right)=\frac{t_{\perp}\left(\left|\text{K}\right|\right)}{A_{u.c.}}\left\{ \left[\begin{array}{cc}
1 & 1\\
1 & 1
\end{array}\right]+\left[\begin{array}{cc}
e^{i\phi} & 1\\
e^{-i\phi} & e^{i\phi}
\end{array}\right]+\left[\begin{array}{cc}
e^{-i\phi} & 1\\
e^{i\phi} & e^{-i\phi}
\end{array}\right]\right\} =\frac{3t_{\perp}\left(\left|\text{K}\right|\right)}{A_{u.c.}}\left[\begin{array}{cc}
0 & 1\\
0 & 0
\end{array}\right].\label{eq:Tmatrix_AB_BLG}
\end{equation}
Comparing Eq. \eqref{eq:Tmatrix_AB_BLG} with the Hamiltonian for
Bernal-stacked BLG, Eq.~(\ref{eq:AB_Hmatrix}), we get
\begin{equation}
t_{\perp}=\frac{3t_{\perp}\left(\left|\text{K}\right|\right)}{A_{u.c.}}.
\end{equation}
This relation fixes the value of $t_{\perp}(K)$ as
\begin{equation}
t_{\perp}\left(\left|\text{K}\right|\right)\simeq0.58\si{\electronvolt\angstrom\squared},
\end{equation}
which is, to a good approximation, consistent with the result obtained
by the numerical calculation (Fig. \ref{fig:tperp(k)}).

The derivation of the interlayer coupling for states close to the
$-\text{K}_{\ell}$ points expansion is completely analogous to the
$+\text{K}_{\ell}$ case. Therefore, we just present the final result,
which reads as
\begin{equation}
\tilde{T}_{12}^{\alpha\beta}\left(\bm{q}_{1},\bm{q}_{2}\right)=\tilde{T}_{\boldsymbol{q}_{b}}^{\alpha\beta}\delta_{\bm{q}_{1}-\bm{q}_{2},\bm{q}_{b}}+\tilde{T}_{\boldsymbol{q}_{tr}}^{\alpha\beta}\delta_{\bm{q}_{1}-\bm{q}_{2},\bm{q}_{tr}}+\tilde{T}_{\boldsymbol{q}_{tl}}^{\alpha\beta}\delta_{\bm{q}_{1}-\bm{q}_{2},\bm{q}_{tl}},
\end{equation}
with
\begin{align}
\tilde{T}_{\boldsymbol{q}_{b}} & =\frac{t_{\perp}\left(\left|\text{K}\right|\right)}{A_{u.c.}}\left[\begin{array}{cc}
1 & 1\\
1 & 1
\end{array}\right],\\
\tilde{T}_{\boldsymbol{q}_{tr}} & =\frac{t_{\perp}\left(\left|\text{K}\right|\right)}{A_{u.c.}}e^{i\bm{g}_{1,2}\cdot\bm{\tau}_{0}}\left[\begin{array}{cc}
e^{-i\phi} & 1\\
e^{i\phi} & e^{-i\phi}
\end{array}\right],\\
\tilde{T}_{\boldsymbol{q}_{tl}} & =\frac{t_{\perp}\left(\left|\text{K}\right|\right)}{A_{u.c.}}e^{i\bm{g}_{1,3}\cdot\bm{\tau}_{0}}\left[\begin{array}{cc}
e^{i\phi} & 1\\
e^{-i\phi} & e^{i\phi}
\end{array}\right].
\end{align}

\subsection{Electronic properties}

Having obtained a model Hamiltonian capable of describing tBLG, we
will now analyze how the dispersion relation of electrons is affected
by the interlayer coupling.

\subsubsection{Renormalization of the Fermi velocity}

We start by studying perturbatively the effect of interlayer coupling
to states close to the Dirac points of one layer. Let us consider
a state of layer $1$, with crystal-momentum $\text{K}_{1}+\bm{q}$.
According to Eq.~(\ref{eq:Dirac_interlayer}), this state will couple
to states of layer $2$ with crystal momentum $\text{K}_{2}+\bm{q}_{2}$,
with three possibilities for $\boldsymbol{q}_{2}$:
\begin{align}
\bm{q}_{2} & =\bm{q}+\bm{q}_{b},\quad\bm{q}_{2}=\bm{q}+\bm{q}_{tr},\quad\text{or}\quad\bm{q}_{2}=\bm{q}+\bm{q}_{tl}.
\end{align}
Considering only these states, we can build the following truncated
Hamiltonian matrix: 
\begin{equation}
H_{4,\text{tBLG}}^{\text{K}}(\bm{q})=\begin{bmatrix}H_{1}^{\text{K}}(\bm{q}) & T_{\bm{q}_{b}} & T_{\bm{q}_{tr}} & T_{\bm{q}_{tl}}\\
T_{\bm{q}_{b}}^{\dagger} & H_{2}^{\text{K}}(\bm{q}+\bm{q}_{b}) & 0 & 0\\
T_{\bm{q}_{tr}}^{\dagger} & 0 & H_{2}^{\text{K}}(\bm{q}+\bm{q}_{tr}) & 0\\
T_{\bm{q}_{tl}}^{\dagger} & 0 & 0 & H_{2}^{\text{K}}(\bm{q}+\bm{q}_{tl})
\end{bmatrix},\label{eq:4-wave_truncated_Hamiltonian}
\end{equation}
which is written in the basis $\left|1,\text{K}_{1}+\bm{q},\alpha\right\rangle $,
$\left|2,\text{K}_{2}+\bm{q}+\bm{q}_{b},\alpha\right\rangle $, $\left|2,\text{K}_{2}+\bm{q}+\bm{q}_{tr},\alpha\right\rangle $,
$\left|2,\text{K}_{2}+\bm{q}+\bm{q}_{tl},\alpha\right\rangle $.

If $\left|\bm{q}\right|\ll\left|\bm{q}_{n}\right|$ and $t_{\perp}\ll v_{F}\hbar\left|\bm{q}_{n}\right|$,
we can integrate out the states from layer $2$ and obtain an effective
Hamiltonian for layer 1. By doing so, we get
\begin{equation}
H_{1,\text{eff}}^{\text{K}}(\bm{q})=H_{1}^{\text{K}}(\bm{q})-\sum_{n=1}^{3}T_{\boldsymbol{q}_{n}}\cdot\left[H_{2}^{\text{K}}(\bm{q}+\bm{q}_{n})\right]^{-1}\cdot T_{\boldsymbol{q}_{n}}^{\dagger}.
\end{equation}
From Eq.~(\ref{eq:rotated_Dirac_Hamiltonian}), it is straightforward
to see that $\left[H_{2}^{\text{K}}(\bm{q})\right]^{-1}=\left(\bm{\sigma}^{\theta}\cdot\bm{q}\right)/\left(v_{F}\hbar\left|\bm{q}\right|^{2}\right)$.
Expanding to lowest order in $\bm{q}$, we obtain 
\begin{equation}
H_{1,\text{eff}}^{\text{K}}(\bm{q})=H_{1}^{\text{K}}(\bm{q})-\frac{1}{v_{F} \hbar \left|\Delta\text{K}\right|^{2}}\sum_{n=1}^{3}\left[T_{\boldsymbol{q}_{n}}\cdot\bm{\sigma}^{\theta}\cdot T_{\boldsymbol{q}_{n}}^{\dagger}\right]\cdot\left(\bm{q}-\bm{q}_{n}+2\bm{q}_{n}\frac{\bm{q}_{n}\cdot\bm{q}}{\bm{q}_{n}^{2}}\right).
\end{equation}
Performing the sum over $\bm{q}_{n}$, we get
\begin{equation}
H_{1,\text{eff}}^{\text{K}}(\bm{q})=v_{F}\hbar\left(1-9\alpha^{2}\right)\left[\begin{array}{cc}
0 & q_{x}-iq_{y}\\
q_{x}+iq_{y} & 0
\end{array}\right]-6v_{F}\hbar\left|\Delta\text{K}\right|\sin\left(\frac{\theta}{2}\right)\alpha^{2}\left[\begin{array}{cc}
1 & 0\\
0 & 1
\end{array}\right],\label{eq:vF_renorm_Hamiltonian}
\end{equation}
where $\alpha=t_{\perp}\left(\left|\text{K}\right|\right)/\left(v_{F}\hbar\left|\Delta\text{K}\right|A_{u.c.}\right)$.
The second term in Eq.~(\ref{eq:vF_renorm_Hamiltonian}) is just
a shift in the zero of energy. The first term leads to a renormalization
of the Fermi velocity \citep{Santos2007}, 
\begin{equation}
\frac{v_{F}^{*}(\theta)}{v_{F}}=1-\left(\frac{t_{\perp}\left(\left|\text{K}\right|\right)}{v_{F}\hbar\left|\text{K}\right|A_{u.c.}}\right)^{2}\frac{1}{4\sin^{2}\left(\theta/2\right)},
\end{equation}
which shows that, depending on the twist angle, the effective Fermi
velocity can take significantly smaller values. Notice that for small
angles, the Fermi velocity can actually become zero. Clearly, for
very small angles we can no longer assume that $t_{\perp}\ll v_{F}\hbar\left|\bm{q}_{n}\right|$,
and the perturbative approach breaks down. However, it is true that
at certain ``magic angles'' (which occur for $\theta\lesssim1.05^{\circ}$)
the Fermi velocity does vanish and flat bands appear at the Fermi
level of tBLG \citep{Bistritzer2011}.

\subsubsection{Band structure, density of states and carrier density profile}

\label{subsection:tBLGmatrix}

In order to obtain an accurate description of the electronic properties
of tBLG, we must go beyond the perturbative approach previously described.
To do so, we must go beyond the truncation employed in the Hamiltonian
of Eq.~(\ref{eq:4-wave_truncated_Hamiltonian}). This Hamiltonian
does not include the fact that each of the states $\left|2,\text{K}_{2}+\bm{q}+\bm{q}_{n},\alpha\right\rangle $
of layer $2$ is also coupled to several states of layer $1$, as
described by Eq.~(\ref{eq:Dirac_interlayer_21}), and so on, for
increasing truncation orders.

It is worth noticing that the three transfered momenta in the interlayer
hopping, Eqs.~(\ref{eq:q_b})-(\ref{eq:q_tl}), can be written as

\begin{equation}
\bm{q}_{b}=\Delta\text{K},\quad\bm{q}_{tr}=\Delta\text{K}+\bm{b}_{2}^{m},\quad\bm{q}_{tl}=\Delta\text{K}-\bm{b}_{1}^{m},
\end{equation}
which tells us that, in the interlayer hopping, there is a transference
of momentum by reciprocal lattice vectors of the moiré pattern. This
motivates us to look for eigenstates of the tBLG structure of the
form 
\begin{equation}
\left|\psi_{\bm{q}}^{m}\right\rangle =\sum_{\ell,\alpha,m_{1},m_{2}}u_{\alpha}^{(\ell)}\left(\bm{q},m_{1},m_{2}\right)\left|\ell,\bm{q},\left(m_{1},m_{2}\right),\alpha\right\rangle ,\label{eq:Bloch-wave_expansion_tBLG}
\end{equation}
where we defined

\begin{equation}
\left|\ell,\bm{q},\left(m_{1},m_{2}\right),\alpha\right\rangle \equiv\left|\ell,\text{K}_{\ell}+\bm{q}+m_{1}\bm{b}_{1}^{m}+m_{2}\bm{b}_{2}^{m},\alpha\right\rangle .
\end{equation}

For example, if we consider the states (we will drop the crystal-momenta
$\bm{q}$ and the sublattice index $\alpha$ for simplicity) $\left|1,\left(0,0\right)\right\rangle ,\,\left|2,(0,0)\right\rangle ,\,\left|2,(0,1)\right\rangle \,\left|2,(-1,0)\right\rangle ,\,\left|1,\left(0,-1\right)\right\rangle ,\,\left|1,\left(1,0\right)\right\rangle ,\,\left|1,\left(0,1\right)\right\rangle ,\,\left|1,\left(1,1\right)\right\rangle ,\,\left|1,\left(-1,0\right)\right\rangle ,\,\left|1,\left(-1,-1\right)\right\rangle $,
we obtain the following matrix: 
\begin{equation}
H_{10,\text{tBLG}}^{\text{K}}(\bm{q})=\left[\begin{array}{cccccccccc}
H_{1}^{\text{K}} & T_{\bm{q}_{b}} & T_{\bm{q}_{tr}} & T_{\bm{q}_{tl}} & 0 & 0 & 0 & 0 & 0 & 0\\
T_{\bm{q}_{b}}^{\dagger} & H_{2}^{\text{K}} & 0 & 0 & T_{\bm{q}_{tr}}^{\dagger} & T_{\bm{q}_{tl}}^{\dagger} & 0 & 0 & 0 & 0\\
T_{\bm{q}_{tr}}^{\dagger} & 0 & H_{2}^{\text{K}} & 0 & 0 & 0 & T_{\bm{q}_{b}}^{\dagger} & T_{\bm{q}_{tl}}^{\dagger} & 0 & 0\\
T_{\bm{q}_{tl}}^{\dagger} & 0 & 0 & H_{2}^{\text{K}} & 0 & 0 & 0 & 0 & T_{\bm{q}_{b}}^{\dagger} & T_{\bm{q}_{tr}}^{\dagger}\\
0 & T_{\bm{q}_{tr}} & 0 & 0 & H_{1}^{\text{K}} & 0 & 0 & 0 & 0 & 0\\
0 & T_{\bm{q}_{tl}} & 0 & 0 & 0 & H_{1}^{\text{K}} & 0 & 0 & 0 & 0\\
0 & 0 & T_{\bm{q}_{b}} & 0 & 0 & 0 & H_{1}^{\text{K}} & 0 & 0 & 0\\
0 & 0 & T_{\bm{q}_{tl}} & 0 & 0 & 0 & 0 & H_{1}^{\text{K}} & 0 & 0\\
0 & 0 & 0 & T_{\bm{q}_{b}} & 0 & 0 & 0 & 0 & H_{1}^{\text{K}} & 0\\
0 & 0 & 0 & T_{\bm{q}_{tr}} & 0 & 0 & 0 & 0 & 0 & H_{1}^{\text{K}}
\end{array}\right],\label{eq:10_wave_expansion}
\end{equation}
where for compactness we have also suppressed the momenta argument
of $H_{1/2}^{\text{K}}$, which should read as $\bm{q}+m_{1}\bm{b}_{1}^{m}+m_{2}\bm{b}_{2}^{m}$
for $H_{1}^{\text{K}}$ and $\bm{q}+\bm{q}_{b}+m_{1}\bm{b}_{1}^{m}+m_{2}\bm{b}_{2}^{m}$
for $H_{2}^{\text{K}}$. By computing the eigenvalues of $H_{10,\text{tBLG}}^{\text{K}}(\bm{q})$
we obtain an approximation for the electronic band structure in the
moiré BZ. The index $10$ means that we are considering ten moiré
reciprocal lattice vectors, which due to the sublattice degree of
freedom, implies that $H_{10,\text{tBLG}}^{\text{K}}(\bm{q})$ is
a $20\times20$ matrix. By considering more and more moiré reciprocal
lattice vectors, this approximation can be improved. In this matrix
construction, we point out the similarities with what we have shown
for the folded band description of SLG in section~\ref{subsection:SLGkspace}.
In fact, if we disregard the interlayer (off-diagonal) hopping terms,
we see that we are basically using a folded band description that
explicitly captures the moiré periodicity to some extent (depending
on the truncation order). In real space, this corresponds to take
an enlarged unit cell with the moiré periodicity.

\begin{figure}
\centering{}\includegraphics[width=0.6\textwidth]{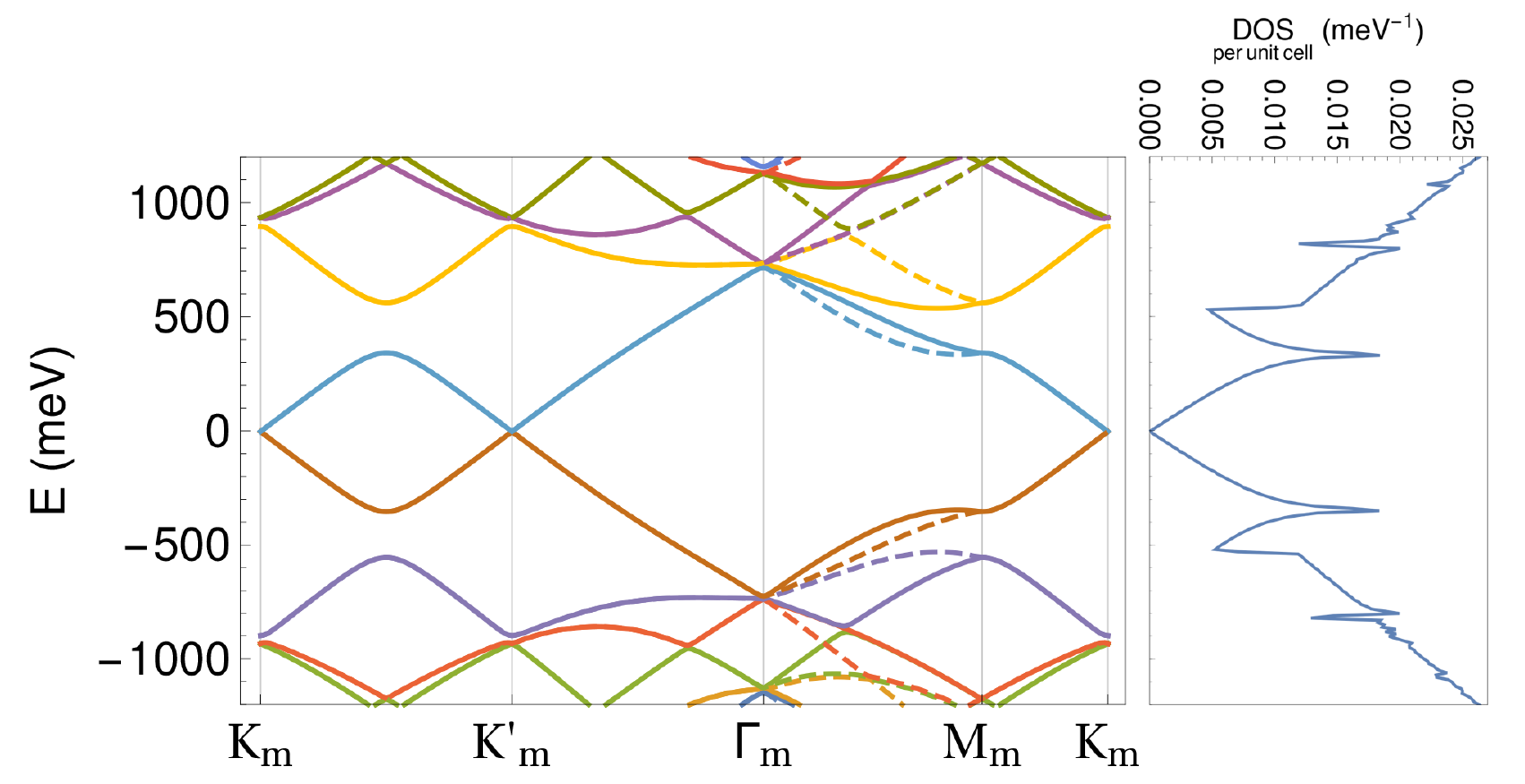}\caption{Electronic spectrum and DOS for tBLG with $\theta=5^{\circ}$. Solid
and dashed lines in the spectrum are for $\text{K}$ and $\text{K}'$
expansions, respectively; the color code clarifies the situation in
which $\text{K}$ and $\text{K}'$ bands are superimposed.}
\label{fig:tBLGspectrumDOS5deg}
\end{figure}

\begin{figure}
\centering{}\includegraphics[width=0.6\textwidth]{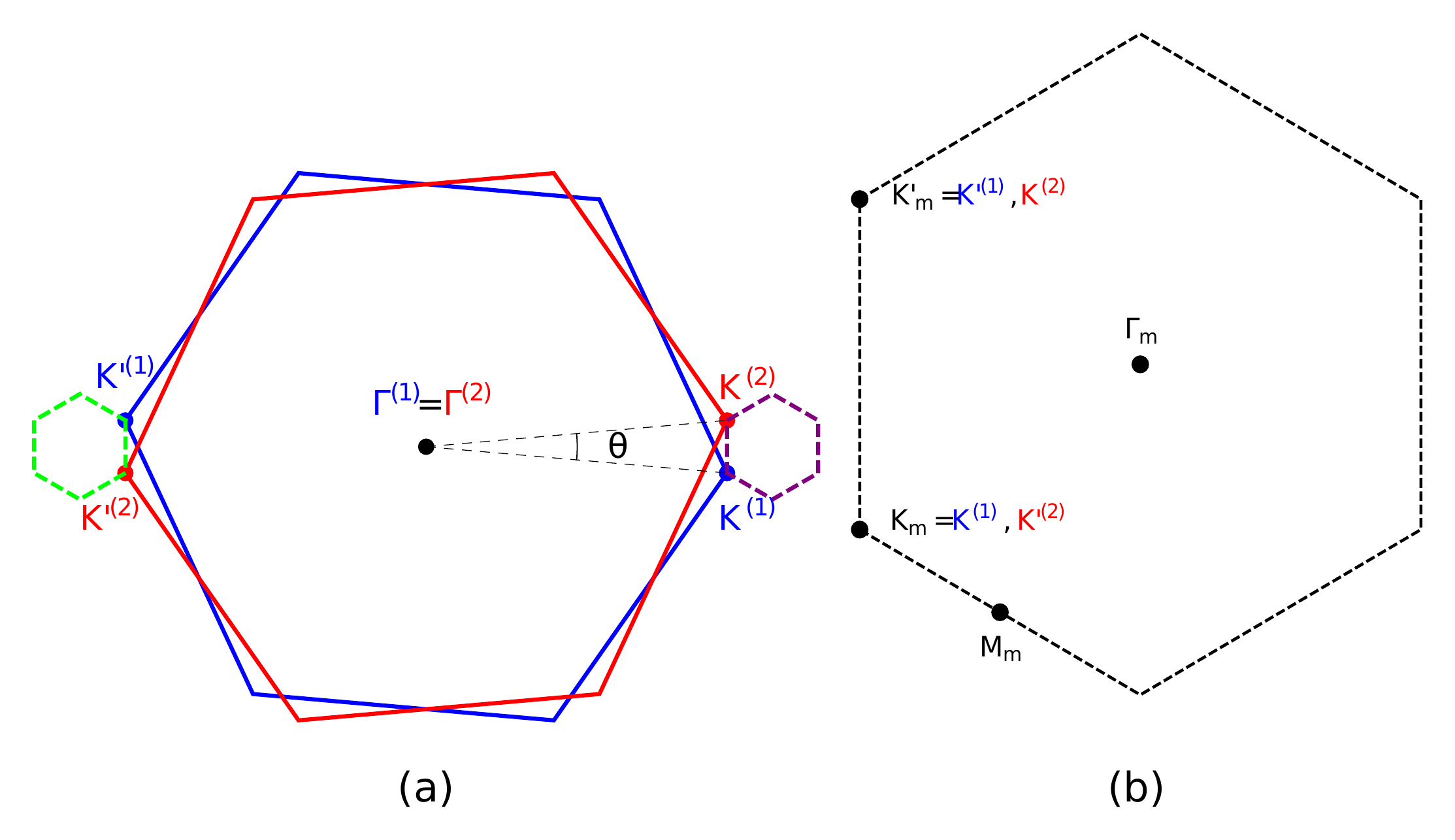}\caption{Picture of $\text{K}$ and $\text{K}^{\prime}$ expansions on a tBLG.
(a) Blue/red hexagons describe the BZs for layers 1/2. Dashed purple/green
hexagons represent moiré unit cells in reciprocal space for $\text{K}$/$\text{K}^{\prime}$
expansions. (b) Moiré BZ with relevant points plotted in it.}
\label{fig:tBLGKbands}
\end{figure}

Results for the electronic spectrum and the DOS, taking contributions
from states close to $\text{K}$ and $\text{K}^{\prime}$, are shown
in Fig.~\ref{fig:tBLGspectrumDOS5deg}. We show how both $\text{K}$
and $\text{K}^{\prime}$ points are folded into the same moiré BZ
in Fig. \ref{fig:tBLGKbands}. It is apparent that, in a $\text{K}$
expansion, the wave vector $\bm{q}$ is measured from $\text{K}_{1}$
($\bm{k}=\text{K}_{1}+\bm{q}$) while, in a $\text{K}^{\prime}$ expansion,
we measure it from $\text{K}_{1}^{\prime}$. Therefore, in order to
match both moiré unit cells in reciprocal space (purple and green),
we identify the points $\text{K}_{1}$ and $\text{K}_{2}^{\prime}$
as the same point in the moiré BZ, such that the paths $\text{K}_{m}\rightarrow\text{K}_{m}^{\prime}\rightarrow\text{M}_{m}\rightarrow\text{K}_{m}$
become equivalent. By doing so, we are making a correspondence $H^{\text{K}}(\bm{q})\leftrightarrow H^{\text{K}^{\prime}}(\bm{q}+\bm{q}_{b})$
in the Hamiltonians obtained within $\text{K}$ and $\text{K}^{\prime}$
expansions.

Looking at the spectrum of Fig. \ref{fig:tBLGspectrumDOS5deg}, we
see that a symmetry for positive and negative bands is apparently
conserved. As predicted in the previous section, we also observe a
renormalization of the Fermi velocity \citep{Santos2007,Bistritzer2011}.
In addition, we verify the emergence of low energy van Hove singularities.
These singularities are due to the avoided crossings of the Dirac
cones of the two layers that occur at the middle point between $\text{K}_{1}$
and $\text{K}_{2}$. Therefore, it is possible to tune the position
in energy of these van Hove singularities by varying the twist angle.
In this chapter, we have avoided the so-called ``magic angles''
\citep{Bistritzer2011}, for which the Fermi velocity vanishes due
to the merging of the two van Hove singularities.

\begin{figure}
\centering{}\includegraphics[width=0.4\textwidth]{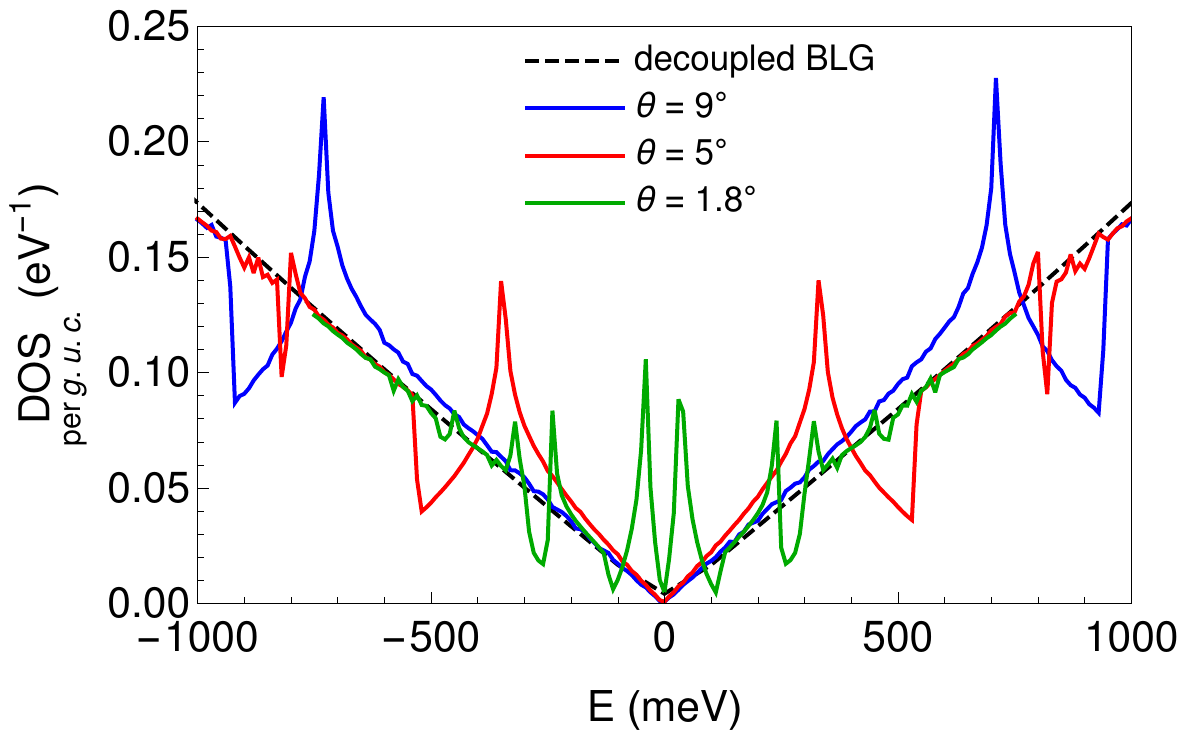}~~\includegraphics[width=0.4\textwidth]{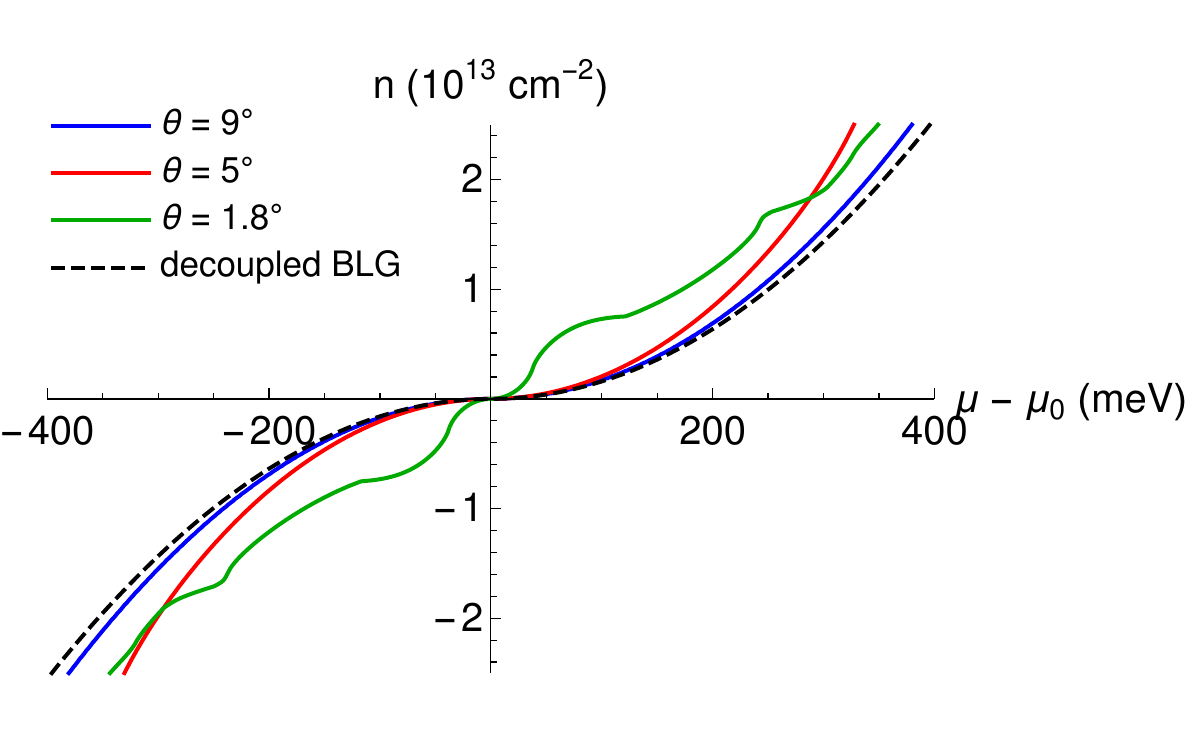}\caption{DOS and carrier density profile for different angles of a tBLG. Since
the size of the unit cells varies with the angle, the DOS is normalized
to the graphene unit cell (g.u.c.). $\mu_{0}$ is the Fermi level
at half filling. Decoupled BLG corresponds to tBLG with $t_{\perp}=0$.}
\label{fig:tBLGDOSandcarrierdensity_variousdeg} 
\end{figure}

The DOS and carrier density profile are plotted in Fig. \ref{fig:tBLGDOSandcarrierdensity_variousdeg}
for different twist angles $\theta$. We confirm that, by varying
the twist angle, van Hove singularities can be brought to experimentally
accessible energies. As for the carrier density profile, we observe
that we start to lose the signature behavior of the decoupled BLG
—tBLG with $t_{\perp}=0$— when we approach small angles.

The method described in this section to evaluate the moiré band structure
of tBLG is analogous to the plane-wave expansion of the form 
\begin{equation}
\psi_{\bm{k},n}(\bm{r})=\sum_{\bm{G}}u_{\bm{k},n}\left(\bm{G}\right)e^{i\left(\bm{k}+\bm{G}\right)\cdot\bm{r}},\label{eq:plane_wave_expansion}
\end{equation}
which can be used to determine the electronic spectrum of a periodic
system with lattice $\left\{ \bm{R}\right\} $ and reciprocal lattice
$\left\{ \bm{G}\right\} $. The main difference is that in the present
case the expansion is made in terms of Bloch waves. There is, yet,
another important difference. The plane-wave expansion in a periodic
system, Eq.~(\ref{eq:plane_wave_expansion}), always contains an
infinite number of $\bm{G}$ vectors, which is then truncated, leading
to electronic bands evaluated with a certain numerical precision.
The expansion for tBLG in Eq.~(\ref{eq:Bloch-wave_expansion_tBLG})
can be either infinite or finite. For a commensurate system, there
will be a certain $\bm{G}^{m}$ that coincides with a reciprocal lattice
vector of both individual layers, such that an expansion of the form
of Eq.~(\ref{eq:Bloch-wave_expansion_tBLG}) becomes finite. For
an incommensurate structure there is never a $\bm{G}^{m}$ that coincides
with reciprocal lattice vectors of both layers and therefore the expansion
in Eq.~(\ref{eq:Bloch-wave_expansion_tBLG}) is formally infinite.
This also has an important physical consequence. In a incommensurate
structure, the electronic properties are independent of the in-plane
translation $\bm{\tau}_{0}$. This can be seen by performing a unitary
transformation in the basis states \citep{Bistritzer2011,Koshino2015},

\begin{equation}
\left|\ell,\boldsymbol{q},(m_{1},m_{2}),\alpha\right\rangle \rightarrow e^{i\left(m_{1}\boldsymbol{b}_{1,1}+m_{2}\boldsymbol{b}_{1,2}\right)\cdot\bm{\tau}_{0}}\left|\ell,\boldsymbol{q},(m_{1},m_{2}),\alpha\right\rangle ,
\end{equation}
which makes the Hamiltonians independent of $\bm{\tau}_{0}$ (check,
for instance, the Hamiltonian of Eq. \eqref{eq:10_wave_expansion}),
allowing us to set $\bm{\tau}_{0}=\bm{0}$ without any loss of generality.
Such transformation does not eliminate the $\bm{\tau}_{0}$ dependency
in a commensurate structure, as can be easily understood by comparing
the Hamiltonians for AA and AB stacked BLG.

Finally, we finish with a discussion about the validity of the model.
The leading corrections involve hopping amplitudes that, due to the
momentum conservation $\text{K}_{1}+\bm{q}_{1}+\bm{G}_{1}=\text{K}_{2}+\bm{q}_{2}+\bm{G}_{2}$,
are negligible when compared to $t_{\perp}(\left|\text{K}\right|)$.
We should also not forget that we are using a Dirac approximation
for the individual layers. Therefore, we expect our model to be accurate
up to energies of $\sim1\si{\electronvolt}$, which can still capture
the first low-energy bands for $\theta\lesssim10^{\circ}$. In case
of larger angles, it is still possible \citep{Koshino2015} to apply
the same technology presented here, but one must consider the general
form of the interlayer coupling, Eq.~(\ref{eq:interlayer_hopping_Bloch}),
and the full tight-binding Hamiltonian of the individual layers, Eq.
\eqref{eq:SLG_hamiltonian_rotated}.

%% file: sections/Section_Optical.tex
\section{Optical response}

\label{chapter:opticalresponse}

The study of light-matter interactions is a topic of interest in science,
with a wide variety of applications, for example in the field of photonics.
For the tBLG system, the response to an applied electromagnetic field
can be characterized by its optical conductivity, which has been measured
experimentally \citep{Wang2010,Zou2013,Cao2016}. On the theoretical
side, we highlight the following works: in Ref. \citep{Tabert2013},
the authors used the simplified model from Ref. \citep{Gail2011}
to study the dynamic (frequency dependent) conductivity at different
levels of chemical potential; tight-binding-based calculations of
the dynamic conductivity were performed by Moon and Koshino \citep{Moon2013};
the real and imaginary parts of the conductivity were calculated by
Stauber et al. \citep{Stauber2013}, using a continuum low-energy
model based on Refs. \citep{Santos2007,Bistritzer2011}.

\begin{figure}
\centering{}\includegraphics[width=0.5\textwidth]{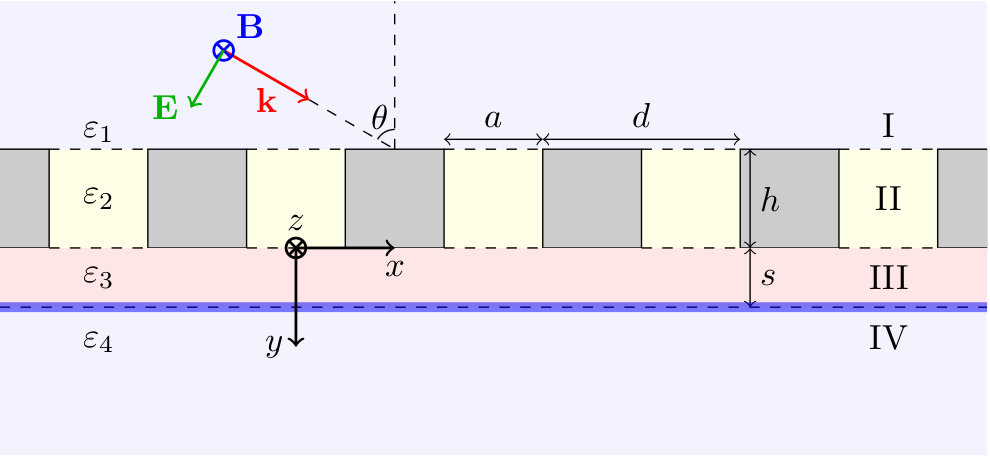}\caption{Possible scheme to excite SPPs in graphene: a graphene layer (blue
line) is located between two dielectric media (III and IV), with a
periodic grid of metallic micro-ribbons placed on top; the polarized
light is shined from the outside (medium I). The periodic grid provides
the momentum that is required to satisfty the energy-momentum conservation
relation between the incident light and the excited SPP.}
\label{fig:GSPPs_scheme} 
\end{figure}

Over the last few years, graphene plasmonics has emerged as a new
research topic. Surface plasmon-polaritons (SPPs) are collective excitations
of coupled charge density modulations and photons, which propagate
along surfaces. The interest on plasmons in graphene picked up after
the experimental excitation of graphene SPPs in the $\si{\tera\hertz}$
spectral range by Ju et al. in 2011 \citep{Ju2011}. The excitation
of graphene SPPs was achieved by shining electromagnetic radiation
onto a periodic grid of graphene micro-ribbons, in a way that the
periodic grid provides the momentum that the light lacks for exciting
the plasmons. Other methods of excitation are also possible, as the
one depicted in Fig.~\ref{fig:GSPPs_scheme}, where a continuous
graphene sheet is shined with light that goes through a metallic grid.
Owing to the 2D nature of the collective excitations, graphene SPPs
are confined much more strongly than those in conventional metals
(particularly in the $\si{\tera\hertz}$ spectral range), making it
a promising candidate for future applications \citep{Jablan2009,Koppens2011,Luo2013}.
In addition, perhaps the most important advantage of using graphene
is the tunability of the graphene SPPs, since carrier densities in
graphene can be easily controlled by electrical gating and doping
\citep{Ju2011,Vakil2011,Fang2012,Fei2012,Chen2012,Grigorenko2012}.
As we will see, within the semi-classical model, the dispersion relation
of SPPs in graphene depends explicitly on the optical conductivity,
wherefore the study of their spectrum follows as a direct application.
For an in-depth introduction to the field of plasmonics in graphene,
we point the interested reader to Ref. \citep{Goncalves2016}.

The goal of this section is to study the response of the tBLG system
to an electromagnetic stimuli. We begin with the optical conductivity,
which we evaluate having as a starting point the electronic Hamiltonian
described in the previous section. Within the linear response theory,
we take the velocity gauge \cite{Passos2018} and derive expressions
that can be implemented when an analytical Hamiltonian matrix is known.
As benchmark, we compute the results for the SLG; then, we apply the
same method to the tBLG. Next, we study the dispersion relation of
SPPs supported by tBLG. We consider few-layer graphene embedded in
dielectric media and derive the equation that describes the propagation
of transverse magnetic waves along the 2D surface, which depends on
the dynamic conductivity. Again, we make the calculations for both
the SLG and the tBLG. We point out that SPPs in tBLG were first studied
by Stauber et al. \citep{Stauber2013}.

\subsection{Conductivity}

\label{section:conductivity}

\subsubsection{Linear response theory}

\label{subsection:conductivity_linearresponsetheory}

\paragraph{\uline{General tight-binding description}\protect \protect \protect \\
 }

We recall that the starting point of our description of tBLG was a
tight-binding Hamiltonian, which in general can be written as in Eq.~(\ref{eq:general_TB_Hamiltonian}),
\begin{equation}
H=\sum_{\bm{R},\bm{\delta},\alpha,\beta}c_{\alpha}^{\dagger}\left(\bm{R}\right)h_{\bm{\delta}}^{\alpha\beta}c_{\beta}\left(\bm{R}+\bm{\delta}\right).
\end{equation}
This Hamiltonian can be coupled to an external electromagnetic field
using a minimal coupling approach \citep{Doughty1990}. For a tight-binding
model, minimal coupling reduces to the Peierls substitution, where
each hopping $h_{\bm{\delta}}^{\alpha\beta}$ is multiplied by the
phase that the electron acquires when hopping from one atomic center
to other. Therefore, the Hamiltonian becomes 
\begin{equation}
H_{A}(t)=\sum_{\bm{R},\bm{\delta},\alpha,\beta}c_{\alpha}^{\dagger}\left(\bm{R}\right)\exp\left[-i\frac{e}{\hbar}\int_{\bm{R}+\bm{\delta}+\bm{\tau}_{\beta}}^{\bm{R}+\bm{\tau}_{\alpha}}d\bm{r}\cdot\bm{A}(\bm{r},t)\right]h_{\bm{\delta}}^{\alpha\beta}c_{\beta}\left(\bm{R}+\bm{\delta}\right),
\end{equation}
where $\bm{A}(\bm{r},t)$ is the electromagnetic vector potential
and $e>0$ is the elementary charge.

In the following, we will be interested in the response to homogeneous
electromagnetic fields, for which the previous expression reduces
to
\begin{equation}
H_{A}(t)=\sum_{\bm{R},\bm{\delta},\alpha,\beta}c_{\alpha}^{\dagger}\left(\bm{R}\right)\exp\left[-i\frac{e}{\hbar}\bm{A}(t)\cdot\left(\bm{\tau}_{\alpha}-\bm{\delta}-\bm{\tau}_{\beta}\right)\right]h_{\bm{\delta}}^{\alpha\beta}c_{\beta}\left(\bm{R}+\bm{\delta}\right).
\end{equation}

 Writing the fermionic operators according to Eq.~(\ref{eq:localized_to_Bloch_operator}),
\begin{equation}
c_{\alpha}^{\dagger}(\bm{R})=\frac{1}{\sqrt{N}}\sum_{\bm{k}}e^{-i\bm{k}\cdot\left(\bm{R}+\bm{\tau}_{\alpha}\right)}c_{\alpha}^{\dagger}(\bm{k}),
\end{equation}
we get
\begin{equation}
H_{A}(t)=\sum_{\bm{k},\alpha,\beta}c_{\alpha}^{\dagger}\left(\bm{k}\right)h_{0}\left(\bm{k}+\frac{e}{\hbar}\bm{A}(t)\right)c_{\beta}\left(\bm{k}\right),\label{eq:Hamiltonian_minimal_coupling}
\end{equation}
where $h_{0}\left(\bm{k}\right)=\sum_{\bm{\delta}}h_{\bm{\delta}}^{\alpha\beta}e^{i\bm{k}\cdot\left(\bm{\delta}+\bm{\tau}_{\beta}-\bm{\tau}_{\alpha}\right)}$
are the matrix elements of the reciprocal-space tight-binding Hamiltonian
in the $\alpha,\beta$ basis. For SLG, $\alpha,\beta$ are sublattice
indeces whereas, for Bernal-stacked BLG, they label both sublattice
and layer. In the case of tBLG, $\alpha,\beta$ run over all the entries
of matrices as the one shown in Eq. \eqref{eq:10_wave_expansion}.
In this case, we also have to separate the sum over $\boldsymbol{k}\in\text{BZ}$
into two sums over $\boldsymbol{q}$ in moiré BZs centered around
$\text{K}$ and $\text{K}^{\prime}$. Finally, we stress that, in
the $\alpha,\beta$ basis, the Hamiltonians are not diagonal.

~

\paragraph{\uline{Perturbative treatment to the minimal coupling}\protect
\protect \protect \\
 }

Starting from the Hamiltonian given by Eq.~(\ref{eq:Hamiltonian_minimal_coupling})
and expanding it in $\bm{A}(t)$, we obtain the standard description
of an unperturbed Hamiltonian $H_{0}$ plus a time-dependent perturbation
$V(t)$, 
\begin{equation}
H_{A}(t)=H_{0}+V(t),
\end{equation}
with

\begin{equation}
H_{0}=\sum_{\bm{k},\alpha,\beta}c_{\alpha}^{\dagger}\left(\bm{k}\right)h_{0}\left(\bm{k}\right)c_{\beta}\left(\bm{k}\right),
\end{equation}

\begin{equation}
V(t)=\sum_{\bm{k},\alpha,\beta}\left(\frac{e}{\hbar}\frac{\partial h_{0}(\bm{k})}{\partial k_{a_{1}}}A_{a_{1}}(t)+\frac{1}{2!}\left(\frac{e}{\hbar}\right)^{2}\frac{\partial^{2}h_{0}(\bm{k})}{\partial k_{a_{1}}\partial k_{a_{2}}}A_{a_{1}}(t)A_{a_{2}}(t)+...\right)c_{\bm{k},\alpha}^{\dagger}c_{\bm{k},\beta}.\label{eq:potential(t)}
\end{equation}
In the equation above, we clarify that we are using Einstein's summation
convention for the mute indices $a_{i}=x,y$. It must also be noted
that, in the Dirac approximation for graphene, only the first term
is non-zero. However, for a general tight-binding Hamiltonian, terms
to all orders in $\bm{A}(t)$ exist.

The homogenous 2D current density operator can be obtained as
\begin{equation}
J_{a_{1}}(t)=-\frac{1}{A}\frac{\partial H_{A}(t)}{\partial A_{a_{1}}}=-\frac{e}{\hbar A}\sum\limits _{\bm{k},\alpha,\beta}\left(\frac{\partial h_{0}(\bm{k})}{\partial k_{a_{1}}}+\frac{e}{\hbar}\frac{\partial^{2}h_{0}(\bm{k})}{\partial k_{a_{1}}\partial k_{a_{2}}}A_{a_{2}}(t)+...\right)c_{\bm{k},\alpha}^{\dagger}c_{\bm{k},\beta},\label{eq:current(t)}
\end{equation}
where $A$ is the area of the 2D system. Using the time-dependent
perturbation theory in the interaction representation, we get, for
the average current density, 
\begin{align}
\braket{J_{a_{1}}^{I}(t)}= & \braket{J_{a_{1}}^{I}(t)}_{0}+\left(-\frac{i}{\hbar}\right)\int_{t_{0}}^{t}dt_{1}\left\langle \left[J_{a_{1}}^{I}(t),V_{I}(t_{1})\right]\right\rangle _{0}\nonumber \\
 & +\left(-\frac{i}{\hbar}\right)^{2}\int_{t_{0}}^{t}dt_{1}\int_{t_{0}}^{t_{1}}dt_{2}\left\langle \left[\left[J_{a_{1}}^{I}(t),V_{I}(t_{1})\right],V_{I}(t_{2})\right]\right\rangle _{0}+...\quad,\label{eq:averagecurrent_interactionrep}
\end{align}
where $\braket{\ }_{0}$ represents a thermal average over unperturbed
states. Here, we are assuming that the initial condition of our system
($t=t_{0})$ is a thermal state of the unperturbed Hamiltonian $H_{0}$.

We now want to write Eqs. (\ref{eq:potential(t)}) and (\ref{eq:current(t)})
in the interaction picture. First, we change to the basis $\left|\lambda\right\rangle $,
which diagonalizes $H_{0}(\bm{k})$ with eigenvalues $\epsilon_{\lambda}(\boldsymbol{k})=\hbar\omega_{\lambda}$
(where we have omitted the momentum dependency of the angular frequencies
$\omega_{\lambda}$), and write the fermionic operators in the interaction
picture as 
\begin{equation}
c_{\bm{k},\alpha}=\sum\limits _{\lambda}\left\langle \alpha|\lambda\right\rangle e^{-i\omega_{\lambda}(t-t_{0})}c_{\bm{k},\lambda},\quad c_{\bm{k},\alpha}^{\dagger}=\sum\limits _{\lambda}\left\langle \lambda|\alpha\right\rangle e^{i\omega_{\lambda}(t-t_{0})}c_{\bm{k},\lambda}^{\dagger}.\label{eq:fermionicop_interactionrep}
\end{equation}
Then, we plug Eq. (\ref{eq:fermionicop_interactionrep}) into Eqs.
(\ref{eq:potential(t)}) and (\ref{eq:current(t)}) and, using the
closure relation, $\sum\limits _{\gamma}\ket{\gamma}\bra{\gamma}=1$,
we obtain

\begin{equation}
V_{I}(t)=\sum\limits _{\bm{k},\lambda,\lambda^{\prime}}\bra{\lambda}\left(\frac{e}{\hbar}\frac{\partial h_{0}(\bm{k})}{\partial k_{a_{1}}}A_{a_{1}}(t)+\frac{1}{2!}\left(\frac{e}{\hbar}\right)^{2}\frac{\partial^{2}h_{0}(\bm{k})}{\partial k_{a_{1}}\partial k_{a_{2}}}A_{a_{1}}(t)A_{a_{2}}(t)+...\right)\ket{\lambda'}e^{i\omega_{\lambda\lambda'}(t-t_{0})}c_{\bm{k},\lambda}^{\dagger}c_{\bm{k},\lambda'},
\end{equation}
\begin{equation}
J_{a_{1}}^{I}(t)=-\frac{e}{\hbar A}\sum\limits _{\bm{k},\lambda,\lambda^{\prime}}\bra{\lambda}\left(\frac{\partial h_{0}(\bm{k})}{\partial k_{a_{1}}}+\frac{e}{\hbar}\frac{\partial^{2}h_{0}(\bm{k})}{\partial k_{a_{1}}\partial k_{a_{2}}}A_{a_{2}}(t)+...\right)\ket{\lambda'}e^{i\omega_{\lambda\lambda'}(t-t_{0})}c_{\bm{k},\lambda}^{\dagger}c_{\bm{k},\lambda'},
\end{equation}
where we have defined $\omega_{\lambda\lambda'}=\omega_{\lambda}-\omega_{\lambda'}$.

~

\paragraph{\uline{Equilibrium current}\protect \protect \protect \\
 }

Collecting the zeroth-order terms (in the fields) from the average
current, Eq. (\ref{eq:averagecurrent_interactionrep}), we obtain
the so-called equilibrium current, 
\begin{equation}
J_{a_{1}}^{0}(t)=-\frac{e}{\hbar A}\sum_{\bm{k},\lambda,\lambda^{\prime}}\left\langle \lambda\left|\frac{\partial h_{0}(\bm{k})}{\partial k_{a_{1}}}\right|\lambda'\right\rangle e^{i\omega_{\lambda\lambda'}(t-t_{0})}\left\langle c_{\bm{k},\lambda}^{\dagger}c_{\bm{k},\lambda'}\right\rangle _{0}.
\end{equation}
In the equation above, the thermal average is trivially computed as
$\braket{c_{\bm{k},\lambda}^{\dagger}c_{\bm{k},\lambda'}}_{0}=\delta_{\lambda,\lambda'}\ n_{F}\left(\epsilon_{\lambda}(\bm{k})\right)$,
where $n_{F}$ stands for the Fermi-Dirac function, $n_{F}\left(\epsilon\right)=\left(e^{\frac{\epsilon-\mu}{k_{B}T}}+1\right)^{-1}$,
in which $k_{B}$ is the Boltzmann constant, $T$ is the absolute
temperature and $\mu$ is the Fermi level.

Taking into account the time inversion symmetry in reciprocal space,
we can show that the equilibrium current is zero, $J_{a_{1}}^{0}(t)=0$.
We point out that, even in our model for the tBLG, time inversion
symmetry is not broken: we can explicitly see that, for every point
$\bm{q}$ in the moiré BZ centered around K ($\bm{k}=\text{K}+\bm{q}$),
we have a completely equivalent point $-\bm{k}=-\text{K}-\bm{q}$,
which corresponds to a point $-\bm{q}$ in the moiré BZ centered around
$\text{K}^{\prime}=-\text{K}$.

~

\paragraph{\uline{Linear response current and conductivity}\protect \protect
\protect \\
 }

We now collect the first order terms, which lead to the following
linear response current: 
\begin{align}
J_{a_{1}}^{1}(t)= & -\frac{e^{2}}{\hbar^{2}A}\sum\limits _{\bm{k},\lambda,\lambda^{\prime}}\left\langle \lambda\left|\frac{\partial^{2}h_{0}}{\partial k_{a_{1}}\partial k_{a_{2}}}\right|\lambda'\right\rangle A_{a_{2}}(t)e^{i\omega_{\lambda\lambda'}(t-t_{0})}\left\langle c_{\bm{k},\lambda}^{\dagger}c_{\bm{k},\lambda'}\right\rangle _{0}\nonumber \\
 & +\frac{ie^{2}}{\hbar^{3}A}\sum\limits _{\bm{k},\lambda_{1},\lambda_{2}}\sum\limits _{\bm{k}^{\prime},\lambda_{3},\lambda_{4}}\left\langle \lambda_{1}\left|\frac{\partial h_{0}}{\partial k_{a_{1}}}\right|\lambda_{2}\right\rangle \left\langle \lambda_{3}\left|\frac{\partial h_{0}}{\partial k_{a_{2}}^{\prime}}\right|\lambda_{4}\right\rangle \left\langle \left[c_{\bm{k},\lambda_{1}}^{\dagger}c_{\bm{k},\lambda_{2}}\ ,\ c_{\bm{k}',\lambda_{3}}^{\dagger}c_{\bm{k}',\lambda_{4}}\right]\right\rangle _{0}\times\nonumber \\
 & \times\int_{t_{0}}^{t}dt_{1}\ e^{i\omega_{\lambda_{1}\lambda_{2}}(t-t_{0})}e^{i\omega_{\lambda_{3}\lambda_{4}}(t_{1}-t_{0})}A_{a_{2}}(t_{1}),\label{eq:J1_full}
\end{align}
where, for simplicity, we have supressed the momentum argument in
$h_{0}$.

Using the fermionic commutation relations, we obtain
\[
\left\langle \left[c_{\bm{k},\lambda_{1}}^{\dagger}c_{\bm{k},\lambda_{2}}\ ,\ c_{\bm{k}',\lambda_{3}}^{\dagger}c_{\bm{k}',\lambda_{4}}\right]\right\rangle _{0}=\delta_{\bm{k},\bm{k}'}\delta_{\lambda_{1},\lambda_{4}}\delta_{\lambda_{2},\lambda_{3}}\left(n_{F}(\epsilon_{\lambda_{1}})-n_{F}(\epsilon_{\lambda_{2}})\right),
\]
where we have also omitted the momentum dependency of the eigenvalues,
and use this to simplify Eq. \eqref{eq:J1_full} into 
\begin{align}
J_{a_{1}}^{1}(t)= & -\frac{e^{2}}{\hbar^{2}A}\sum\limits _{\bm{k},\lambda}\left\langle \lambda\left|\frac{\partial^{2}h_{0}}{\partial k_{a_{1}}\partial k_{a_{2}}}\right|\lambda\right\rangle n_{F}(\epsilon_{\lambda})A_{a_{2}}(t)\ \nonumber \\
 & +\frac{ie^{2}}{\hbar^{3}A}\sum\limits _{\bm{k},\lambda_{1},\lambda_{2}}\left\langle \lambda_{1}\left|\frac{\partial h_{0}}{\partial k_{a_{1}}}\right|\lambda_{2}\right\rangle \left\langle \lambda_{2}\left|\frac{\partial h_{0}}{\partial k_{a_{2}}}\right|\lambda_{1}\right\rangle \left(n_{F}(\epsilon_{\lambda_{1}})-n_{F}(\epsilon_{\lambda_{2}})\right)\int_{t_{0}}^{t}dt_{1}\ e^{i\omega_{\lambda_{1}\lambda_{2}}(t-t_{1})}A_{a_{2}}(t_{1}).\label{eq:J1_simplified}
\end{align}

At this point, we apply a Fourier transform to the magnetic vector
potential, 
\begin{equation}
A_{a}(t)=\int_{\mathbb{R}}\frac{d\omega}{2\pi}A_{a}(\omega)e^{-i\omega t},
\end{equation}
where $\omega$ is the angular frequency of the light, and use the
relation between the Fourier amplitude of the magnetic vector potential,
$A_{a}(\omega)$, and the Fourier amplitude of the electric field,
$E_{a}(\omega)$, 
\begin{equation}
A_{a}(\omega)=\frac{E_{a}(\omega)}{i\omega},
\end{equation}
to obtain 
\begin{equation}
A_{a}(t)=\int_{\mathbb{R}}\frac{d\omega}{2\pi}\frac{E_{a}(\omega)}{i\omega}e^{-i\omega t},\label{eq:AFTE}
\end{equation}
where, in the adiabatic regime, we make $\omega\rightarrow\omega+i\gamma,\ \gamma\rightarrow0^{+}$,
meaning that we switch on the electromagnetic fields very slowly.

Substituting Eq. (\ref{eq:AFTE}) into Eq. (\ref{eq:J1_simplified}),
we get 
\begin{align}
J_{a_{1}}^{1}(t)= & \int_{\mathbb{R}}\frac{d\omega}{2\pi}\ \left(\frac{e^{2}}{\hbar^{2}A}\sum\limits _{\bm{k},\lambda}\left\langle \lambda\left|\frac{\partial^{2}h_{0}}{\partial k_{a_{1}}\partial k_{a_{2}}}\right|\lambda\right\rangle \frac{in_{F}(\epsilon_{\lambda})}{\omega}\right)E_{a_{2}}(\omega)e^{-i\omega t}\nonumber \\
 & +\frac{ie^{2}}{\hbar^{3}A}\sum\limits _{\bm{k},\lambda_{1},\lambda_{2}}\left\langle \lambda_{1}\left|\frac{\partial h_{0}}{\partial k_{a_{1}}}\right|\lambda_{2}\right\rangle \left\langle \lambda_{2}\left|\frac{\partial h_{0}}{\partial k_{a_{2}}}\right|\lambda_{1}\right\rangle \left(n_{F}(\epsilon_{\lambda_{1}})-n_{F}(\epsilon_{\lambda_{2}})\right)\times\nonumber \\
 & \times\int_{t_{0}}^{t}dt_{1}\ e^{i\omega_{\lambda_{1}\lambda_{2}}(t-t_{1})}\int_{\mathbb{R}}\frac{d\omega}{2\pi}\frac{E_{a_{2}}(\omega)}{i\omega}e^{-i\omega t_{1}}.
\end{align}
We can compute the integral in time, 
\begin{equation}
\int_{t_{0}}^{t}dt_{1}\ e^{i\omega_{\lambda_{1}\lambda_{2}}(t-t_{1})}e^{-i\omega t_{1}}=e^{i\omega_{\lambda_{1}\lambda_{2}}t}\int_{t_{0}}^{t}dt_{1}\ e^{-i\left(\omega_{\lambda_{1}\lambda_{2}}+\omega\right)t_{1}}=\frac{ie^{-i\omega t}}{\omega_{\lambda_{1}\lambda_{2}}+\omega}+...,
\end{equation}
where we have eliminated the last the term by making $t_{0}\rightarrow-\infty$,
which means that we have waited long enough for the transient terms
to be negligible. Using this result, we simplify the expression for
the linear current density into
\begin{align}
J_{a_{1}}^{1}(t)= & \int_{\mathbb{R}}\frac{d\omega}{2\pi}\ \left(\frac{ie^{2}}{\hbar^{2}A}\sum\limits _{\bm{k},\lambda}\left\langle \lambda\left|\frac{\partial^{2}h_{0}}{\partial k_{a_{1}}\partial k_{a_{2}}}\right|\lambda\right\rangle \frac{n_{F}(\epsilon_{\lambda})}{\omega}\right)E_{a_{2}}(\omega)e^{-i\omega t}\ \nonumber \\
 & +\int_{\mathbb{R}}\frac{d\omega}{2\pi}\ \left(\frac{ie^{2}}{\hbar^{3}A}\sum\limits _{\bm{k},\lambda_{1},\lambda_{2}}\left\langle \lambda_{1}\left|\frac{\partial h_{0}}{\partial k_{a_{1}}}\right|\lambda_{2}\right\rangle \left\langle \lambda_{2}\left|\frac{\partial h_{0}}{\partial k_{a_{2}}}\right|\lambda_{1}\right\rangle \frac{n_{F}(\epsilon_{\lambda_{1}})-n_{F}(\epsilon_{\lambda_{2}})}{\omega(\omega_{\lambda_{1}\lambda_{2}}+\omega)}\right)E_{a_{2}}(\omega)e^{-i\omega t}.
\end{align}

Taking a closer look at the last expression, we identify the conductivity
(rank-2) tensor, $\sigma$, which leads to the current that arises
in response to an electric field ($\bm{J}=\sigma\bm{E}$ in matrix
form), as 
\begin{equation}
\sigma_{a_{1}a_{2}}(\omega)=\frac{i4\sigma_{0}}{NA_{u.c.}}\sum\limits _{\bm{k},\lambda_{1}}\left(\left\langle \lambda_{1}\left|\frac{\partial^{2}h_{0}}{\partial k_{a_{1}}\partial k_{a_{2}}}\right|\lambda_{1}\right\rangle \frac{n_{\lambda_{1}}^{F}}{\hbar\omega}+\sum\limits _{\lambda_{2}\neq\lambda_{1}}\left\langle \lambda_{1}\left|\frac{\partial h_{0}}{\partial k_{a_{1}}}\right|\lambda_{2}\right\rangle \left\langle \lambda_{2}\left|\frac{\partial h_{0}}{\partial k_{a_{2}}}\right|\lambda_{1}\right\rangle \frac{n_{\lambda_{1}}^{F}-n_{\lambda_{2}}^{F}}{\hbar\omega(\epsilon_{\lambda_{1}\lambda_{2}}+\hbar\omega)}\right),
\end{equation}
where $\sigma_{0}=e^{2}/(4\hbar)$ is the graphene universal conductivity,
$\epsilon_{\lambda_{1}\lambda_{2}}=\epsilon_{\lambda_{1}}-\epsilon_{\lambda_{2}}$
and $n_{F}(\epsilon_{\lambda})\equiv n_{\lambda}^{F}$. We recall
that we have omitted the sum over spin; therefore, since we do not
have any spin dependency, we should add a factor of 2 to the conductivity,
which is taken into account in the subsequent sections.

~

\paragraph{\uline{Drude and regular conductivity}\protect \protect \protect \\
 }

It is usual to split the conductivity in a Drude contribution plus
a regular term. Making 
\begin{equation}
\frac{1}{\hbar\omega\left(\epsilon_{\lambda_{1}\lambda_{2}}+\hbar\omega\right)}=\frac{1}{\hbar\omega\ \epsilon_{\lambda_{1}\lambda_{2}}}-\frac{1}{\epsilon_{\lambda_{1}\lambda_{2}}\left(\epsilon_{\lambda_{1}\lambda_{2}}+\hbar\omega\right)},\quad\epsilon_{\lambda_{1}\lambda_{2}}\neq0,
\end{equation}
we write the conductivity as the sum of two terms 
\begin{equation}
\sigma_{a_{1}a_{2}}(\omega)=\sigma_{a_{1}a_{2}}^{D}(\omega)+\sigma_{a_{1}a_{2}}^{reg}(\omega),
\end{equation}
where $\sigma_{a_{1}a_{2}}^{D}(\omega)$ is the Drude conductivity,
\begin{equation}
\sigma_{a_{1}a_{2}}^{D}(\omega)=\frac{8\sigma_{0}i}{NA_{u.c.}\hbar\omega}\sum\limits _{\bm{k},\lambda_{1}}\left(\left\langle \lambda_{1}\left|\frac{\partial^{2}h_{0}}{\partial k_{a_{1}}\partial k_{a_{2}}}\right|\lambda_{1}\right\rangle n_{\lambda_{1}}^{F}+\sum\limits _{\lambda_{2}\neq\lambda_{1}}\left\langle \lambda_{1}\left|\frac{\partial h_{0}}{\partial k_{a_{1}}}\right|\lambda_{2}\right\rangle \left\langle \lambda_{2}\left|\frac{\partial h_{0}}{\partial k_{a_{2}}}\right|\lambda_{1}\right\rangle \frac{n_{\lambda_{1}}^{F}-n_{\lambda_{2}}^{F}}{\epsilon_{\lambda_{1}\lambda_{2}}}\right),
\end{equation}
and the $\sigma_{a_{1}a_{2}}^{reg}(\omega)$ is the regular conductivity,
\begin{equation}
\sigma_{a_{1}a_{2}}^{reg}(\omega)=\frac{-8\sigma_{0}i}{NA_{u.c.}}\sum_{\bm{k},\lambda_{1}\neq\lambda_{2}}\left\langle \lambda_{1}\left|\frac{\partial h_{0}}{\partial k_{a_{1}}}\right|\lambda_{2}\right\rangle \left\langle \lambda_{2}\left|\frac{\partial h_{0}}{\partial k_{a_{2}}}\right|\lambda_{1}\right\rangle \frac{n_{\lambda_{1}}^{F}-n_{\lambda_{2}}^{F}}{\epsilon_{\lambda_{1}\lambda_{2}}\left(\epsilon_{\lambda_{1}\lambda_{2}}+\hbar\omega\right)}.\label{eq:regcond}
\end{equation}

Using the mathematical relation, 
\begin{equation}
\frac{1}{x\pm i\eta}=P\left(\frac{1}{x}\right)\mp i\pi\delta(x),\quad\eta\rightarrow0^{+},\label{eq:1/(x+i0)}
\end{equation}
where $P$ stands for the Cauchy principal value, we can rewrite the
expression for the Drude conductivity in the adiabatic regime as 
\begin{equation}
\sigma_{a_{1}a_{2}}^{D}(\omega)=\frac{i}{\pi}\frac{D_{a_{1}a_{2}}}{\hbar\omega+i\Gamma}\rightarrow D_{a_{1}a_{2}}\left(\delta(\hbar\omega)+P\left(\frac{i}{\pi\hbar\omega}\right)\right),\label{eq:Drudecond}
\end{equation}
where $\Gamma=\hbar\gamma\rightarrow0^{+}$ and the Drude weight is
given by 
\begin{equation}
D_{a_{1}a_{2}}=\frac{8\pi\sigma_{0}}{NA_{u.c.}}\sum\limits _{\bm{k},\lambda_{1}}\left(\left\langle \lambda_{1}\left|\frac{\partial^{2}h_{0}}{\partial k_{a_{1}}\partial k_{a_{2}}}\right|\lambda_{1}\right\rangle n_{\lambda_{1}}^{F}+\sum\limits _{\lambda_{2}\neq\lambda_{1}}\left\langle \lambda_{1}\left|\frac{\partial h_{0}}{\partial k_{a_{1}}}\right|\lambda_{2}\right\rangle \left\langle \lambda_{2}\left|\frac{\partial h_{0}}{\partial k_{a_{2}}}\right|\lambda_{1}\right\rangle \frac{n_{\lambda_{1}}^{F}-n_{\lambda_{2}}^{F}}{\epsilon_{\lambda_{1}\lambda_{2}}}\right).\label{eq:Drudeweight}
\end{equation}
We can thus see that the real part of $\sigma^{D}$ corresponds to
the typical Drude peak for $\omega=0$, characteristic of metals \citep{AshcroftMermin1976}.
Therefore, we interpret this contribution as an intraband term (where
momentum is not conserved), which reflects the response of the electrons
to a static applied electric field. Consequently, the regular conductivity
is understood as an interband term, which corresponds to electronic
band transitions (within the same $\bm{k}$) with energy $\hbar\omega$,
induced by an applied harmonic electric field, $\bm{E}\sim e^{-i\omega t}$.
We also note that we can empirically account for disorder effects
by considering a finite $\Gamma$, which is a broadening parameter
(usually interpreted as a scattering rate) that may depend on intrinsic
and extrinsic aspects, such as impurities, for example.

At this point, we already have expressions to compute the optical
conductivity. Let us clarify the numerical computations by expressing
all the dependencies that were omitted before. For the Drude conductivity,
we compute the Drude weight by Eq. (\ref{eq:Drudeweight}), 
\begin{align}
D_{a_{1}a_{2}}= & \frac{8\pi\sigma_{0}}{NA_{u.c.}}\sum\limits _{\bm{k},\lambda_{1}}\Bigg[\left\langle \lambda_{1},\bm{k}\left|\frac{\partial^{2}h_{0}(\bm{k})}{\partial k_{a_{1}}\partial k_{a_{2}}}\right|\lambda_{1},\bm{k}\right\rangle n_{F}\left(\epsilon_{\lambda_{1}}(\bm{k})\right)\nonumber \\
 & +\sum\limits _{\lambda_{2}\neq\lambda_{1}}\left\langle \lambda_{1},\bm{k}\left|\frac{\partial h_{0}(\bm{k})}{\partial k_{a_{1}}}\right|\lambda_{2},\bm{k}\right\rangle \left\langle \lambda_{2},\bm{k}\left|\frac{\partial h_{0}(\bm{k})}{\partial k_{a_{2}}}\right|\lambda_{1},\bm{k}\right\rangle \frac{n_{F}\left(\epsilon_{\lambda_{1}}(\bm{k})\right)-n_{F}\left(\epsilon_{\lambda_{2}}(\bm{k})\right)}{\epsilon_{\lambda_{1}}(\bm{k})-\epsilon_{\lambda_{2}}(\bm{k})}\Bigg],\label{eq:Drude1}
\end{align}
and then apply Eq. (\ref{eq:Drudecond}) with a finite $\Gamma$,
\begin{equation}
\sigma_{a_{1}a_{2}}^{D}(\omega)=\frac{i}{\pi}\frac{D_{a_{1}a_{2}}}{\hbar\omega+i\Gamma}.
\end{equation}
For the regular conductivity, we use Eq. (\ref{eq:regcond}) with
$\hbar\omega\rightarrow\hbar\omega+i\Gamma$, 
\begin{align}
\sigma_{a_{1}a_{2}}^{reg}(\omega)=\frac{-8\sigma_{0}i}{NA_{u.c.}}\sum_{\bm{k},\lambda_{1}\neq\lambda_{2}}\left\langle \lambda_{1},\bm{k}\left|\frac{\partial h_{0}(\bm{k})}{\partial k_{a_{1}}}\right|\lambda_{2},\bm{k}\right\rangle \left\langle \lambda_{2},\bm{k}\left|\frac{\partial h_{0}(\bm{k})}{\partial k_{a_{2}}}\right|\lambda_{1},\bm{k}\right\rangle \times\nonumber \\
\times\frac{n_{F}\left(\epsilon_{\lambda_{1}}(\bm{k})\right)-n_{F}\left(\epsilon_{\lambda_{2}}(\bm{k})\right)}{\left[\epsilon_{\lambda_{1}}(\bm{k})-\epsilon_{\lambda_{2}}(\bm{k})\right]\left[\epsilon_{\lambda_{1}}(\bm{k})-\epsilon_{\lambda_{2}}(\bm{k})+\hbar\omega+i\Gamma\right]}.\label{eq:regcond2}
\end{align}

These expressions must work when we have the complete Hamiltonian
defined in the full BZ. However, for effective Hamiltonians, they
might not be appropriate. In particular, when computing the Drude
weight, we expect that all the dependency comes from the electrons
near the Fermi level, which are the ones that can flow in response
to the static applied electric field. Yet, this is not explicit in
our expression, which indicates that there should be an underlying
annulment of the other terms. For this reason, we will work Eq. (\ref{eq:Drudeweight})
into a more convenient form. Regarding the regular conductivity, we
observe that the real part is strongly constrained to eigenstates
within $\hbar\omega$ of the Fermi level; therefore, this computation
should not be problematic and we will keep this method. For the imaginary
part, we see that, even for small $\omega$, we do not have an argument
to avoid a summation over all the bands; we will thus make use of
the Kramers-Kronig (KK) relations to compute the imaginary part using
the results obtained for the real part.

~

\paragraph{\uline{Drude weight — 2nd method}\protect \protect \protect \\
 }

Here, we derive an alternative expression to evaluate the Drude weight.
Using that

\begin{equation}
\left\langle \lambda\left|\frac{\partial h_{0}}{\partial k_{a_{1}}}\right|\lambda'\right\rangle =\frac{\partial\epsilon_{\lambda}}{\partial k_{a_{1}}}\delta_{\lambda,\lambda'}+\epsilon_{\lambda\lambda'}\left\langle \lambda\mid\partial_{k_{a_{1}}}\lambda'\right\rangle ,
\end{equation}
the second term in Eq. (\ref{eq:Drudeweight}), can be written as
\begin{equation}
\sum\limits _{\lambda_{1}\neq\lambda_{1}}\left\langle \lambda_{1}\left|\frac{\partial h_{0}}{\partial k_{a_{1}}}\right|\lambda_{2}\right\rangle \left\langle \lambda_{2}\left|\frac{\partial h_{0}}{\partial k_{a_{2}}}\right|\lambda_{1}\right\rangle \frac{n_{\lambda_{1}}^{F}-n_{\lambda_{2}}^{F}}{\epsilon_{\lambda_{1}\lambda_{2}}}=\sum\limits _{\lambda_{1}\neq\lambda_{1}}\left\langle \lambda_{1}\left|\frac{\partial h_{0}}{\partial k_{a_{1}}}\right|\lambda_{2}\right\rangle \left\langle \lambda_{2}\mid\partial_{k_{a_{2}}}\lambda_{1}\right\rangle \left(n_{\lambda_{1}}^{F}-n_{\lambda_{2}}^{F}\right).
\end{equation}
Clearly, the last sum can be extended to the case where $\lambda_{2}=\lambda_{1}$.
We then proceed with the following manipulations: 
\begin{align}
\sum\limits _{\lambda_{1},\lambda_{2}} & \left\langle \lambda_{1}\left|\frac{\partial h_{0}}{\partial k_{a_{1}}}\right|\lambda_{2}\right\rangle \left\langle \lambda_{2}\left|\partial_{k_{a_{2}}}\right.\lambda_{1}\right\rangle \left(n_{\lambda_{1}}^{F}-n_{\lambda_{2}}^{F}\right)\nonumber \\
 & =\sum\limits _{\lambda_{1}}\left\langle \lambda_{1}\left|\frac{\partial h_{0}}{\partial k_{a_{1}}}\right|\partial_{k_{a_{2}}}\lambda_{1}\right\rangle n_{\lambda_{1}}^{F}-\sum\limits _{\lambda_{1},\lambda_{2}}\left\langle \lambda_{1}\left|\frac{\partial h_{0}}{\partial k_{a_{1}}}\right|\lambda_{2}\right\rangle \left\langle \lambda_{2}\mid\partial_{k_{a_{2}}}\lambda_{1}\right\rangle n_{\lambda_{2}}^{F}\nonumber \\
 & =\sum\limits _{\lambda_{1}}\left\langle \lambda_{1}\left|\frac{\partial h_{0}}{\partial k_{a_{1}}}\right|\partial_{k_{a_{2}}}\lambda_{1}\right\rangle n_{\lambda_{1}}^{F}+\sum\limits _{\lambda_{1},\lambda_{2}}\left\langle \lambda_{1}\left|\frac{\partial h_{0}}{\partial k_{a_{1}}}\right|\lambda_{2}\right\rangle \left\langle \partial_{k_{a_{2}}}\lambda_{2}\mid\lambda_{1}\right\rangle n_{\lambda_{2}}^{F}\nonumber \\
 & =\sum\limits _{\lambda_{1}}\left(\left\langle \lambda_{1}\left|\frac{\partial h_{0}}{\partial k_{a_{1}}}\right|\partial_{k_{a_{2}}}\lambda_{1}\right\rangle +\left\langle \partial_{k_{a_{2}}}\lambda_{1}\left|\frac{\partial h_{0}}{\partial k_{a_{1}}}\right|\lambda_{1}\right\rangle \right)n_{\lambda_{1}}^{F}.
\end{align}

Collecting all terms, the Drude weight yields: 
\begin{align}
D_{a_{1}a_{2}} & =\frac{8\pi\sigma_{0}}{NA_{u.c.}}\sum\limits _{\bm{k},\lambda_{1}}\left(\left\langle \lambda_{1}\left|\frac{\partial^{2}h_{0}}{\partial k_{a_{1}}\partial k_{a_{2}}}\right|\lambda_{1}\right\rangle +\left\langle \lambda_{1}\left|\frac{\partial h_{0}}{\partial k_{a_{1}}}\right|\partial_{k_{a_{2}}}\lambda_{1}\right\rangle +\left\langle \partial_{k_{a_{2}}}\lambda_{1}\left|\frac{\partial h_{0}}{\partial k_{a_{1}}}\right|\lambda_{1}\right\rangle \right)n_{\lambda_{1}}^{F}\nonumber \\
 & =\frac{8\pi\sigma_{0}}{NA_{u.c.}}\sum\limits _{\bm{k},\lambda_{1}}n_{\lambda_{1}}^{F}\frac{\partial}{\partial k_{a_{2}}}\left\langle \lambda_{1}\left|\frac{\partial h_{0}}{\partial k_{a_{1}}}\right|\lambda_{1}\right\rangle \nonumber \\
 & =\frac{8\pi\sigma_{0}}{NA_{u.c.}}\sum\limits _{\bm{k},\lambda_{1}}\frac{\partial^{2}\epsilon_{\lambda_{1}}}{\partial k_{a_{1}}\partial k_{a_{2}}}n_{\lambda_{1}}^{F}\nonumber \\
 & =\frac{8\pi\sigma_{0}}{NA_{u.c.}}\sum\limits _{\bm{k},\lambda_{1}}\left(\frac{\partial}{\partial k_{a_{1}}}\left(\frac{\partial\epsilon_{\lambda_{1}}}{\partial k_{a_{2}}}n_{\lambda_{1}}^{F}\right)-\frac{\partial\epsilon_{\lambda_{1}}}{\partial k_{a_{2}}}\frac{\partial n_{\lambda_{1}}^{F}}{\partial k_{a_{1}}}\right).
\end{align}
The first term in the last line of the previous equation can be shown
to give zero, since it is a total derivative of a periodic quantity
that is being integrated over the whole BZ.

The final expression therefore reads as
\begin{align}
D_{a_{1}a_{2}} & =-\frac{8\pi\sigma_{0}}{NA_{u.c.}}\sum\limits _{\bm{k},\lambda_{1}}\frac{\partial\epsilon_{\lambda_{1}}}{\partial k_{a_{2}}}\frac{\partial n_{\lambda_{1}}^{F}}{\partial k_{a_{1}}}\nonumber \\
 & =-\frac{8\pi\sigma_{0}}{NA_{u.c.}}\sum\limits _{\bm{k},\lambda_{1}}\frac{\partial\epsilon_{\lambda}(\bm{k})}{\partial k_{a_{1}}}\frac{\partial\epsilon_{\lambda}(\bm{k})}{\partial k_{a_{2}}}\frac{\partial n_{F}\left(\epsilon_{\lambda}(\bm{k})\right)}{\partial\epsilon}.\label{eq:Drude2}
\end{align}
As foreseen, this expression only takes into account electrons near
the Fermi level. This can be directly detected by the presence of
the derivative of the Fermi-Dirac function with respect to the energy.

~

\paragraph{\uline{Imaginary part of the regular conductivity — 2nd method}\protect
\protect \protect \\
 }

The KK relations relate the real part of a response function with
its imaginary part. They enable us to find one of the components if
we know the other at all frequencies. In our case, we want to compute
the imaginary part of the regular conductivity. The appropriate relation
is \citep{Kittel1966} 
\begin{equation}
\text{Im}\{\sigma^{reg}(\omega)\}=-\frac{2\omega}{\pi}P\int_{0}^{+\infty}ds\ \frac{\text{Re}\{\sigma^{reg}(s)\}}{s^{2}-\omega^{2}}.\label{eq:KKimag}
\end{equation}
Looking at this expression, there is apparently no advantage in using
this method for effective models, since the integral extends to infinity.
Moreover, this integral is ill defined, since at high frequencies
the continuum model for the tBLG is expected to yield a constant $\text{Re}\{\sigma^{reg}(\omega)\}=2\sigma_{0}$.

Following Ref. \citep{Stauber2013}, we can thus perform a regularization
of Eq. (\ref{eq:KKimag}) by invoking the following property: 
\begin{equation}
P\int_{0}^{+\infty}ds\frac{1}{s^{2}-\omega^{2}}=0.
\end{equation}
The final regularized definition then reads as
\begin{equation}
\text{Im}\{\sigma^{reg}(\omega)\}=-\frac{2\omega}{\pi}P\int_{0}^{+\infty}ds\ \frac{\text{Re}\{\sigma^{reg}(s)\}-2\sigma_{0}}{s^{2}-\omega^{2}},\label{eq:KKregularized}
\end{equation}
which we can now evaluate by introducing a finite cutoff $\Lambda$
for which $\text{Re}\{\sigma^{reg}(\Lambda)\}\simeq2\sigma_{0}$.

\subsubsection{Results for single layer graphene}

\label{subsection:SLGconductivity}

\begin{figure}
\centering{}\includegraphics[width=0.45\textwidth]{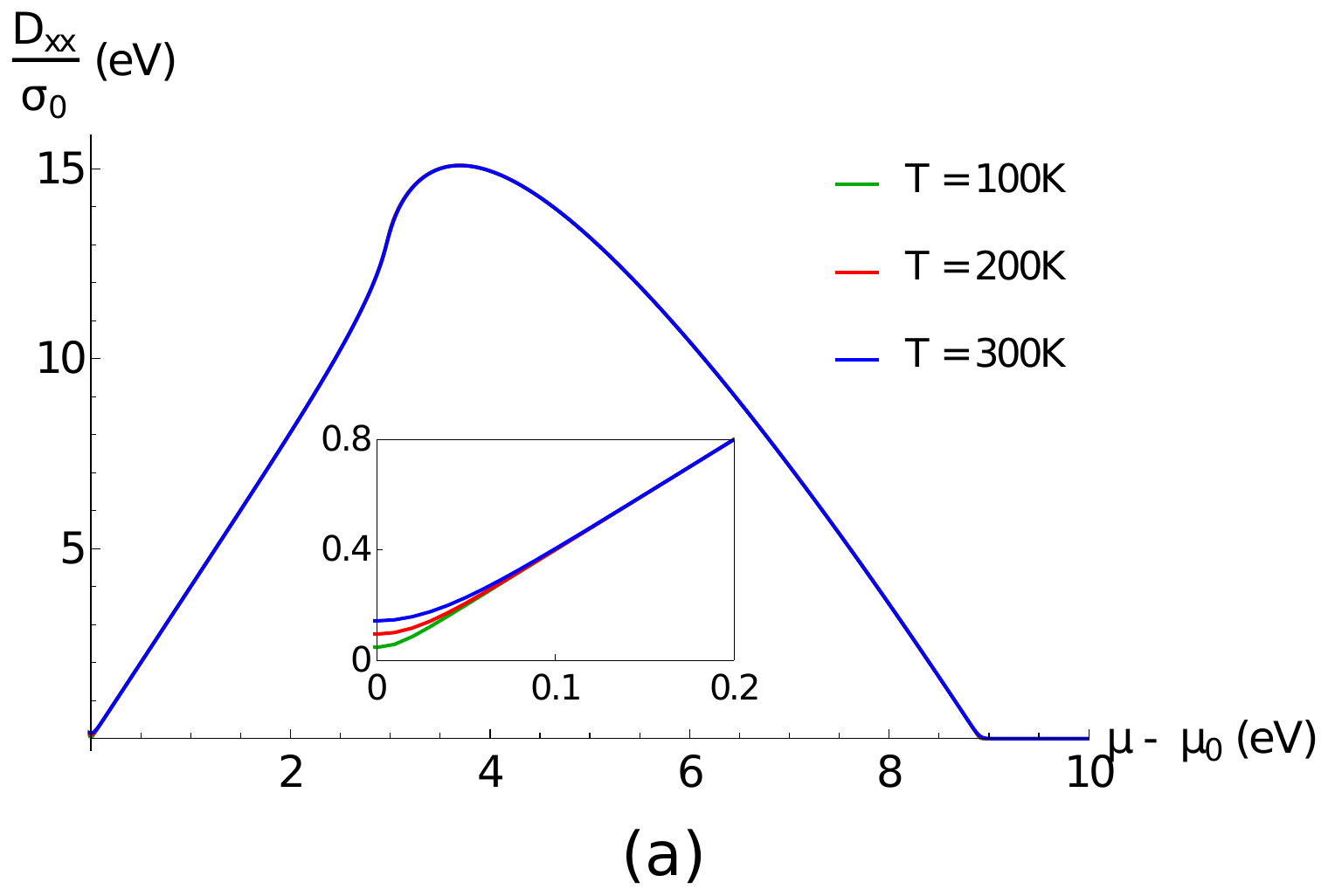}~~\includegraphics[width=0.45\textwidth]{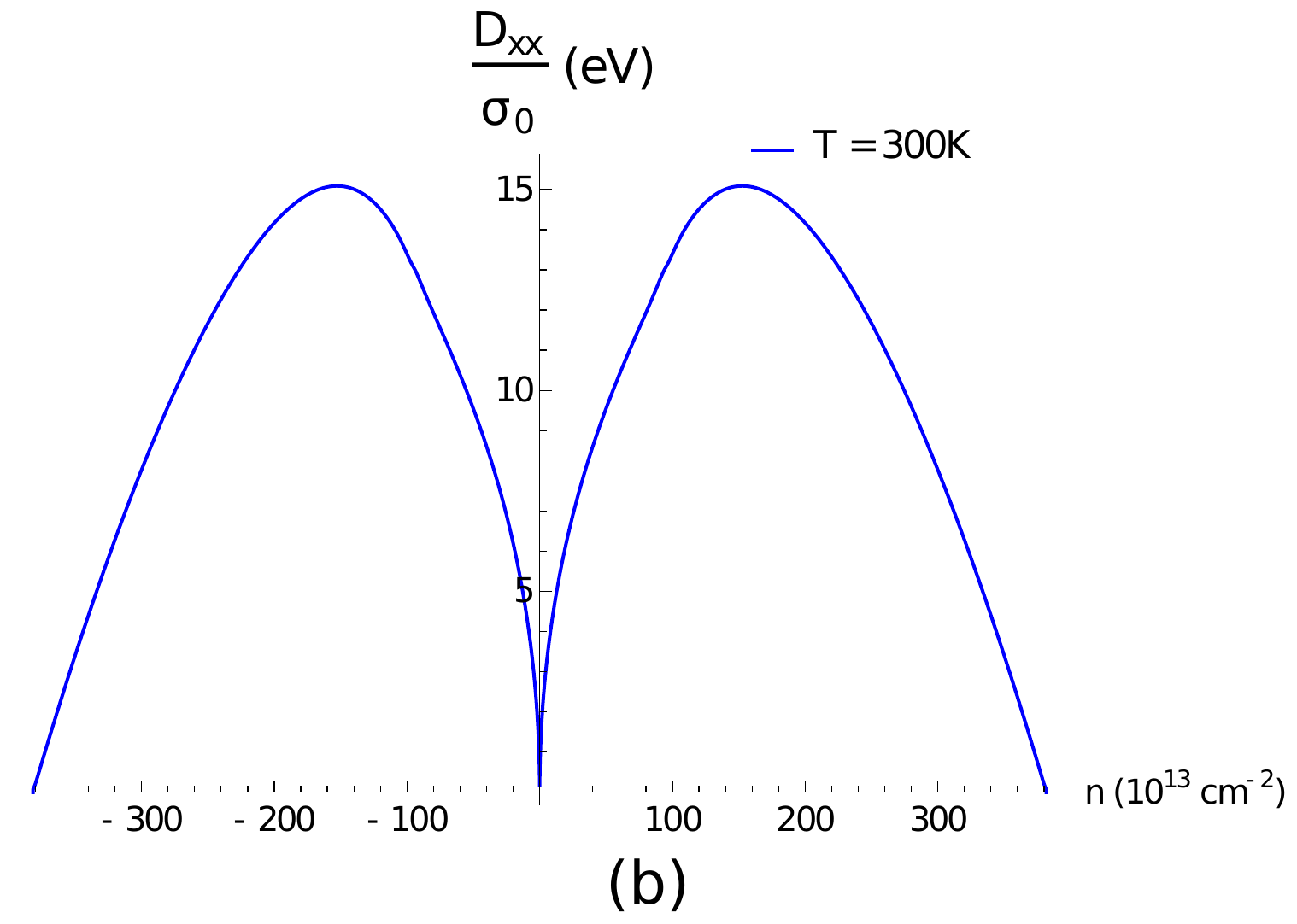}\caption{Drude weight results for SLG: (a) as a function of the Fermi level;
(b) as a function of the carrier density. In (a), $\mu_{0}$ is the
Fermi level at half filling.}
\label{fig:SLGDrude_carrierandvariousT} 
\end{figure}

As benchmark, we apply the expressions obtained in the previous section
to the SLG system. We start with the Drude weight. The results are
presented in Fig. \ref{fig:SLGDrude_carrierandvariousT}. We stress
that both methods —Eqs. (\ref{eq:Drude1}) and (\ref{eq:Drude2})—
give the same output and yield $D_{xx}=D_{yy},\ D_{xy}=0$. The low-energy
results for $D_{xx}=D_{yy}\equiv D$ are in agreement with the theoretical
predictions for the Drude conductivity at $T=0\si{\kelvin}$ \citep{Stauber2008,Goncalves2016},
\begin{equation}
\sigma^{D}(\omega)=4\sigma_{0}\frac{i}{\pi}\frac{\mu}{\hbar\omega+i\Gamma}.
\end{equation}
From this expression, we recognize the Drude weight as $D/\sigma_{0}=4\mu$,
which we compare to the inset of Fig. \ref{fig:SLGDrude_carrierandvariousT}(a).
The smoothed behavior near $\mu\sim\mu_{0}$ (where $\mu_{0}$ is
the Fermi level at half filling, which we set as zero) is explained
by the finite temperature: we have electrons available for the transport
due to thermal activation.

\begin{figure}
\centering{}\includegraphics[width=0.4\textwidth]{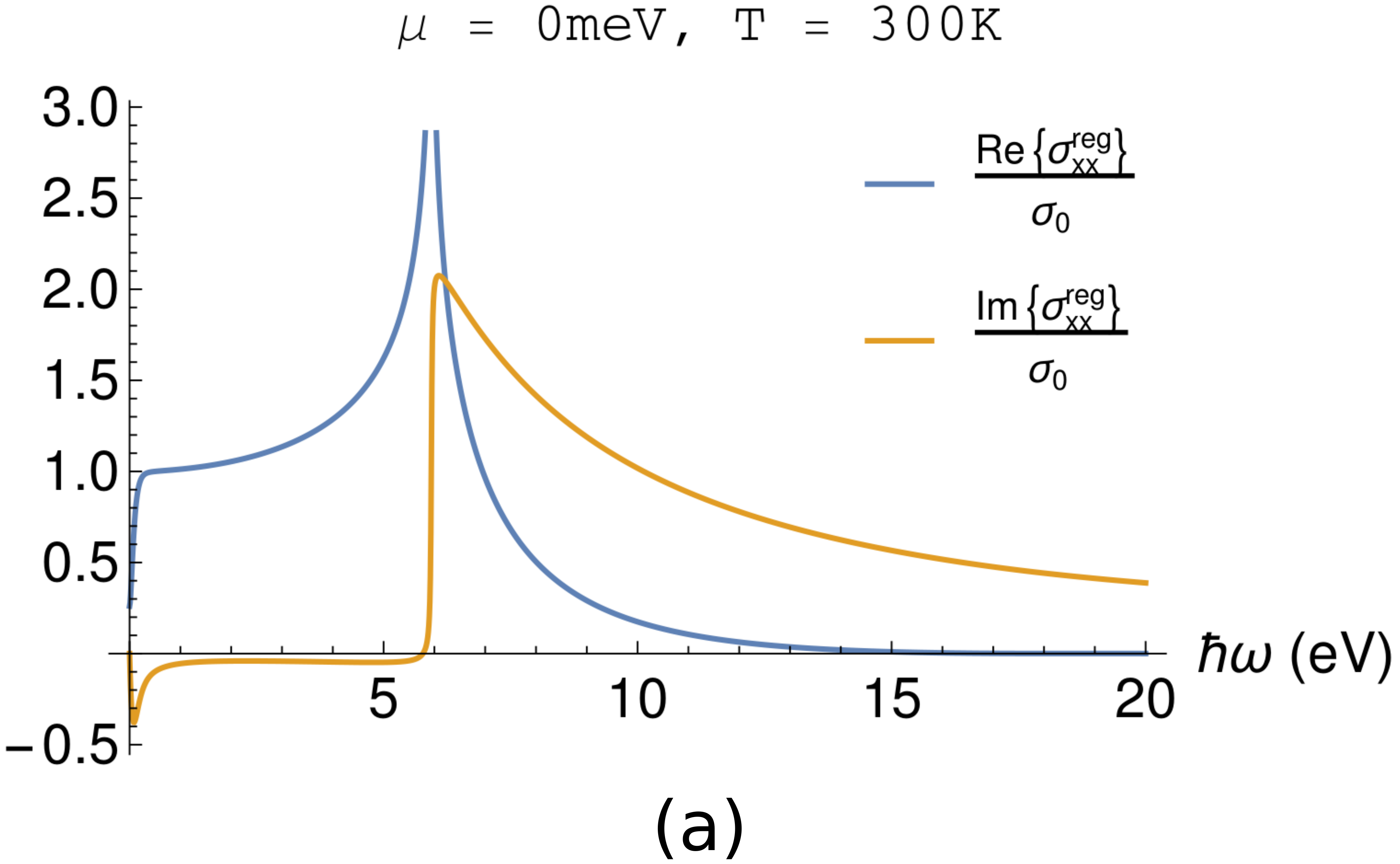}~~\includegraphics[width=0.35\textwidth]{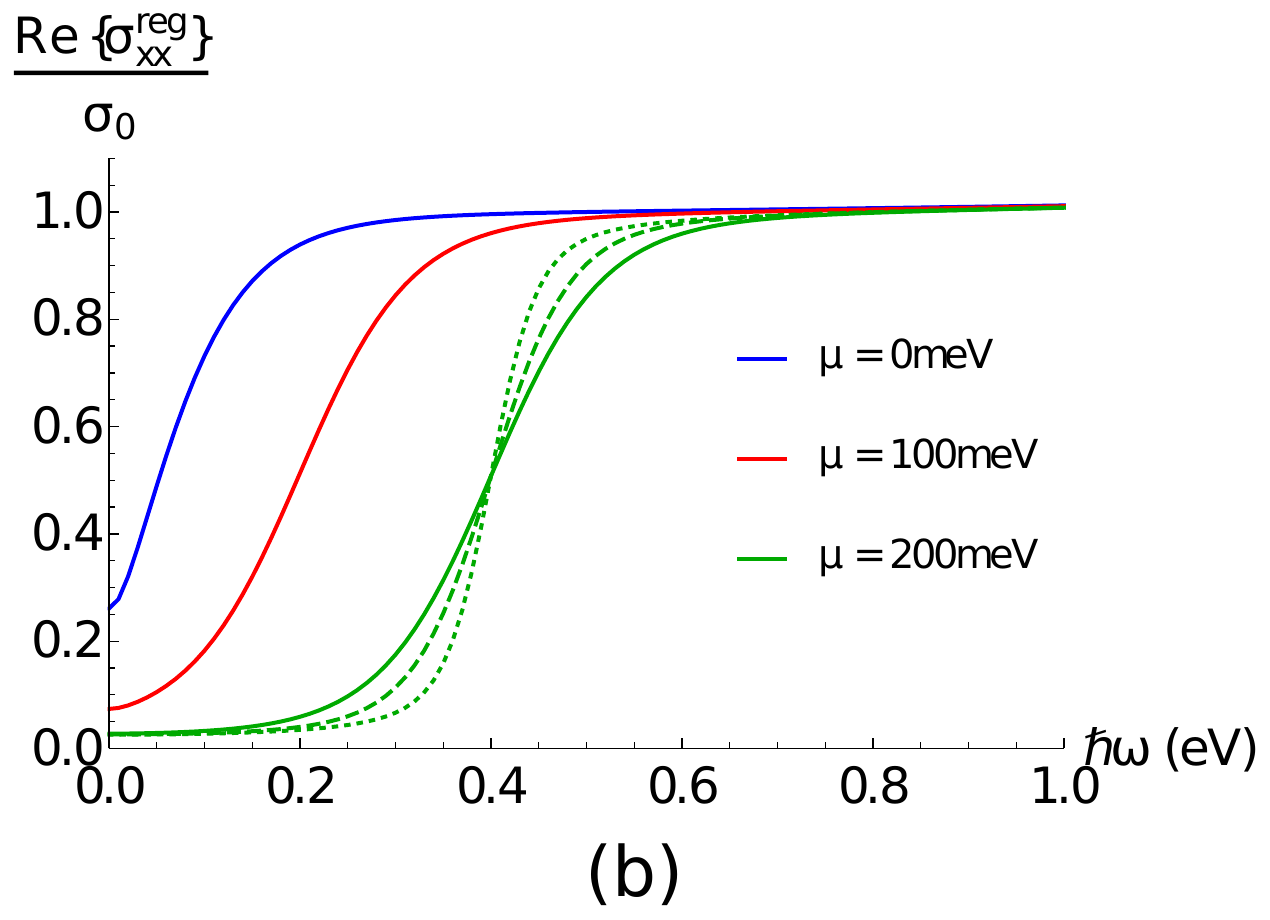}\caption{(a) and (b) show results for the regular conductivity in SLG. In (b),
the dotted line corresponds to $T=100\si{\kelvin}$, the dashed line
to $T=200\si{\kelvin}$ and the solid lines to $T=300\si{\kelvin}$.}
\label{fig:SLGregcond_Ef0mevandvariousEf} 
\end{figure}

\begin{figure}
\centering{}\includegraphics[width=0.35\textwidth]{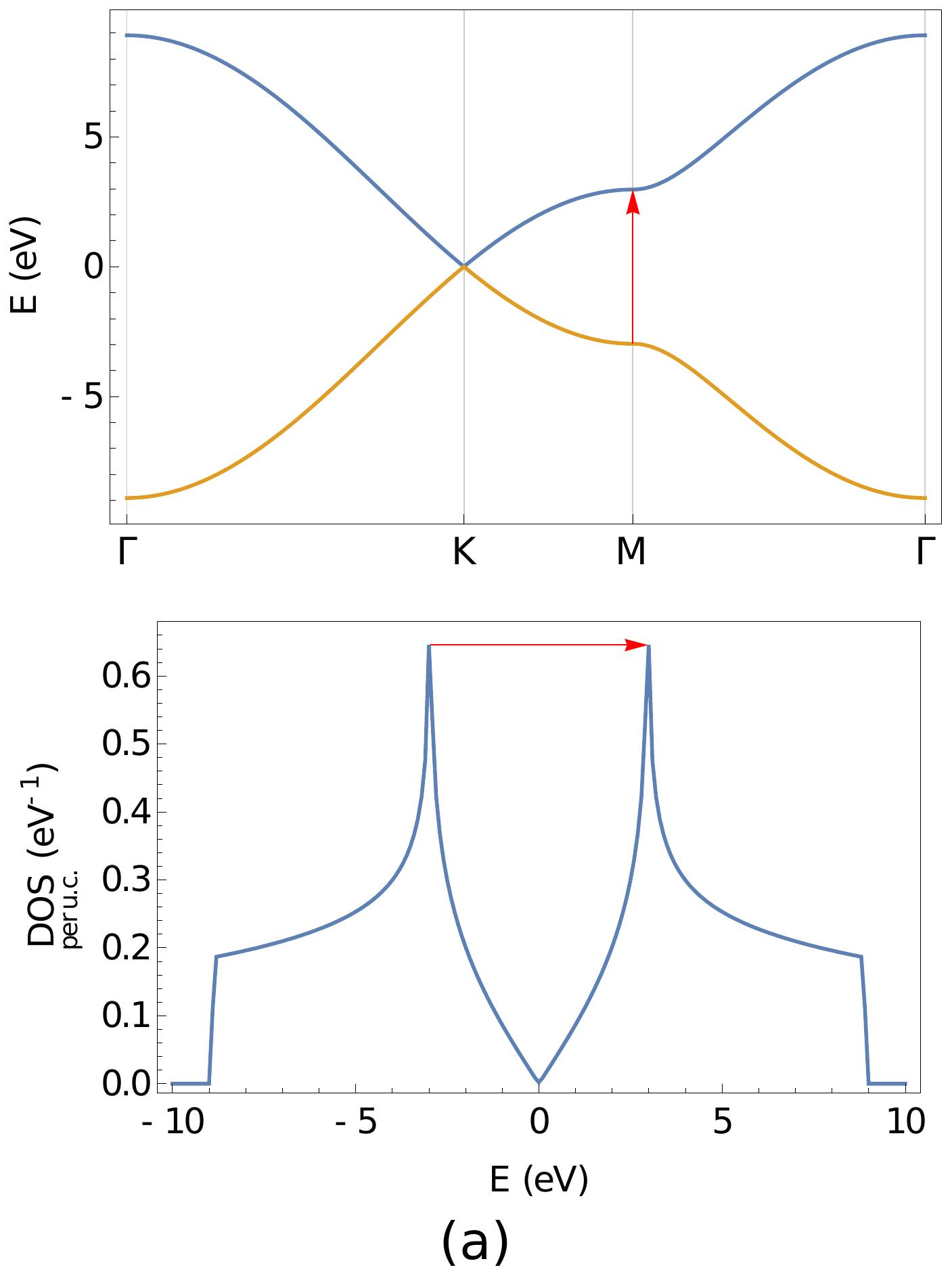}~~\includegraphics[width=0.4\textwidth]{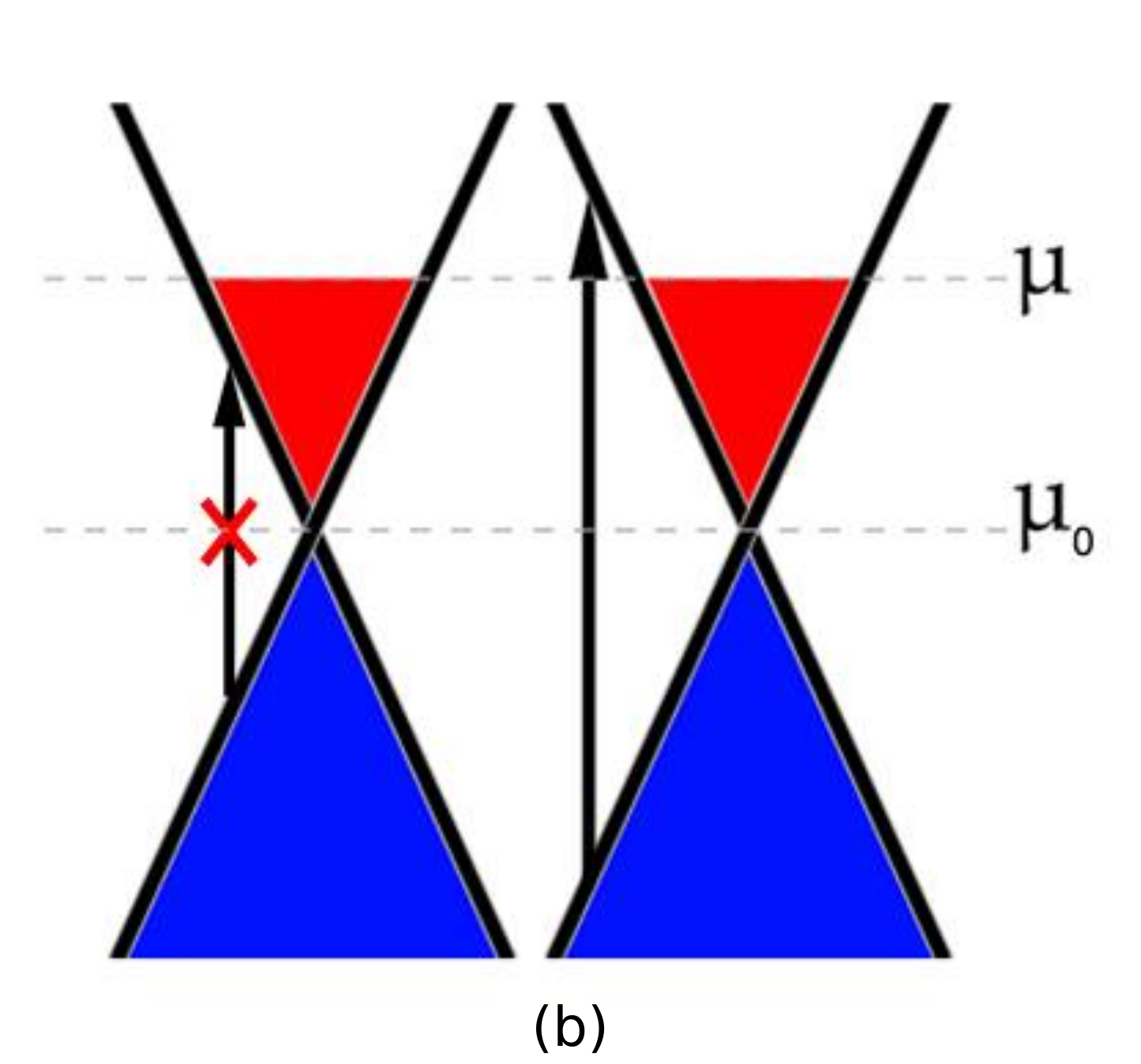}\caption{Picture of interband transitions in SLG. In (a), the spectrum and
the DOS are plotted, showing the dominant transition that occurs between
the van Hove singularities. In (b), the Pauli exclusion principle
is depicted for the low-energy regime, where the Dirac cone picture
is valid.}
\label{fig:SLGinterbandtransitionspic.} 
\end{figure}

We now move to the regular conductivity. The results are shown in
Fig. \ref{fig:SLGregcond_Ef0mevandvariousEf}. Here, and in what follows,
we set $\Gamma=16\si{\milli\electronvolt}$ in agreement with Ref.
\citep{Ju2011}. Once again, both methods —Eq. (\ref{eq:regcond2})
for both real and imaginary parts or Eq. (\ref{eq:regcond2}) for
the real part along with Eq. (\ref{eq:KKregularized}) for the imaginary
part— yield the same results and $\sigma_{xx}^{reg}=\sigma_{yy}^{reg},\ \sigma_{xy}^{reg}=0$.
The fact that we obtain an isotropic (total) conductivity is an expected
result from group theory since the system has hexagonal symmetry \citep{Nowick1995}.
Analyzing Fig. \ref{fig:SLGregcond_Ef0mevandvariousEf}(a), we interpret
the peak at $\hbar\omega=2t\simeq6\si{\electronvolt}$ as electronic
transitions from the van Hove singularity of the valence band to the
van Hove singularity of the conduction band, as depicted by the red
arrows in Fig. \ref{fig:SLGinterbandtransitionspic.}(a). These transitions
are enhanced because there is a peak in the number of electrons that
can occupy the initial and final energy states. Regarding Fig. \ref{fig:SLGregcond_Ef0mevandvariousEf}(b),
we also infer that transitions with $\hbar\omega<2\mu$ are forbidden,
which has been observed experimentally \citep{Li2008}. The explanation
is sketched in Fig. \ref{fig:SLGinterbandtransitionspic.}(b). When
we increase the Fermi level up to $\mu>\mu_{0}$, states with $E<\mu$
become occupied. Therefore, transitions for those states are blocked
due to the Pauli exclusion principle. Since we have particle-hole
symmetry, we conclude that we can only have transitions when $\hbar\omega>2\mu$.
The increase of the temperature is verified to smooth out this behavior
(see Fig. \ref{fig:SLGregcond_Ef0mevandvariousEf}(b)).

\subsubsection{Results for twisted bilayer graphene}

\label{subsection:tBLGconductivity}

\begin{figure}
\centering{}\includegraphics[width=0.45\textwidth]{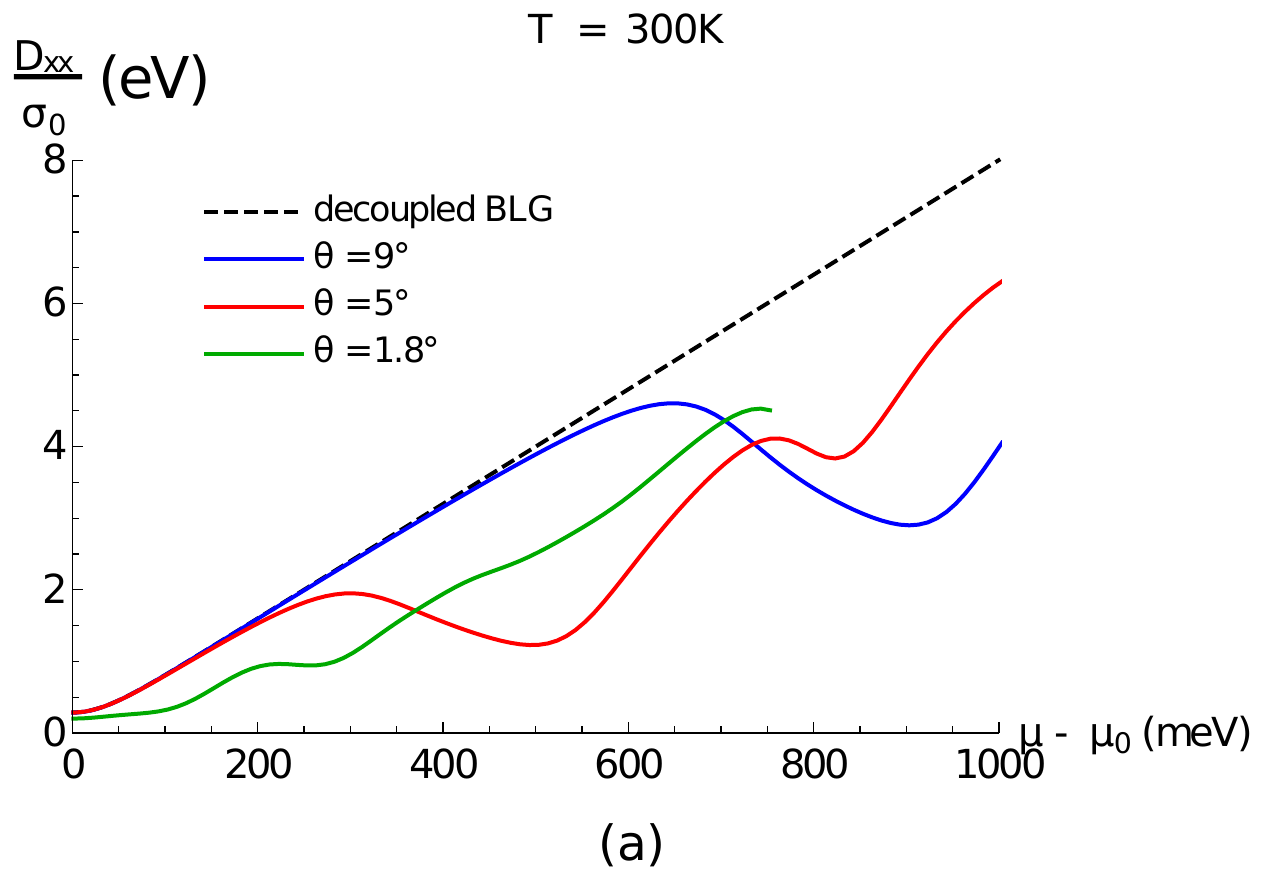}~~\includegraphics[width=0.4\textwidth]{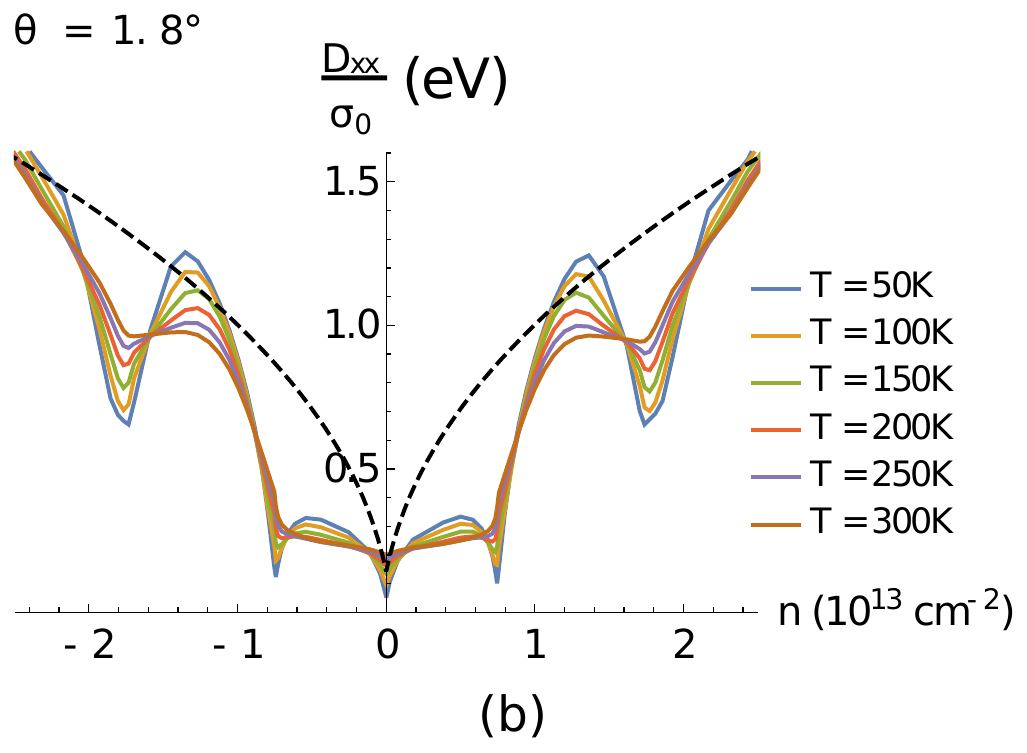}\caption{Drude weight results for tBLG ($2^{\text{nd}}$ method): (a) as a
function of the Fermi level, for different twist angles; (b) as a
function of the carrier density, for different temperatures. The outcomes
were isotropic. In (b), the black dashed line is for decoupled BLG
at $T=300\si{\kelvin}$. The results for decoupled BLG (tBLG model
with $t_{\perp}=0$) were verified to match the results for SLG multiplied
by 2.}
\label{fig:tBLGDrude2}
\end{figure}

A summary of the Drude weight results for tBLG is provided in Fig.
\ref{fig:tBLGDrude2}. We stress that only the 2\textsuperscript{nd}
method was verified to work well for these computations. This happens
because we are working with an effective Hamiltonian, as discussed
before. Similarly to what we have seen for the SLG, we observe symmetric
results for electron or hole doping; this reflects the apparent symmetry
in the valence and conduction bands mentioned in section \ref{subsection:tBLGmatrix}.
By looking at Fig. \ref{fig:tBLGDrude2}(a), along with Fig. \ref{fig:tBLGDOSandcarrierdensity_variousdeg}(a),
we conclude that the Drude weight curve changes drastically (compared
to SLG or decoupled BLG) when we cross van Hove singularities. This
tendency coincides with what was found in Ref. \citep{Stauber2013}
and the drops in the curves are attributed to the fact that the first
derivatives of the energy go to zero when we cross van Hove singularities.
The effect of increasing the temperature is, as usual, the smoothing
of this behavior.

\begin{figure}
\centering{}\includegraphics[width=0.3\textwidth]{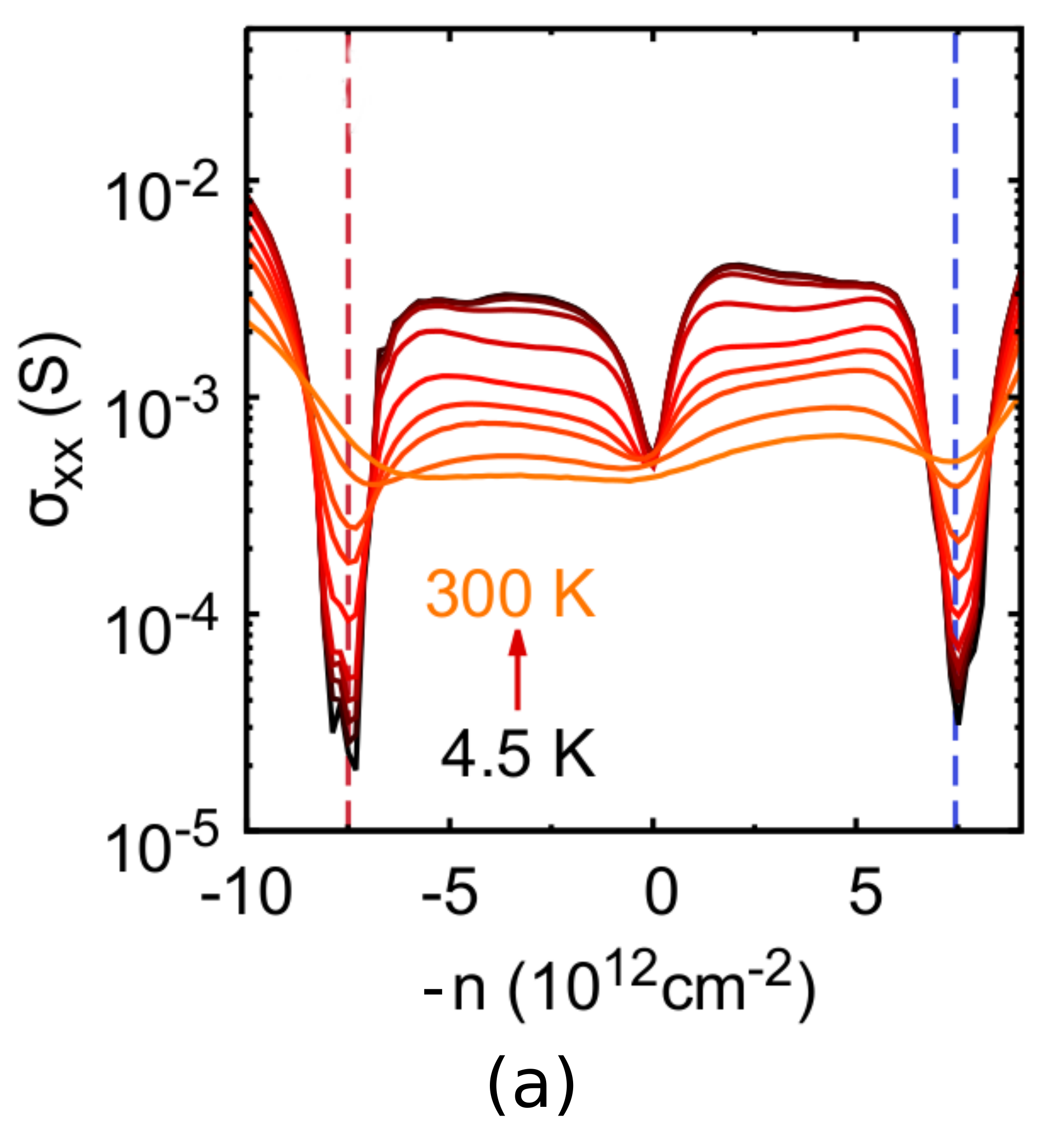}~~~~~\includegraphics[width=0.4\textwidth]{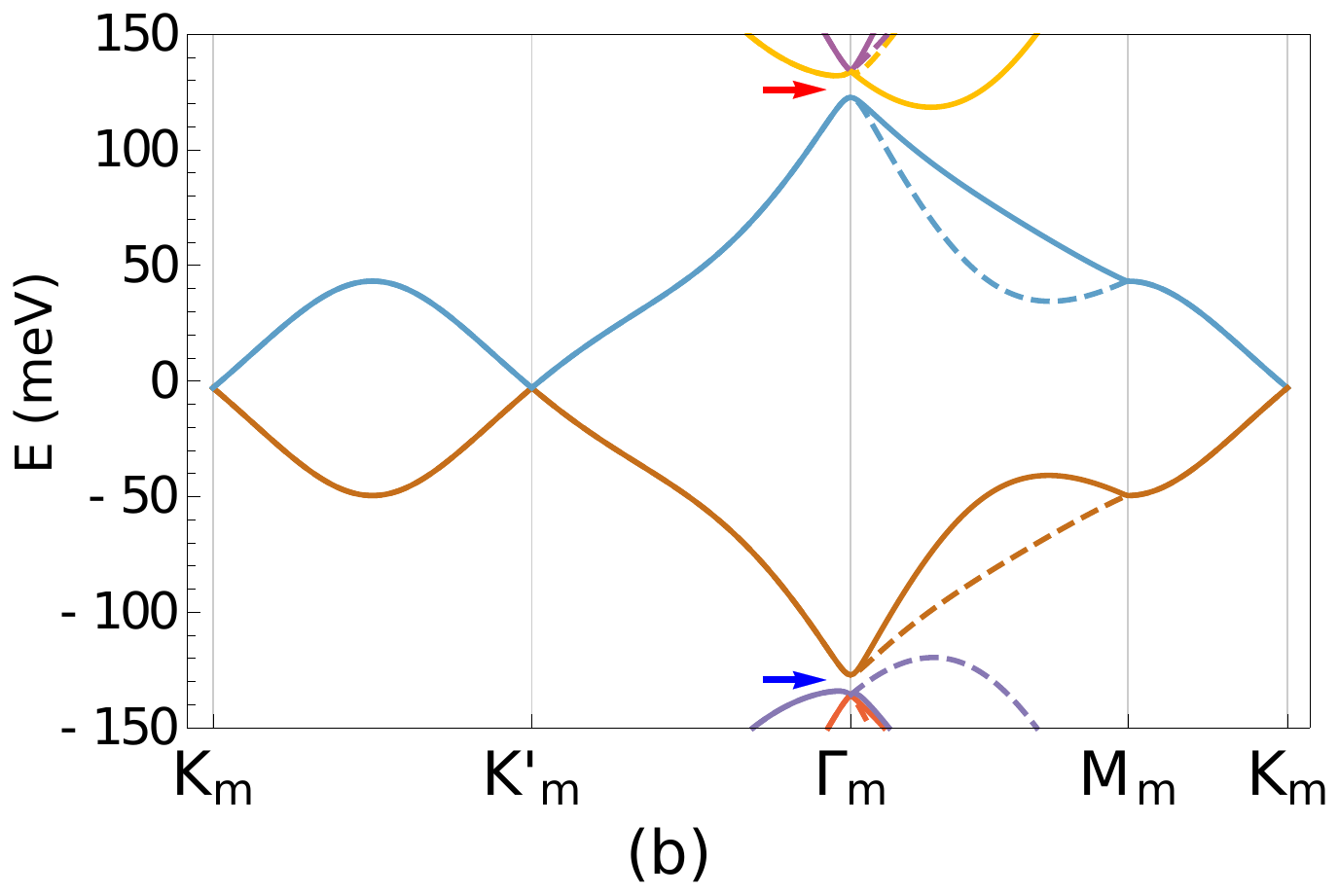}\caption{tBLG with $\theta=1.8^{\circ}$: (a) experimental results for the
DC optical conductivity (source: Ref. \citep{Cao2016}); (b) electronic
band structure.}
\label{fig:tBLG1dot8deg_expDrudeandspectrum} 
\end{figure}

In Fig. \ref{fig:tBLG1dot8deg_expDrudeandspectrum}(a), we show recent
experimental results of the DC optical conductivity in the tBLG, obtained
in Ref. \citep{Cao2016}. We observe the expected symmetry for doping
with electrons and holes. Moreover, since for $\omega=0$ the conductivity
is dominated by the Drude contribution, we may compare the experimental
results with the theoretical calculations from Fig. \ref{fig:tBLGDrude2}(b).
We note the need of including the disorder broadening $\Gamma$ in
order to obtain a quantitative agreement with the experiment. In addition,
we see that the experimental drop in the conductivity, at $|n|\sim7.5\times10^{12}\si{\per\centi\meter\squared}$,
is in agreement with what our model predicts. These insulating states
are interpreted as the gaps occurring at the $\Gamma_{m}$ point in
the electronic spectrum (Fig. \ref{fig:tBLG1dot8deg_expDrudeandspectrum}(b)).
However, we immediately verify that the insulating behavior is much
more pronounced in the experimental results. In the work done in Ref.
\citep{Cao2016}, the authors estimated a band gap of $50-60\si{\milli\electronvolt}$,
which is much larger than what we observe in the electronic spectrum.
This effect might be due to the electron-electron interactions not
considered in the model. Finally, we also note the disagreement between
the experimental results and the theory near the Dirac point (which
corresponds to $n=0$), in particular the fact that the conductivity
is not very sensitive to $T$ below some value $T_{max}$ (see Fig.
\ref{fig:tBLG1dot8deg_expDrudeandspectrum}(a)). This is an expected
feature observed in graphene and the explanation is that it occurs
due to inhomogeneities in the system (extrinsic disorder, ripples,
etc...), which make the Dirac point inaccessible \citep{Morozov2008}.

\begin{figure}
\centering{}\includegraphics[width=0.4\textwidth]{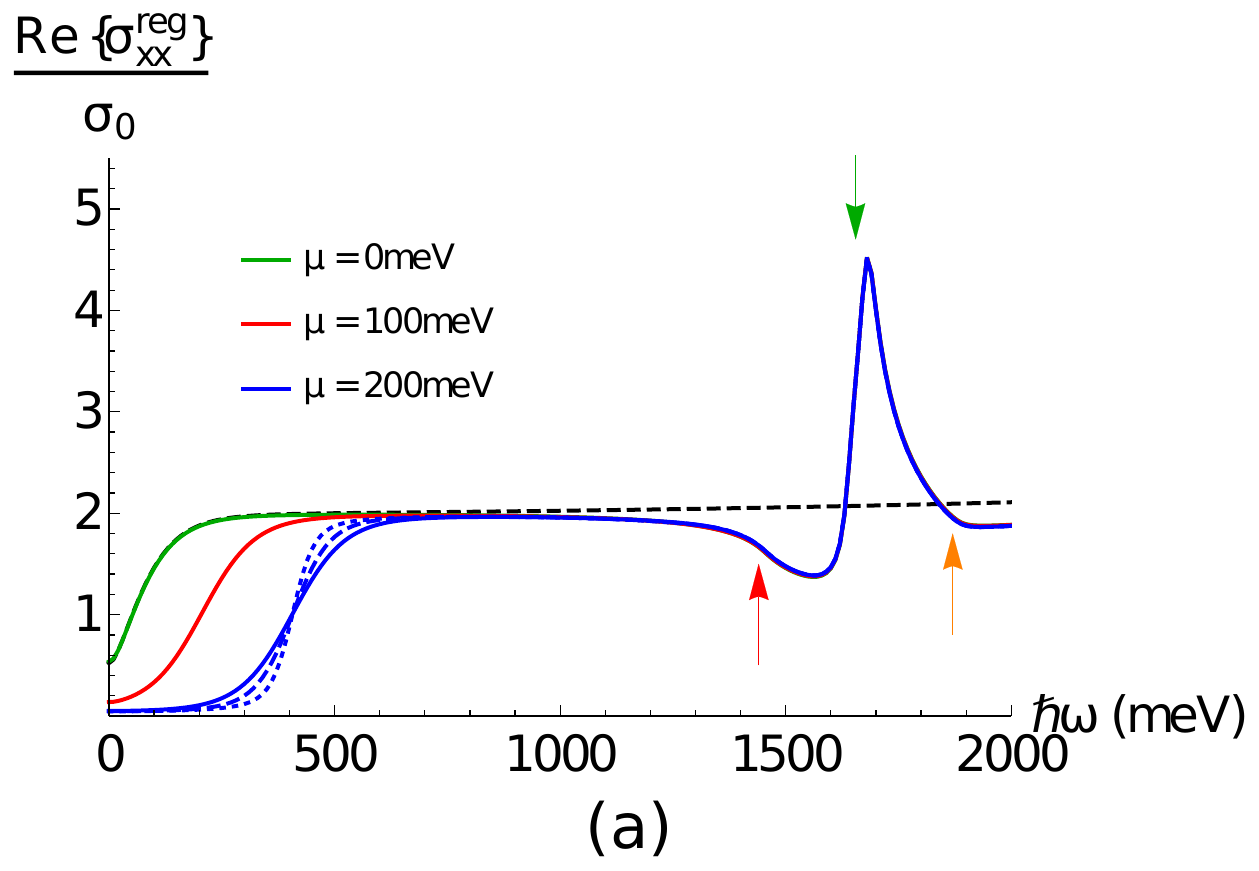}~~\includegraphics[width=0.4\textwidth]{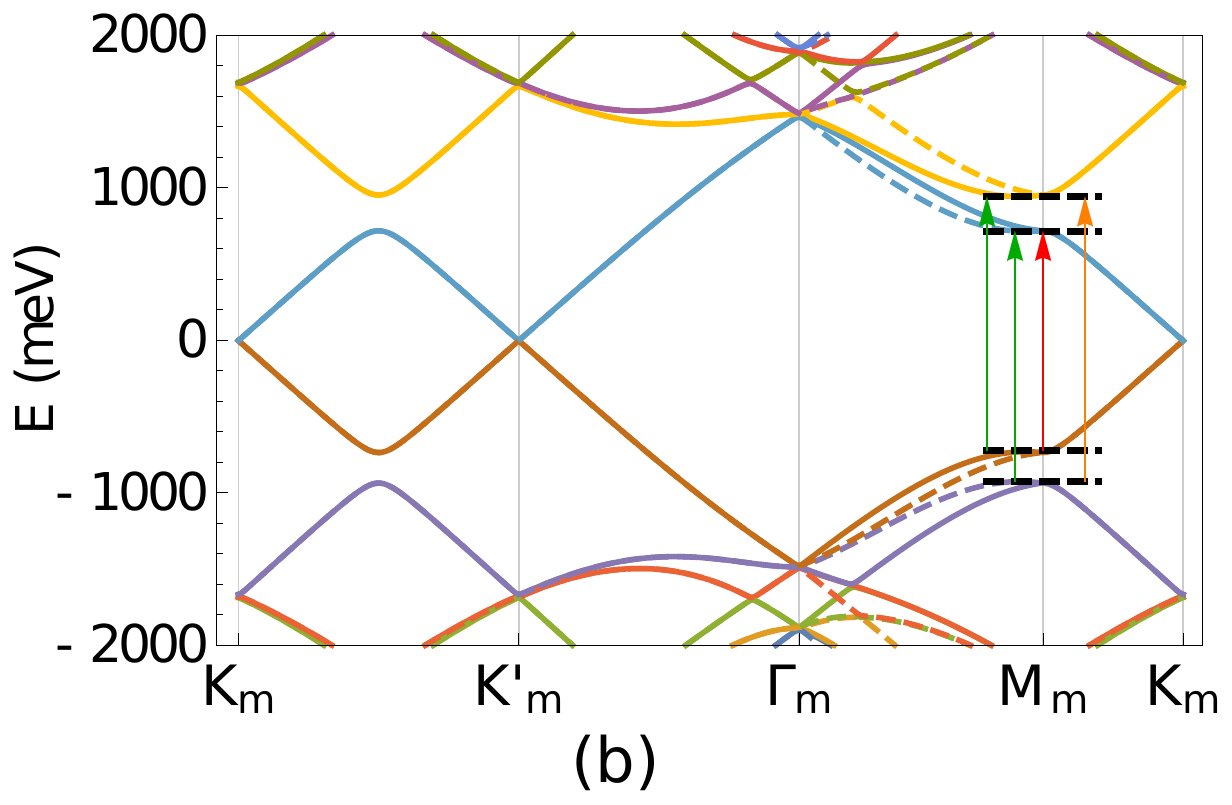}
\caption{tBLG with $\theta=9^{\circ}$: (a) real part of the regular conductivity;
(b) electronic spectrum. In (a), the dotted blue line corresponds
to $T=100\si{\kelvin}$, the dashed blue line to $T=200\si{\kelvin}$
and the solid lines to $T=300\si{\kelvin}$; the black dashed line
is for decoupled tBLG (or SLG multiplied by 2) at $T=300\si{\kelvin}$
and $\mu=0$.}
\label{fig:tBLG9deg_regcondandspectrum}
\end{figure}

\begin{figure}
\centering{}\includegraphics[width=0.4\textwidth]{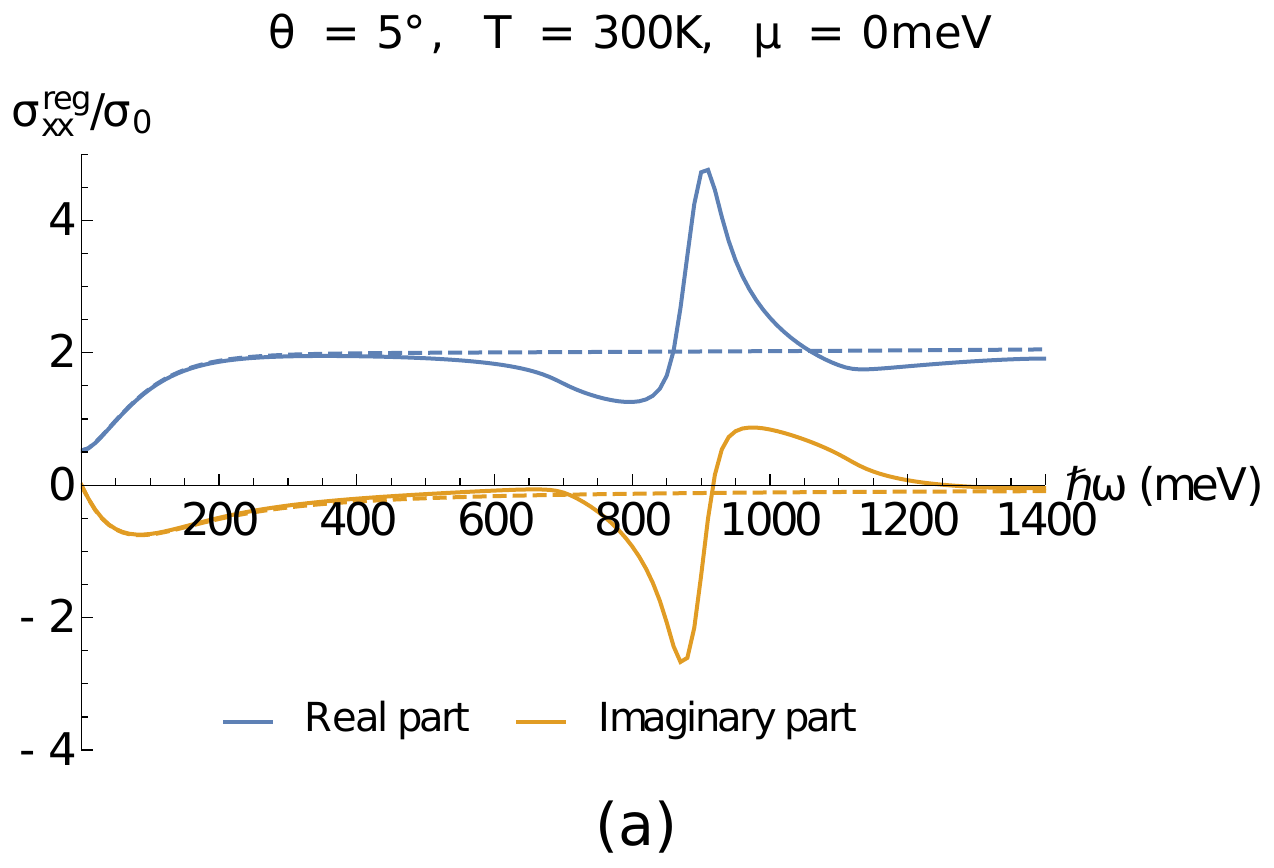}~~\includegraphics[width=0.4\textwidth]{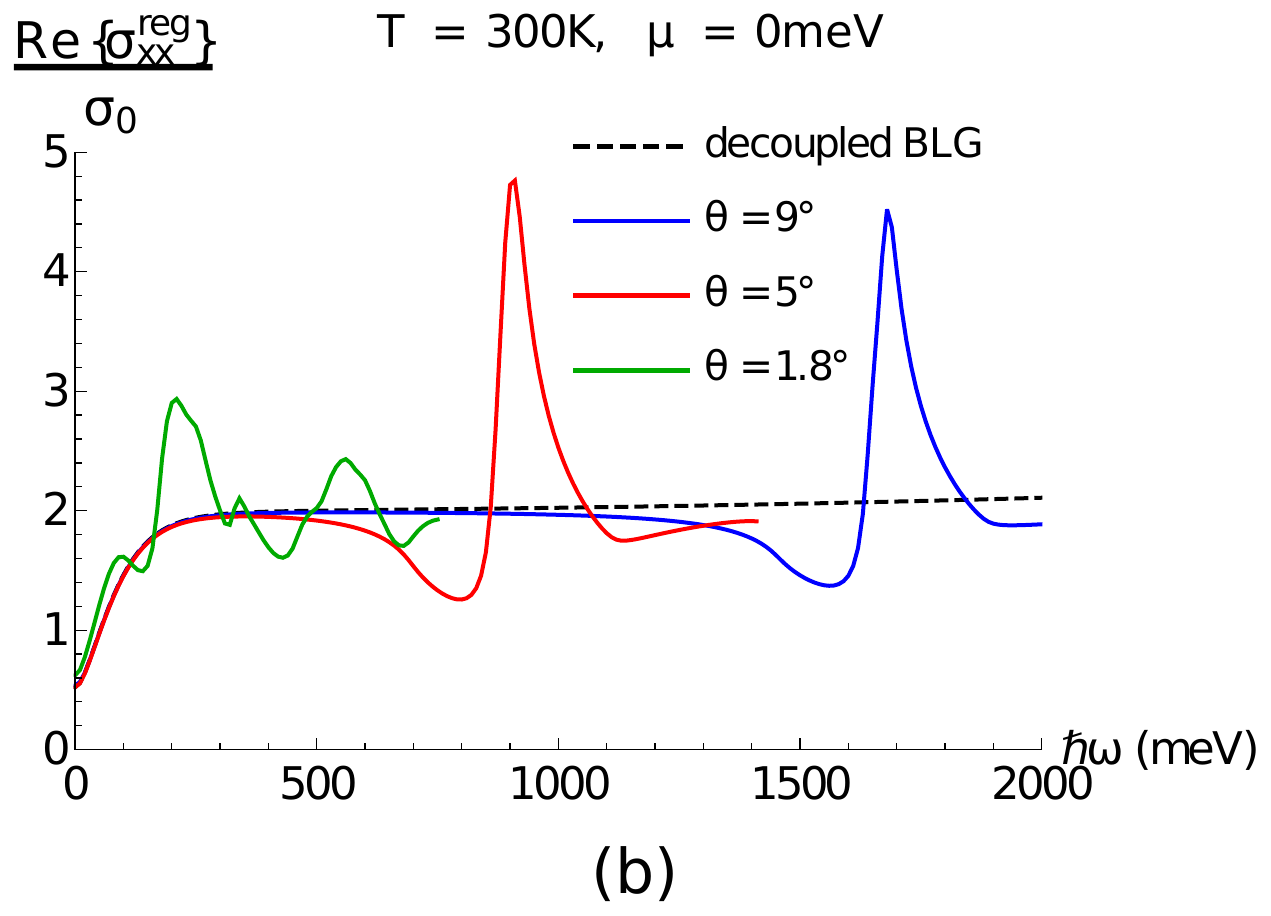}\caption{Regular conductivity results for tBLG at $\mu=0$: (a) real and imaginary
parts for $\theta=5^{\circ}$; (b) real part for different $\theta$.
In both (a) and (b), all dashed lines are for decoupled BLG or SLG
multiplied by 2.}
\label{fig:tBLGregcond_5degandvariousdeg}
\end{figure}

In Figs. \ref{fig:tBLG9deg_regcondandspectrum} and \ref{fig:tBLGregcond_5degandvariousdeg},
we show representative results which allow us to analyze the regular
conductivity in tBLG systems. All conductivity results obtained were
isotropic. Just like in the SLG, this feature is expected from group
theory since the tBLG also has an hexagonal symmetry (moiré pattern).
Before discussing the results, we give a word about the numerical
implementation of the real part (the imaginary part is straightforwardly
computed from the real part by using the regularized KK relation).
In contrast with the other calculations, where we only need to consider
bands with $|E|\lesssim1\si{\electronvolt}$ (which are well described
by the model), in this case we see that, for a given $\mu$ and for
a given $\omega$, all bands with energy respecting $|E-\mu|\lesssim\hbar\omega$
contribute. Therefore, depending on the Fermi level $\mu$ and, most
importantly, on the energy $\hbar\omega$ of the interband transition
that we want to capture, we may need to consider higher energy bands
that are not well described by the model. Still, this does not constitute
a big concern because these bands lead to the well established constant
value $2\sigma_{0}$, typical of the Dirac cone approximation.

Looking at Fig. \ref{fig:tBLG9deg_regcondandspectrum}(a), we first
notice the already discussed dependency on both the Fermi level and
temperature. In addition, we observe a low-energy peak (marked with
a green arrow), which we interpret as the dominant transitions shown
in Fig. \ref{fig:tBLG9deg_regcondandspectrum}(b). Notice that there
are other transitions (red and orange arrows) which we would also
expect to be dominant, since they connect different van Hove singularities;
however, these transitions are optically inactive, in agreement with
what was found in Refs. \citep{Tabert2013,Moon2013}. This optical
selection rule occurs due to a symmetry in the effective Hamiltonian
which makes the matrix elements from Eq. (\ref{eq:regcond2}) null
for bands with symmetric energies at the $\text{M}_{m}$ points \citep{Moon2013}.
From Fig. \ref{fig:tBLGregcond_5degandvariousdeg}(a), we highlight
the fact that the results obtained for the decoupled tBLG —tBLG with
a null interlayer hopping parameter, $t_{\perp}=0$— match perfectly
the results for SLG multiplied by 2. Although this was trivially expected,
it was only achieved when we used the second method for computing
the imaginary part of the regular conductivity; therefore, this served
as a benchmark test for the validity of the computational methods.
Moreover, we remark that we now have a region with a big deep on $\text{Im}\{\sigma^{reg}(\omega)\}$
occurring at lower frequencies, which will be an important feature
in the next section. Regarding Fig. \ref{fig:tBLGregcond_5degandvariousdeg}(b),
we emphasize that, for small angles, we start to lose the ``signature''
behavior of the curves because of the presence of multiple low-energy
van Hove singularities.

\subsection{Spectrum of surface plasmon-polaritons}

\label{section:GSPPs_spectrum}

\subsubsection{Dispersion relation — transverse magnetic modes}

\label{subsection:dispersionTM}

\begin{figure}
\centering{}\includegraphics[width=0.5\textwidth]{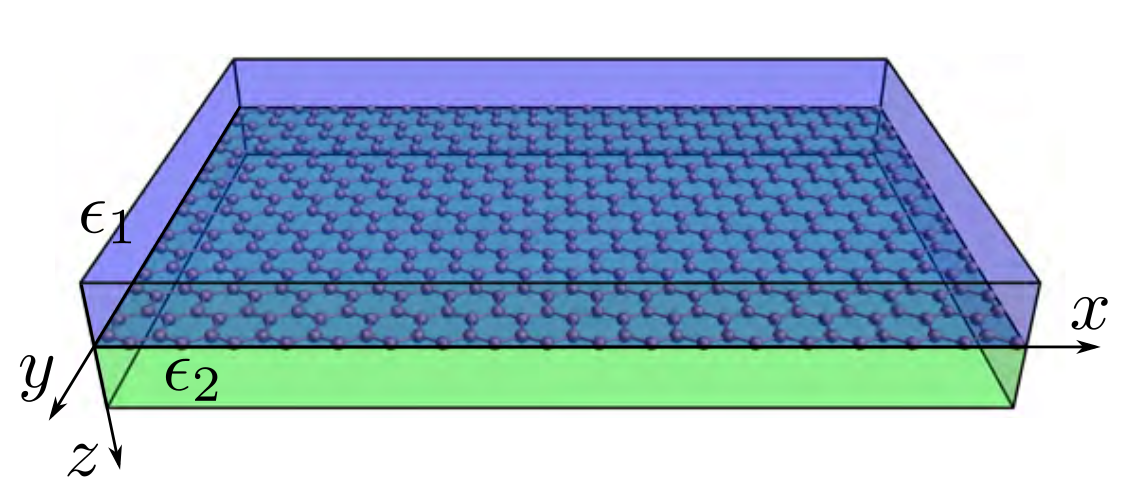}\caption[Illustration of a single graphene sheet sandwiched between two semi-infinite
insulators.]{Illustration of a single graphene sheet sandwiched between two semi-infinite
insulators with relative permittivities $\epsilon_{i}\equiv\varepsilon_{i}^{r}$
(in our notation). Medium 1 occupies the $z<0$ half-space and medium
2 the $z>0$; the graphene sheet is located at the $z=0$ plane. Source:
Ref. \citep{Goncalves2016}.}
\label{fig:systemGSPPs_monolayer} 
\end{figure}

For this derivation, we will closely follow Ref. \citep{Goncalves2016}.
We consider a system consisting of a single graphene sheet clad between
two semi-infinite dielectric media, characterized by the real dielectric
constants (relative permittivities) $\varepsilon_{1}^{r}$ and $\varepsilon_{2}^{r}$,
as depicted in Fig. \ref{fig:systemGSPPs_monolayer}. We stress that,
although the tBLG is not truly a 2D surface, its thickness is still
negligible and we can view it as a monolayer for these purposes \footnote{Typically, the 2D nature is still predominant for less than 10 layers
\citep{GeimNovoselov2007}.}.

Let us assume a solution of Maxwell's equations in the form of a transverse
magnetic (TM) wave. We use the following ansatz for the electric and
magnetic fields in the medium $j=1,2$: 
\begin{equation}
\bm{E}_{j}=(E_{j,x}\bm{\hat{x}}+E_{j,z}\bm{\hat{z}})e^{iqx}e^{-\kappa_{j}|z|},\quad\bm{B}_{j}=B_{j,y}e^{iqx}e^{-\kappa_{j}|z|}\bm{\hat{y}}.\label{eq:TMfields}
\end{equation}
This ansatz describes an electromagnetic wave (TM mode) which is confined
to the neighborhood of the graphene sheet (with damping parameter
$\kappa_{j}$ such that $\text{Re}\{\kappa_{j}\}>0$) and propagates
along the $x$ direction. Due to translational invariance symmetry,
the linear momentum along the propagation direction must be conserved,
enabling us to write $q\equiv q_{1}=q_{2}$, where $q_{1/2}$ is the
momentum of the electromagnetic wave propagating in medium $1/2$.
Moreover, we note that we are just writing the spatial components
of the fields; the time dependency, in what follows, is assumed to
be of the typical harmonic form, i.e., $e^{-i\omega t}$.

We now make use of Maxwell's equations. For each one of the media,
Faraday's law of induction and Ampère's law read, respectively, 
\begin{equation}
\bm{\nabla}\times\bm{E}_{j}=-\frac{\partial\bm{B}_{j}}{\partial t},\label{eq:Faradaylaw}
\end{equation}
\begin{equation}
\bm{\nabla}\times\bm{H}_{j}=\bm{J}_{j}^{f}+\frac{\partial\bm{D}_{j}}{\partial t}.
\end{equation}
Considering isotropic linear dielectric media, we can write the electric
displacement as $\bm{D}_{j}=\varepsilon_{0}\varepsilon_{j}^{r}\bm{E}_{j}$,
where $\varepsilon_{0}$ is the vacuum permittivity. Assuming isotropic
linear magnetic media with unitary relative permeability, we may also
write the magnetic field strength as $\bm{H}_{j}=\frac{\bm{B}_{j}}{\mu_{0}}$,
where $\mu_{0}$ is the vacuum permeability. Finally, if the free
current density is zero, $\bm{J}_{j}^{f}=\bm{0}$, we rewrite Ampère's
law as 
\begin{equation}
\bm{\nabla}\times\bm{B}_{j}=\frac{\varepsilon_{j}^{r}}{c^{2}}\frac{\partial\bm{E}_{j}}{\partial t},\label{eq:Amperelaw}
\end{equation}
where $c=1/\sqrt{\mu_{0}\varepsilon_{0}}$ is the speed of light in
vacuum. Introducing the fields given by Eq. (\ref{eq:TMfields}) into
Eqs. (\ref{eq:Faradaylaw}) and (\ref{eq:Amperelaw}), we obtain the
following useful relations: 
\begin{equation}
-\sgn(z)\kappa_{j}E_{j,x}-iqE_{j,z}=i\omega B_{j,y},
\end{equation}
\begin{equation}
\sgn(z)\kappa_{j}B_{j,y}=-i\omega\frac{\varepsilon_{j}^{r}}{c^{2}}E_{j,x},
\end{equation}
\begin{equation}
iqB_{j,y}=-i\omega\frac{\varepsilon_{j}^{r}}{c^{2}}E_{j,z}.
\end{equation}
From these, we can deduce 
\begin{equation}
E_{j,x}=i\sgn(z)\frac{\kappa_{j}c^{2}}{\omega\varepsilon_{j}^{r}}B_{j,y},\label{eq:Ejx}
\end{equation}
\begin{equation}
E_{j,z}=-\frac{qc^{2}}{\omega\varepsilon_{j}^{r}}B_{j,y},
\end{equation}
\begin{equation}
\kappa_{j}^{2}=q^{2}-\frac{\omega^{2}\varepsilon_{j}^{r}}{c^{2}}.
\end{equation}

Within the linear response regime, the boundary conditions linking
the electromagnetic fields at $z=0$ read as
\begin{equation}
E_{1,x}(x,z=0)=E_{2,x}(x,z=0),\label{eq:boundaryE}
\end{equation}
\begin{equation}
B_{1,y}(x,z=0)-B_{2,y}(x,z=0)=\mu_{0}J_{x}(x)=\mu_{0}\sigma_{xx}E_{2,x}(x,z=0),\label{eq:boundaryB}
\end{equation}
which assure the continuity of the tangential component of the electric
field across the interface and relate the discontinuity of the tangential
component of the magnetic field to the surface current density. We
emphasize that the conductivity of graphene is taken into account
in the boundary conditions only. For unstrained graphene (and for
the systems in focus), the conductivity is isotropic and frequency-dependent
and so we write $\sigma(\omega)\equiv\sigma_{xx}=\sigma_{yy}$. From
Eqs. (\ref{eq:boundaryE}) and (\ref{eq:Ejx}), we get 
\begin{equation}
B_{1,y}=-\frac{\kappa_{2}}{\kappa_{1}}\frac{\varepsilon_{1}^{r}}{\varepsilon_{2}^{r}}B_{2,y},
\end{equation}
which we insert into Eq. (\ref{eq:boundaryB}) to obtain 
\begin{equation}
\frac{\varepsilon_{1}^{r}}{\kappa_{1}(q,\omega)}+\frac{\varepsilon_{2}^{r}}{\kappa_{2}(q,\omega)}+i\frac{\sigma(\omega)}{\omega\varepsilon_{0}}=0.\label{eq:TMdispersionrelation}
\end{equation}
This last equation describes the dispersion relation, $\omega(q)$,
of TM SPPs. Notice that this is an implicit equation, so it needs
to be solved numerically. Nonetheless, we can see that it can only
have solutions when $\text{Im}\left\{ \sigma(\omega)\right\} >0$.

\subsubsection{Results for single layer graphene}

\label{subsection:spectumGSPPs_SLG}

\begin{figure}
\centering{}\includegraphics[width=0.4\textwidth]{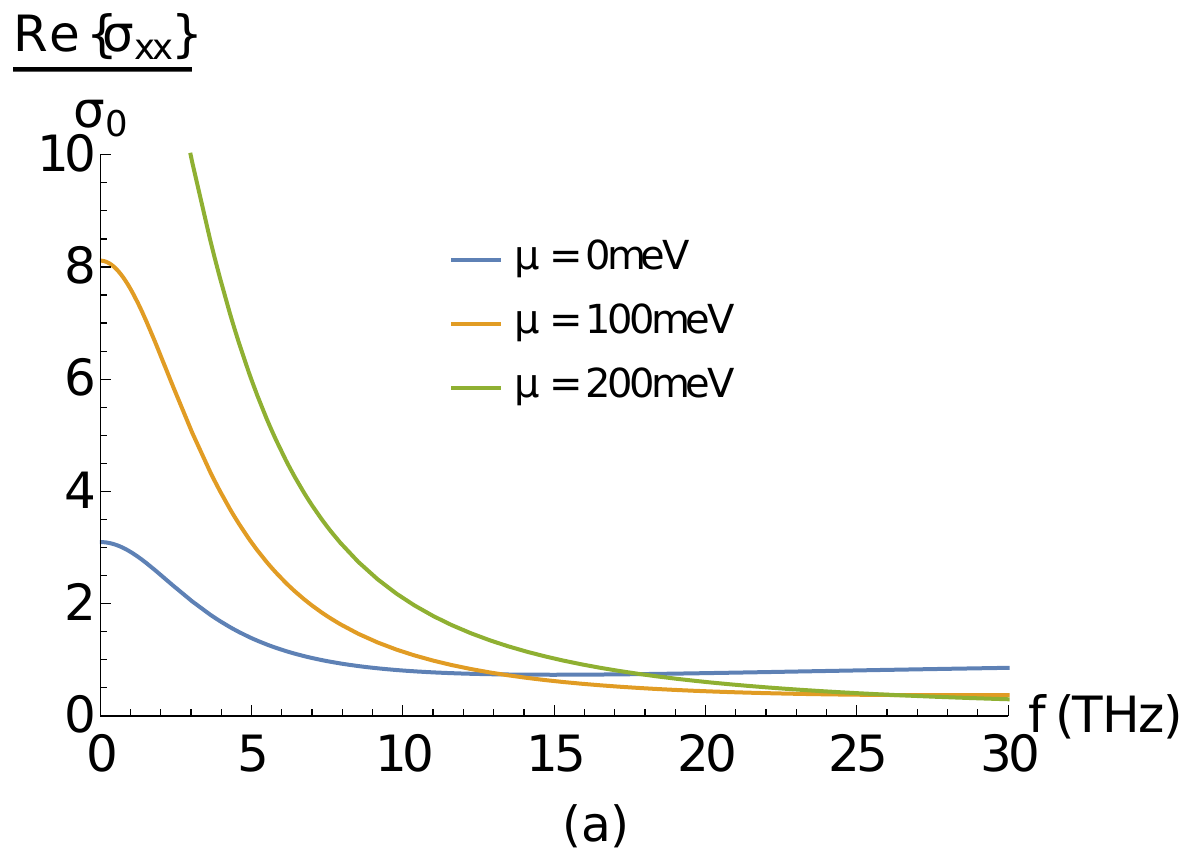}~~~~~\includegraphics[width=0.4\textwidth]{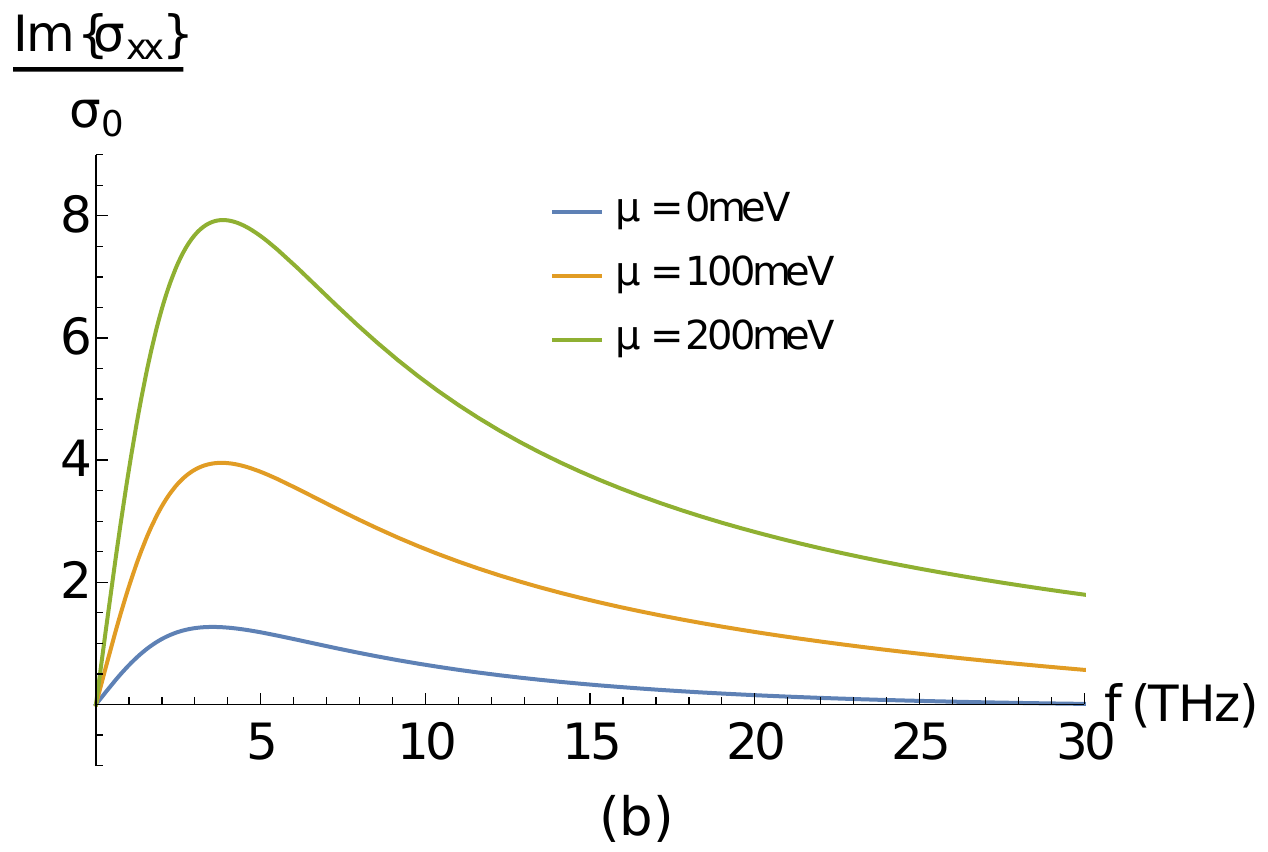}\caption{Total conductivity in SLG: (a) real part; (b) imaginary part.}
\label{fig:SLGtotalcond_variousEf} 
\end{figure}

\begin{figure}
\centering{}\includegraphics[width=0.7\textwidth]{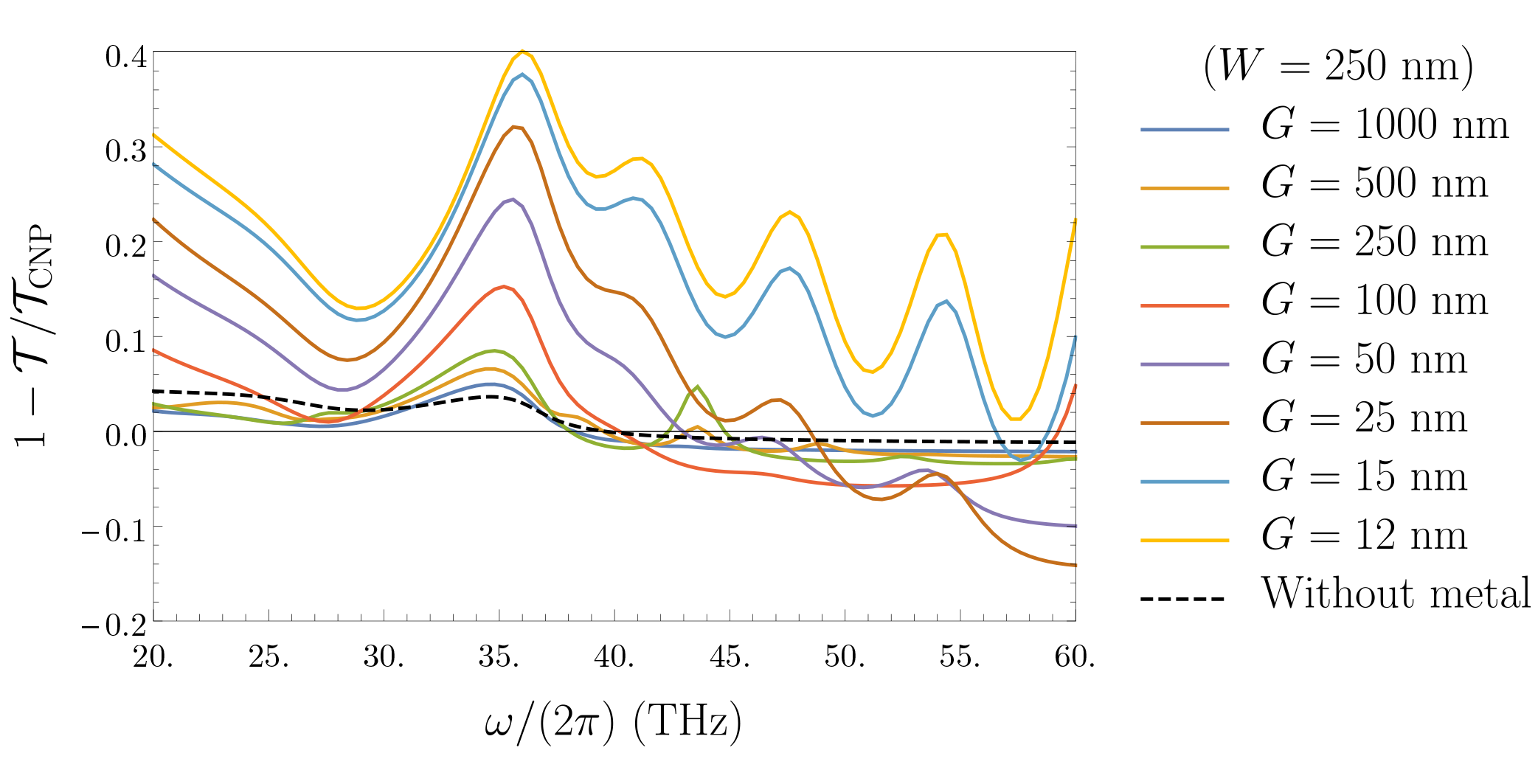}
\caption{Evidence of surface optical phonons arising from the $\text{SiO}_{2}$
substrate. In this plot, the quantity in focus is the optical transmittance,
$\mathcal{T}$, with $\mathcal{T}_{\text{CNP}}$ being the optical
transmittance at the charge neutrality point. This quantity was obtained
within a theoretical calculation, which takes into account the full
dielectric function of the $\text{SiO}_{2}$. The parameters $W$
and $G$ correspond to $a$ and $d-a$, respectively, in the scheme
of Fig. \ref{fig:GSPPs_scheme}. The peak that starts at $f\sim30\,\text{THz}$
was interpreted as a contribution from the $\text{SiO}_{2}$ optical
phonons. Source: Ref. \citep{EduardoDias_tese}.}
\label{fig:plasmons_SiO2} 
\end{figure}

In Fig. \ref{fig:SLGtotalcond_variousEf}, we present our results
for the total conductivity (Drude plus regular terms) in SLG, as a
function of the frequency, $f=\omega/(2\pi)$, across the spectral
region where we are interested to study the spectrum of graphene SPPs
—from the THz up to the mid-infrared. We recall that we have set $\Gamma=16\si{\milli\electronvolt}$;
moreover, in the following results, we will always be considering
room temperature, $T=300\si{\kelvin}$. We also stress that we avoided
exceeding frequencies of $30\si{\tera\hertz}$ because of the surface
polar phonons that arise from the $\text{SiO}_{2}$ —the typical substrate
used as medium 2—, according to Ref. \citep{EduardoDias_tese} (Fig.
\ref{fig:plasmons_SiO2}).

\begin{figure}
\centering{}\includegraphics[width=0.4\textwidth]{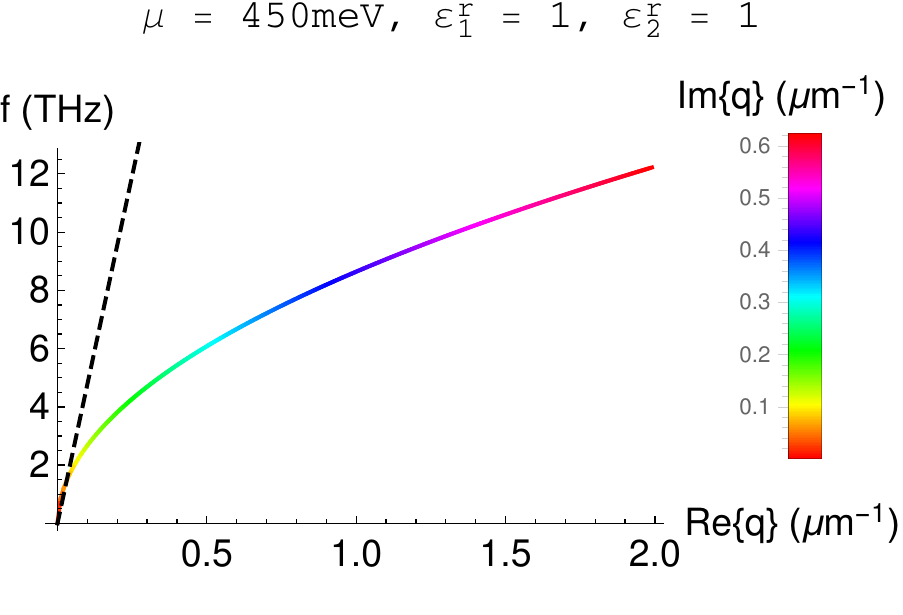}
\caption{Dispersion relation of TM SPPs in SLG. The dashed line corresponds
to the light dispersion, $\omega=cq$, where $c$ is the speed of
the light in the medium (in this case, air).}
\label{fig:SLGTM_Ef450meV} 
\end{figure}

Given the total conductivity at a given Fermi level, we can obtain
the dispersion curve by solving Eq. (\ref{eq:TMdispersionrelation})
numerically. Notice that, if we consider only the Drude contribution
with $\Gamma=0$, we have a pure imaginary conductivity and we can
solve this equation with real $q$. If not, we have to consider a
complex-valued wave vector, whose imaginary part characterizes the
attenuation of the SPPs \citep{BludovFerreiraPeresEtAl2013,Goncalves2016}.
In Fig. \ref{fig:SLGTM_Ef450meV}, we present the spectrum of SPPs
in SLG for $\mu=450\si{\milli\electronvolt}$, which we obtained by
taking into account the total conductivity. This curve is in agreement
with the results obtained in Ref. \citep{Goncalves2016}, namely with
Fig. 4.2, where the authors considered only the Drude contribution
(which is the dominant term in this case) with no absorption ($\Gamma=0$),
and Fig. 4.3, where they verified that the consideration of absorption
($\Gamma\neq0$) only affects the spectrum in the region of low wave
vectors. Analyzing the spectrum, we see that the dispersion curve
lies to the right of the light line, which indicates, as we have mentioned
before, that we cannot excite graphene SPPs simply by directly shining
electromagnetic radiation \footnote{In fact, if we look closely, we see that the dispersion curve crosses
the light line at some point (this happens only because we are considering
a non-zero $\Gamma$, which is the more realistic situation). However,
this point falls within the overdamped regime, $\omega_{SPP}/\gamma<1$,
in which SPPs cannot be sustained \citep{Goncalves2016}.}. It is now clear why we need to use a setup with a periodic grid,
like the one described in section \ref{chapter:opticalresponse} (see
Fig. \ref{fig:GSPPs_scheme}).

\begin{figure}
\centering{}\includegraphics[width=0.4\textwidth]{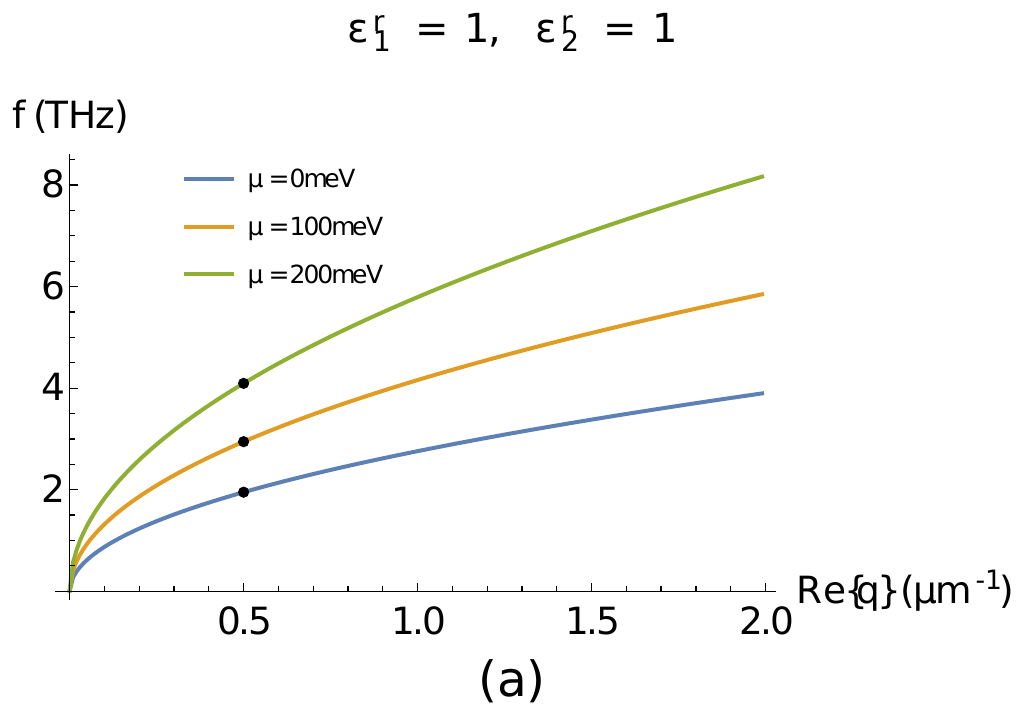}~~~~~\includegraphics[width=0.4\textwidth]{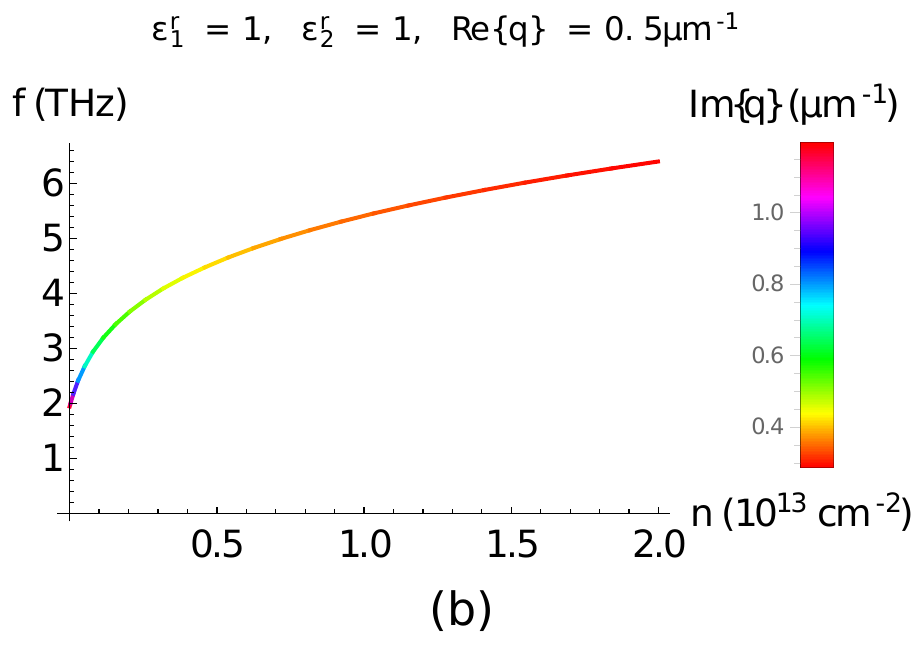}
\caption{Spectrum of TM SPPs in SLG: dependency on the Fermi level/carrier
density. Panel (a) schematically shows the procedure used to obtain
panel (b).}
\label{fig:SLGTM_variousEfandro} 
\end{figure}

At last, we can fix a wave vector —physically, if we take the light
line as roughly vertical, this corresponds to fixing a gap in the
periodic grid— and study the dependency of the dispersion curves on
the Fermi level (or the carrier density). The results are shown in
Fig. \ref{fig:SLGTM_variousEfandro}. We stress that we do not obtain
$f(n=0)=0$ in Fig. \ref{fig:SLGTM_variousEfandro}(a) because of
the finite temperature.

\subsubsection{Results for twisted bilayer graphene}

\label{subsection:spectumGSPPs_tBLG}

\begin{figure}
\centering{}\includegraphics[width=0.4\textwidth]{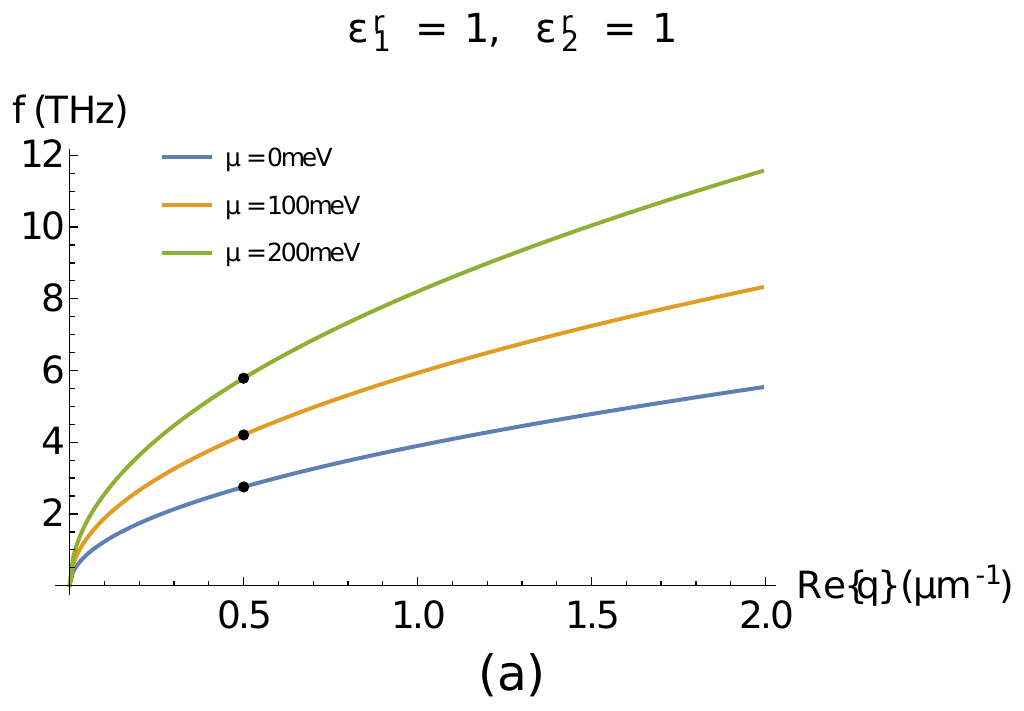}~~~~~\includegraphics[width=0.4\textwidth]{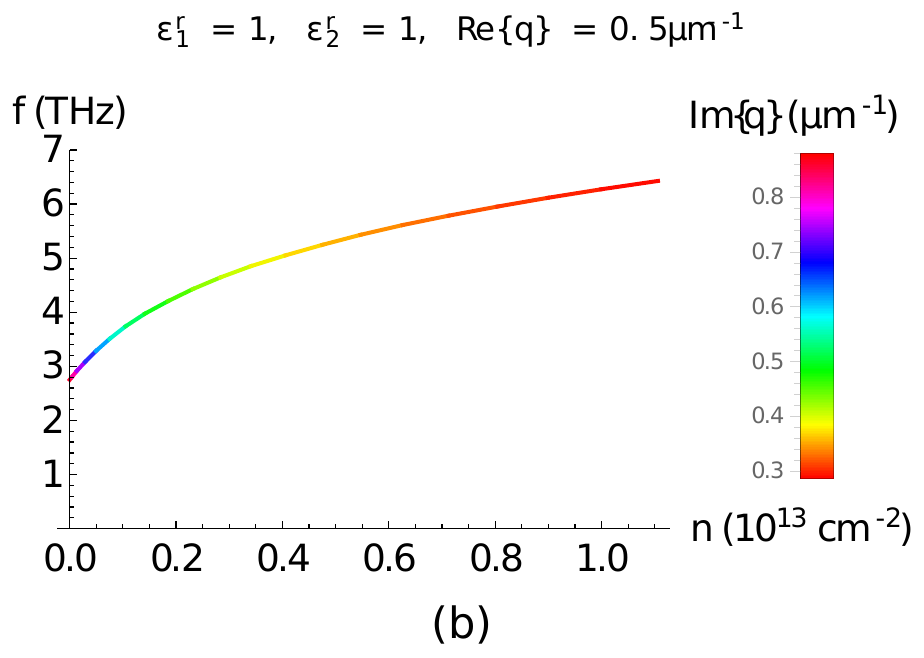}\caption{Spectrum of TM SPPs in tBLG with $\theta=9^{\circ}$: dependency on
the Fermi level/carrier density. The black dots in (a) mark the fixed
parameter in (b).}
\label{fig:tBLG9degTM_variousEfandro}
\end{figure}

For the tBLG, we repeat the previous analysis, namely the last 2 plots
from Fig. \ref{fig:SLGTM_variousEfandro}, for two different twist
angles. We start with $\theta=9^{\circ}$ (Fig. \ref{fig:tBLG9degTM_variousEfandro}).
For this angle, we see that the signature of the curves do not differ
a lot from those of the SLG. This happens due to two main reasons: 
\begin{itemize}
\item Within this range of frequency, $f\lesssim30\si{\tera\hertz}\Leftrightarrow\hbar\omega\lesssim124\si{\milli\electronvolt}$,
the regular conductivity is basically twice the value obtained for
the SLG (see Fig. \ref{fig:tBLG9deg_regcondandspectrum}(a)). Moreover,
for $\mu\lesssim250\si{\milli\electronvolt}$, the Drude weight is
also twice the value obtained for the SLG (see Fig. \ref{fig:tBLGDrude2}(a)).
Therefore, we do not capture any hybridization effect and we just
recover the total conductivity of a decoupled BLG.
\item As we can see in Fig. \ref{fig:tBLGDOSandcarrierdensity_variousdeg}(b),
the curves $n(\mu)$ for $\theta=9^{\circ}$ and for decoupled BLG
are also very close within the range in focus, $\mu\lesssim250\si{\milli\electronvolt}\Leftrightarrow n\lesssim1\times10^{13}\si{\per\centi\meter\squared}$.
\end{itemize}

\begin{figure}
\centering{}\includegraphics[width=0.4\textwidth]{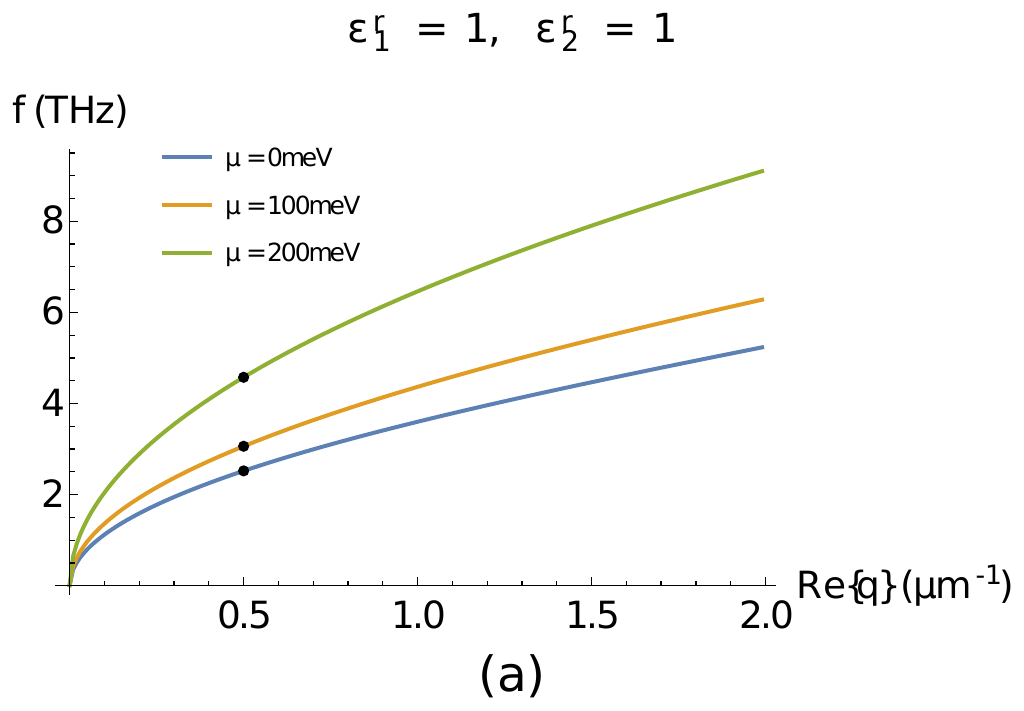}~~~~~\includegraphics[width=0.4\textwidth]{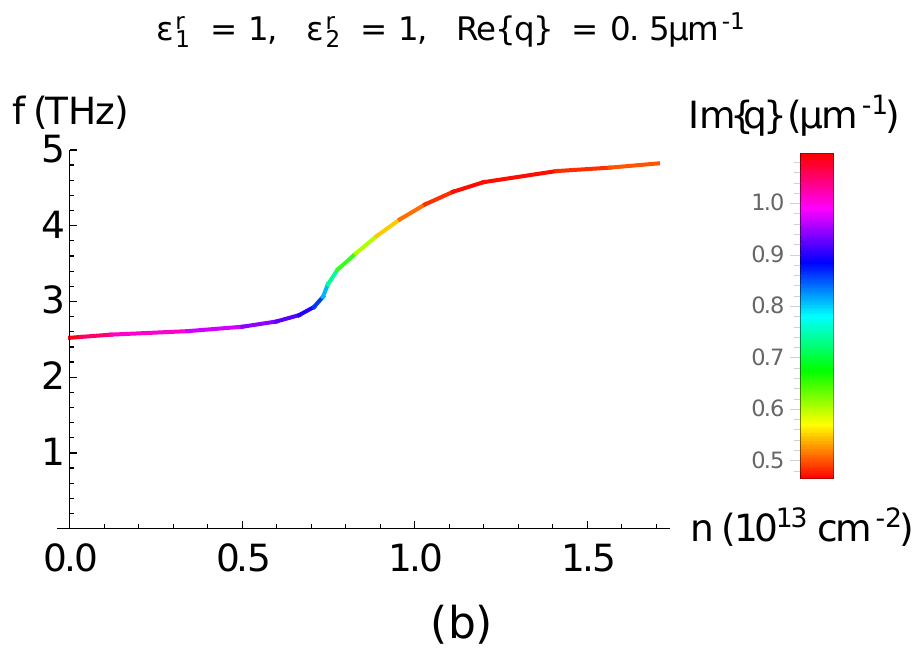}\caption{Spectrum of TM SPPs in tBLG with $\theta=1.8^{\circ}$: dependency
on the Fermi level/carrier density. The black dots in (a) mark the
fixed parameter in (b).}
\label{fig:tBLG1dot8degTM_variousEfandro}
\end{figure}

We move on to the angle $\theta=1.8^{\circ}$ (Fig. \ref{fig:tBLG1dot8degTM_variousEfandro}).
In this case, not only the total conductivity is different but also
the relation $n(\mu)$ changes drastically. This leads to the plot
of Fig. \ref{fig:tBLG1dot8degTM_variousEfandro}(b), which we highlight
since it is totally different from all the results obtained before.
As an immediate application, we can think of using these results as
an alternative method for determining the twist angle. Nevertheless,
a more extensive study on the behavior of these curves with the variation
of $\theta$ remains to be done, in order to investigate more promising
applications.

As future work, we shall also consider a frequency range for which
the big deep on $\text{Im}\left\{ \sigma^{reg}(\omega)\right\} $
of tBLG (seen in Fig. \ref{fig:tBLGregcond_5degandvariousdeg}(a)
for $\theta=5^{\circ}$) plays a role. By varying the twist angle,
this deep can be brought to arbitrary low frequencies. However, at
some point, the contribution from the Drude conductivity will dominate.
In between, we can find a regime for which the total conductivity
has a negative imaginary part. In this case, Eq. \eqref{eq:TMdispersionrelation}
has no solutions and therefore we cannot sustain TM modes. Instead,
we will have transverse electric modes, which were not addressed in
this chapter. With this, we can explore the use of tBLG as a polarizer
for waveguide modes.

%% file: sections/Section_Conclusions.tex
\section{Conclusions}

\label{chapter:conclusions}

In this chapter we provided a pedagogical introduction to the electronic
and optical properties of twisted bilayer graphene. We started, in
section~\ref{chapter:model}, with a brief description of the tight-binding
models for single layer graphene and perfectly aligned Bernal-stacked
bilayer graphene. We explicitly showed how the electronic bands of
single layer graphene fold when larger than the minimal unit cells
are considered, and how to obtain the corresponding Hamiltonian directly
in reciprocal space. From this, we moved to the description of twisted
bilayer graphene in section~\ref{chapter:tBLG}. We saw how the interference
between the periodicities of the two misaligned layers leads to the
emergence of a moiré pattern. We also saw how the properties of the
electrons moving in this moiré pattern can be described starting from
a tight-binding Hamiltonian, expressing the electronic wave function
in terms of Bloch-waves of each layer and considering generalized
umklapp processes. With this machinery, we studied how the interlayer
coupling leads to a twist-angle-dependent renormalization of the Fermi
velocity, in the weak coupling limit. We also showed how to describe
the electronic spectrum beyond the weak coupling limit, by including
a greater number of generalized umklapp processes. With the spectrum
reconstruction established, we computed the profiles of the density
of states and the carrier density. We observed that the electronic
spectrum is strongly modified by varying the twist angle, namely by
bringing van Hove singularities to lower energies and thus making
them easily accessible via electrostatic doping. Having established
the electronic spectrum of the twisted bilayer graphene, we addressed
its optical properties in section~\ref{chapter:opticalresponse}.
Within the linear response theory, we presented a derivation of general
tight-binding-based expressions to compute both the regular and the
Drude contributions to the homogeneous optical conductivity, which
are suitable for low-energy effective models. Applying these expressions
to the twisted bilayer graphene, we observed that the conductivity
profiles can be drastically modified by varying the twist angle. We
then used these results to study the dispersion relation of surface
plasmon-polaritons in twisted bilayer graphene.

We expect that the research in twisted bilayer graphene will continue
to reveal new and interesting physics, with potential applications,
and that these findings will further guide research into other kinds
of van der Waals structures.

At the time of writing, electron-electron interactions in twisted
bilayer graphene were an underdeveloped topic. This suddenly changed
after the experimental findings of correlated insulator behavior \cite{Cao2018a}
and superconductivity \cite{Cao2018} in doped magic-angle twisted
bilayer graphene.
\begin{acknowledgments}
B. A. received funding from the European Union's Horizon 2020 research
and innovation programme under grant agreement No. 706538.
\end{acknowledgments}